%% file: thesis.tex
\documentclass[11pt,a4paper,oneside]{memoir}

\usepackage[T1]{fontenc}

\usepackage[english]{babel}

\usepackage[utf8]{inputenc}

\usepackage{tikz}
\usetikzlibrary{positioning, arrows.meta}

\usepackage{amsmath,amssymb,amsfonts,mathrsfs}

\usepackage[amsmath,thmmarks]{ntheorem}

\usepackage{graphicx}

\usepackage{soul}

\usepackage{pdfpages}
\usepackage[mathscr]{eucal}

\input{extrapackages}

\input{layoutsetup}

\input{theoremsetup}

\input{macrosetup}

\input{commands}

\definecolor{citeForestGreen}{rgb}{0.2, 0.5, 0.2}    

\usepackage[linkcolor=citeForestGreen,colorlinks=true,citecolor=citeForestGreen,filecolor=black]{hyperref}  

\newtagform{green}[\color{citeForestGreen}]{(}{)}
\usetagform{green}

\usepackage{multirow}

\usepackage{cleveref}

\crefname{theorem}{theorem}{theorems}
\Crefname{theorem}{Theorem}{Theorems}

\crefname{lemma}{lemma}{lemmas}
\Crefname{lemma}{Lemma}{Lemmas}

\crefname{corollary}{corollary}{corollaries}
\Crefname{corollary}{Corollary}{Corollaries}

\crefname{proposition}{proposition}{propositions}
\Crefname{proposition}{Proposition}{Propositions}

\crefname{claim}{claim}{claims}
\Crefname{claim}{Claim}{Claims}

\crefname{definition}{definition}{definitions}
\Crefname{definition}{Definition}{Definitions}

\crefname{example}{example}{examples}
\Crefname{example}{Example}{Examples}

\crefname{remark}{remark}{remarks}
\Crefname{remark}{Remark}{Remarks}

\crefname{Rule}{rule}{rules}
\Crefname{Rule}{Rule}{Rules}

\usepackage{thmtools}
\usepackage{thm-restate}

\usepackage{algorithm}        
\usepackage{algorithmicx}
\usepackage{algpseudocode}

\title{Self-Confirming Equilibrium under Model Misspecification}
\author{Riccardo Formenti}
\thesistype{Master Thesis}
\advisors{Prof.\ Simone Cerreia-Vioglio (Università Bocconi)\\
          Prof.\ Dylan Possamaï (ETH Zürich)}
\department{Department of Mathematics}
\date{Tuesday 4\textsuperscript{th} August, 2026}

\begin{document}

\frontmatter

\begin{titlingpage}
  \calccentering{\unitlength}
  \begin{adjustwidth*}{\unitlength-24pt}{-\unitlength-24pt}
    \maketitle
  \end{adjustwidth*}
\end{titlingpage}

\input{abstract.tex}

\chapter*{Acknowledgments}
\input{acknowledgments.tex}

\pagebreak

\cleartorecto

{
\hypersetup{linkcolor=black}
\tableofcontents
}
\mainmatter

\input{sections_manager}

\backmatter

\bibliographystyle{plainnat-mine.bst}
\bibliography{refs}

\appendix
\setcounter{chapter}{1}
\input{Chapters/appendix_contribution}

\setcounter{chapter}{2}

\input{Chapters/appendix_newtheorems_proofs}

\setcounter{chapter}{3}

\input{Chapters/appendix_uncertainty_aversion}

\includepdf[pages={-}]{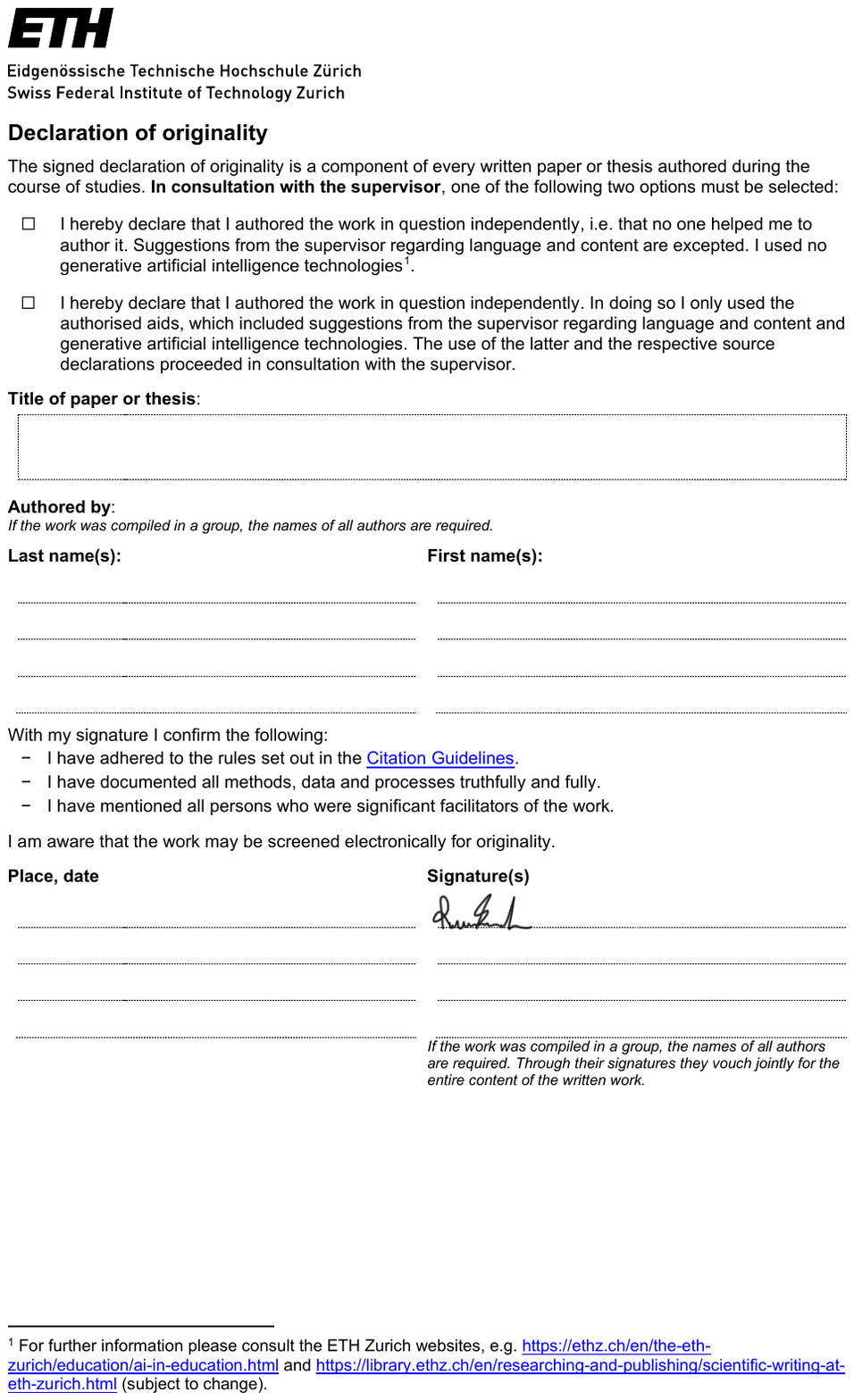}

\end{document}

%% file: extrapackages.tex
\usepackage{tikz}
\usetikzlibrary{arrows.meta,positioning,fit,calc,backgrounds}

\usepackage{datetime}

\usepackage{mathtools}

\usepackage{array}

\usepackage{listings}
\usepackage{microtype}

\usepackage{booktabs}

\usepackage{mleftright}

\usepackage{xspace}

\usepackage{xcolor}

\usepackage[authoryear,longnamesfirst, round]{natbib}

%% file: layoutsetup.tex
\usepackage{ETHlogo}

\nonzeroparskip
\defaultlists

\makeatletter

\if@twoside
  \copypagestyle{chapter}{Ruled}
\else
  \copypagestyle{chapter}{ruled}
\fi
\makeoddhead{chapter}{}{}{}
\makeevenhead{chapter}{}{}{}
\makeheadrule{chapter}{\textwidth}{0pt}
\copypagestyle{abstract}{empty}

\makechapterstyle{bianchimod}{%
  \chapterstyle{default}
  \renewcommand*{\chapnamefont}{\normalfont\Large\sffamily}
  
  \renewcommand*{\printchaptername}{%
    \chapnamefont\centering\@chapapp}

  }

\chapterstyle{bianchimod}

\setsecheadstyle{\Large\bfseries\sffamily}
\setsubsecheadstyle{\large\bfseries\sffamily}
\setsubsubsecheadstyle{\bfseries\sffamily}
\setparaheadstyle{\normalsize\bfseries\sffamily}
\setsubparaheadstyle{\normalsize\itshape\sffamily}
\setsubparaindent{0pt}

\captionnamefont{\rmfamily\bfseries\footnotesize}
\captiontitlefont{\rmfamily\footnotesize}

\setsecnumdepth{subsection}
\settocdepth{subsection}

\pretitle{\vspace{0pt plus 0.7fill}\begin{center}\HUGE\sffamily\bfseries}
\posttitle{\end{center}\par}
\preauthor{\par\begin{center}\let\and\\\Large\sffamily}
\postauthor{\end{center}}
\predate{\par\begin{center}\Large\sffamily}
\postdate{\end{center}}

\def\@advisors{}
\newcommand{\advisors}[1]{\def\@advisors{#1}}
\def\@department{}
\newcommand{\department}[1]{\def\@department{#1}}
\def\@thesistype{}
\newcommand{\thesistype}[1]{\def\@thesistype{#1}}

\renewcommand{\maketitlehookb}{\vspace{1in}%
  \par\begin{center}\Large\sffamily\@thesistype\end{center}}

\renewcommand{\maketitlehookd}{%
  \vfill\par
  \begin{flushright}
    \sffamily
    \@advisors\par
    \@department, ETH Z\"urich
  \end{flushright}
}

\checkandfixthelayout

\makeatother

\theoremstyle{plain}


%% file: theoremsetup.tex
\numberwithin{equation}{chapter}

\theoremseparator{.}
\newtheorem{theorem}{Theorem}[chapter]
\newtheorem{example}[theorem]{Example}

\newtheorem{corollary}[theorem]{Corollary}
\newtheorem{definition}[theorem]{Definition}
\newtheorem{lemma}[theorem]{Lemma}

\theoremstyle{nonumberplain}
\theorembodyfont{\normalfont}
\theoremsymbol{\ensuremath{\square}}
\newtheorem{proof}{Proof}

%% file: macrosetup.tex
\DeclarePairedDelimiter{\Paren}{(}{)}
\DeclarePairedDelimiter{\Bracket}{[}{]}
\DeclarePairedDelimiter{\Brace}{\{}{\}}

\NewDocumentCommand{\Oh}{som}{%
  \ensuremath{\mathcal{O}\IfBooleanTF{#1}{%
    \Paren*{#3}
  }{%
    \IfNoValueTF{#2}{%
      \Paren{#3}
    }{%
      \Paren[#2]{#3}
    }%
  }%
}}

\NewDocumentCommand{\Om}{som}{%
\ensuremath{\Omega\IfBooleanTF{#1}{%
    \Paren*{#3}%
  }{%
    \IfNoValueTF{#2}{%
      \Paren{#3}%
    }{%
      \Paren[#2]{#3}%
    }%
  }%
}}

\NewDocumentCommand{\Th}{som}{%
\ensuremath{\Theta\IfBooleanTF{#1}{%
    \Paren*{#3}%
  }{%
    \IfNoValueTF{#2}{%
      \Paren{#3}%
    }{%
      \Paren[#2]{#3}%
    }%
  }%
}}

\NewDocumentCommand{\oh}{som}{%
\ensuremath{o\IfBooleanTF{#1}{%
    \Paren*{#3}%
  }{%
    \IfNoValueTF{#2}{%
      \Paren{#3}%
    }{%
      \Paren[#2]{#3}%
    }%
  }%
}}

\NewDocumentCommand{\om}{som}{%
\ensuremath{\omega\IfBooleanTF{#1}{%
    \Paren*{#3}%
  }{%
    \IfNoValueTF{#2}{%
      \Paren{#3}%
    }{%
      \Paren[#2]{#3}%
    }%
  }%
}}

\NewDocumentCommand{\Prob}{som}{%
\ensuremath{\mathbb{P}
  \IfBooleanTF{#1}{%
    \Bracket*{#3}%
  }{%
    \IfNoValueTF{#2}{%
      \Bracket{#3}%
    }{%
      \Bracket[#2]{#3}%
    }%
  }%
}}

\NewDocumentCommand{\E}{som}{%
\ensuremath{\mathbb{E}
  \IfBooleanTF{#1}{%
    \Bracket*{#3}%
  }{%
    \IfNoValueTF{#2}{%
      \Bracket{#3}%
    }{%
      \Bracket[#2]{#3}%
    }%
  }%
}}

\NewDocumentCommand{\Max}{s o m m}{%
\ensuremath{\max
  \IfBooleanTF{#1}{%
    \Brace*{#3,\, #4}%
  }{%
    \IfNoValueTF{#2}{%
      \Brace{#3,\, #4}%
    }{%
      \Brace[#2]{#3,\, #4}
    }%
  }%
}}

\NewDocumentCommand{\Min}{s o m m}{%
\ensuremath{\min
  \IfBooleanTF{#1}{%
    \Brace*{#3,\, #4}%
  }{%
    \IfNoValueTF{#2}{%
      \Brace{#3,\, #4}%
    }{%
      \Brace[#2]{#3,\, #4}
    }%
  }%
}}

\NewDocumentCommand{\Exp}{som}{%
\ensuremath{\exp
  \IfBooleanTF{#1}{%
    \Paren*{#3}%
  }{%
    \IfNoValueTF{#2}{%
      \Paren{#3}%
    }{%
      \Paren[#2]{#3}%
    }%
  }%
}}

\NewDocumentCommand{\Log}{som}{%
\ensuremath{\log
  \IfBooleanTF{#1}{%
    \Paren*{#3}%
  }{%
    \IfNoValueTF{#2}{%
      \Paren{#3}%
    }{%
      \Paren[#2]{#3}%
    }%
  }%
}}

\NewDocumentCommand{\Var}{som}{%
\ensuremath{\mathrm{Var}
  \IfBooleanTF{#1}{%
    \Bracket*{#3}%
  }{%
    \IfNoValueTF{#2}{%
      \Bracket{#3}%
    }{%
      \Bracket[#2]{#3}%
    }%
  }%
}}

\NewDocumentCommand{\SetBuilder}{s o m m}{%
\ensuremath{\IfBooleanTF{#1}{%
    \Brace*{\,#3 : #4\,}%
  }{%
    \IfNoValueTF{#2}{%
      \Brace{\,#3 : #4\,}%
    }{%
      \Brace[#2]{\,#3 : #4\,}%
    }%
  }%
}}

\definecolor{darkred}{rgb}{0.7, 0, 0}

%% file: commands.tex
\newcommand{\R}{\ensuremath{\mathbb{R}}}

\newcommand{\mi}{\ensuremath{_{-i}}}

\usepackage[multiple]{footmisc}

\newcommand{\RNE}{\ensuremath{\text{\textnormal{R-NE}}}}
\newcommand{\SCEM}{\ensuremath{\text{\textnormal{SCEM}}}}

%% file: abstract.tex
\begin{abstract}

We study the steady states of games played recurrently within large anonymous populations of players who receive partial ex-post feedback about the play of their opponents. Departing from the classical Bayesian account, players do not trust  their conjectures as a correct description of the strategic environment but consider them as approximations of a reality that is only partially modeled and partially observed. Building on the axiomatization of decision making under model misspecification of \citet{cerreia2026making}, we introduce games with feedback and uncertainty attitudes, in which the normal-form structure is enriched with a profile of feedback functions and, for each player, a preference functional that encodes aversion to misspecification and to ambiguity. The induced misspecification sensitive criterion contains the variational, maxmin, and smooth-ambiguity evaluations as special cases and collapses to expected utility when beliefs are unambiguous and the player is neutral to misspecification. The appropriate steady-state notion for such games, which we introduce, is the Self-Confirming Equilibrium under Model Misspecification (SCEM), inspired by the Self-Confirming Equilibrium of \citet{battigalli1987thesis} and \citet{fudenberg1993self}, which requires each player to best respond to a set of structured conjectures that the feedback generated by equilibrium play cannot falsify. In contrast with the classical theory and with its ambiguity-averse extensions, mixed-strategy Nash Equilibria need not be self-confirming. We therefore develop a companion notion of Robust Nash Equilibrium (R-NE), in which every supported action remains optimal under a variational evaluation anchored at the correct reference conjecture. Robust Nash equilibria refine SCEM, exist whenever the statistical distances are continuous, thereby delivering existence of SCEM, and coincide with the latter under fully informative feedback. Two comparative-statics exercises follow: more informative feedback shrinks the set of SCEM action profiles, whereas monotonicity with respect to uncertainty aversion fails in general. Finally, we identify two additional assumptions, one behavioral (observationally consistent divergences) and one informational (observable payoffs), under which the notions of R-NE and SCEM reduce respectively to Nash Equilibrium and to Self-Confirming Equilibrium under Ambiguity (SCEA), a generalization of the smooth self-confirming equilibrium of \citet{battigalli2015self}, thus reconnecting SCEM to the literature.

\end{abstract}

%% file: acknowledgments.tex
I would like to take this space to thank all the people who made this work possible, my parents in primis, Simona and Pericle, without whom I would be nothing but a speck of dust, my family, friends and partner, who supported me through the highs and lows, my advisors, Professor Cerreia-Vioglio and Professor Possamaï, whose ideas and conversations were insightful and guided me throughout, and all the great scholars of the past and present, named and unnamed, on whose shoulders I modestly stand today.

%% file: sections_manager.tex
\newpage

\input{Chapters/introduction}

\input{Chapters/contributions1}

%% file: Chapters/introduction.tex
\chapter{Introduction}\label{ch:paper_introduction}

This work studies the steady states of games played recurrently within large anonymous populations, when players observe only partial ex-post feedback about opponents' play and, aware that their conjectures are approximations of the strategic environment, are averse to their possible misspecification. 

The equilibrium notion appropriate to recurrent play under partial feedback is the self-confirming equilibrium, introduced independently by \citet{battigalli1987thesis}, under the name of conjectural equilibrium, and by \citet{fudenberg1993self}; in equilibrium, each player best responds to conjectures that are confirmed by, hence indistinguishable from, the feedback her own equilibrium play generates. \citet{battigalli2015self} extend the notion to ambiguity-averse players with their smooth self-confirming equilibrium (SSCE); their players, however, are not represented as entertaining the possibility that the posited set of conjectures fails to contain the truth. In decision theory, meanwhile, \citet{cerreia2026making} have axiomatized preferences sensitive to precisely this concern. The decision maker is endowed with a set of structured models she can defend on substantive grounds, but fears that none of them is correct, and evaluates acts by aggregating, across structured models, robust evaluations that penalize departures from each model through a statistical distance. Our contribution is as follows: we define games with feedback and uncertainty attitudes, propose self-confirming equilibrium under model misspecification (SCEM) as their equilibrium notion, establish its existence and its relations with the literature, and develop its comparative statics with respect to feedback informativeness and uncertainty aversion.

This thesis is structured as follows:
\begin{itemize}
  \item Section~\ref{sec:litrev} surveys the literature the present work builds upon, in particular the decision-theoretic treatment of uncertainty and its uses in game theory.
  \item Section~\ref{sec:math_definitions_chapter1} collects mathematical background and notation used throughout.
  \item Chapter~\ref{ch:contributions1} contains the original contribution of this work. It introduces games with feedback and uncertainty attitudes, defines SCEM, and establishes its existence via a companion notion of robust Nash equilibrium; it then develops the taxonomy relating SCEM to Nash equilibrium and two comparative-statics exercises, monotonicity with respect to feedback informativeness and a failure of monotonicity with respect to uncertainty aversion, and finally shows that, under observationally consistent divergences and observable payoffs, monotonicity is restored and the framework reduces to self-confirming equilibrium under ambiguity, reconnecting with SSCE.
  \item Appendices~\hyperref[ch:appendix_newtheorems]{A}, \hyperref[ch:appendix_contribution]{B}, \hyperref[ch:appendix_uncertainty]{C} contain respectively auxiliary results, the proofs omitted
  from the main text, and the decision-theoretic foundations of our
  notion of comparative uncertainty aversion in games.
\end{itemize}

\section{Uncertainty, from Decisions to Games}\label{sec:litrev}

This section surveys the two literatures whose intersection the present work occupies. We first trace the decision-theoretic treatment of uncertainty from the Bayesian paradigm of Savage through Ellsberg's paradox, to the modern axiomatizations of ambiguity and, most recently, of model misspecification. We then turn to strategic interaction, reviewing how non-Bayesian attitudes have been imported into games.

\subsection*{Ambiguity and Misspecification}

On the subjectivist reading, the probability assigned to a proposition is identified with a subjective \emph{degree of belief} in that proposition, an expression of the epistemic uncertainty of a decision maker (DM) reflecting her private information. What \citet{ramsey1926truth} and \citet{de1931sul} added to this picture is the methodological thesis that such degrees of belief are not directly accessible by introspection but can be operationally defined and made transparent only through the DM's willingness to act on them. What it is for one belief to be held more firmly than another must be understood in terms of how the DM would (or should) act on these beliefs; and the "old-established" way of eliciting such comparisons, in Ramsey's phrase, is through the DM's \emph{preferences} over uncertain wagers. 

This program faces an immediate difficulty. The DM's propensity to accept a bet depends as much on her beliefs about the bet-relevant contingencies as on her preferences over the wagered prizes; beliefs and preferences are, in fact, jointly manifested in choice behavior. Ramsey and de Finetti each resolve this identification problem by axiomatizing the elicitation of subjective probabilities under auxiliary assumptions on utility; working in the opposite direction, \citet{von1947theory} axiomatize expected utility while taking probabilities as objectively given. It is only with Savage's~\citeyearpar{savage1954} fundamental contribution, however, that the two approaches are unified in a coherent presentation in which beliefs and tastes are jointly and uniquely recovered from the DM's preferences over uncertain acts. The result is what would become the canonical formulation of the Bayesian paradigm; we refer the reader to the insightful work of \citet{gilboa2016ambiguity}, and the references therein, for further details on the history and philosophical interpretation of decision theory. 

Savage's work is built on three primitives. The first is a set $S$ of \emph{states of the world}, whose elements $s \in S$ represent the exhaustive contingencies relevant to the decision problem at hand; the elements of a distinguished algebra $\Sigma$ of subsets of $S$ are called events. The second is a set $X$ of \emph{consequences}. The third is a collection of (Savage) \emph{acts}, wagers in the language of early contributors, that is, functions $f \colon S \to X$, where for every state of the world $s$, the value $f(s)$ represents the outcome that realizes after all uncertainty has been resolved toward $s$. Savage's contribution is to recognize that a DM's preferences over acts satisfy certain prima facie reasonable axioms if and only if there exist a unique subjective probability $p \in \Delta$, where $\Delta$ denotes the set of probabilities over $(S, \Sigma)$, and a unique (up to positive affine transformations) utility $u \colon X \to \mathbb{R}$ which jointly represent these preferences via the expected utility criterion
\begin{equation*}
    \int u(f) \, dp.
\end{equation*}
The argument is especially powerful from a normative perspective: notwithstanding the philosophical predispositions of the individual, anyone who accepts Savage's axioms ought to act as if she held a well defined and unique belief. 

It is precisely this uniqueness claim that we now examine. \citet{ellsberg1961risk} argued persuasively that the variety of situations encountered in human affairs cannot be accommodated by a single probabilistic belief, and not merely as a descriptive matter, in view of mere humans and their mistakes and biases, but normatively, in the sense that even a fully rational DM ought not to act as a subjective expected utility maximizer in such situations. We briefly present Ellsberg's three-color urn thought experiment, which we believe to be both convincing and representative of the broader argument. For acts $f, h$ and an event $A$, we denote by $f_A h$ the act equal to $f$ on $A$ and to $h$ otherwise. Consider an urn containing $90$ balls, of which $30$ are known to be red, while the remaining $60$ are split in unknown proportions between black and yellow. A single ball is to be drawn, and we consider the following wagers in Table~\ref{table1}.
\begin{table}[htbp]

  \centering

  \caption{Acts in the three-colour urn example}

  \label{table1}

  \setlength{\tabcolsep}{12pt}

  \begin{tabular}{@{} l ccc @{}}

    \toprule

    Act & $R$ & $B$ & $Y$ \\

    \midrule

    $1_R0$                  & 1 & 0 & 0 \\

    $1_B0$                  & 0 & 1 & 0 \\

    $1_{R \, \cup \, Y}0$   & 1 & 0 & 1 \\

    $1_{B \, \cup \, Y}0$   & 0 & 1 & 1 \\

    \bottomrule

  \end{tabular}

  \par\medskip

  {\footnotesize A ball is drawn at random, and for each event $E$ the act $1_E0$ pays $1$ if the drawn ball lies in $E$ and $0$ otherwise.\par}

\end{table}

A sensible DM might strictly prefer $1_R0$ to $1_B0$ and $1_{B \, \cup \, Y}0$ to $1_{R \, \cup \, Y}0$, as in each pair, the preferred wager bets on an event of unambiguous probability, while its counterpart is inherently ambiguous. Such preferences are well documented empirically, and we contend they are also normatively defensible, at least when the stakes are sufficiently high. They do, however, directly violate one of Savage's axioms, the \textit{sure-thing principle} ($\mathrm{P2}$), which states that for any acts $f, g, h, h'$ and event $A$, $f_A h$ is preferred to $g_Ah$ if and only if $f_Ah'$ is preferred to $g_Ah'$. Such behavior is well attested as a robust empirical finding across a wide range of decision settings; see, e.g., \citet{camerer1992recent} and \citet{trautmann2015ambiguity} for empirical surveys.

We can venture still deeper into the fog of uncertainty. In the Ellsberg urn experiment, the DM does not hold a unique probability belief over the draw; nevertheless, she knows that any sensible belief belongs to an explicit one-dimensional parameter family. In this sense, the structure of the decision environment confines the DM's ignorance within a well-specified parametric class, which she is confident contains the "correct"\footnote{The scare quotes signal that "correct" is not to be read as invoking an objective probability standing behind the DM's beliefs. We defer a precise account of its subjectivist content to Footnote~\ref{fn:subjectivist-truth}, where it can be given against the backdrop of the taxonomy introduced below.} belief. This is a luxury not generally available in substantive decision problems, those of war and peace, of health and disease, of economic boom or bust. There, any probabilistic belief is at best a stylized approximation of a complex environment that is only partially modeled and partially observed, and the DM might be unwilling to assert that the correct belief lies within the particular family of such beliefs she has been able to entertain or formulate. Her models are simplifications and the relevant epistemic concern is precisely that all of them might be wrong in ways she cannot, in advance, parametrize away. A more detailed critique of Savage's axioms, equivalently of the Bayesian paradigm, in such contexts has been articulated by \citet{gilboa2008probability,gilboa2009always, gilboa2012rationality}. Their argument is that on a normative reading, Savage's framework is too demanding in substantive decision problems: where no relative-frequency data and no structural symmetry are available to anchor a prior, Laplace's principle of indifference fails to discriminate between symmetry grounded in evidence and symmetry imposed by ignorance, and any specific prior one adopts is to that extent, arbitrary. In such problems, they argue, it may be more rational to acknowledge that one's information underdetermines any single prior than to commit to one for the sake of internal coherence.

Following the tripartite taxonomy proposed by \citet{hansen2016ambiguity}, we will use the following terminology throughout the rest of this work:
\begin{itemize}
  \item[(i)] \emph{Risk} denotes the DM's subjective uncertainty over outcomes conditional on a single, fully specified probabilistic belief $p \in \Delta$.
  \item[(ii)] \emph{Ambiguity} denotes the DM's uncertainty over which element of a posited set $\mathrm{Q} \subseteq \Delta$ of probabilistic beliefs is the correct\footnote{\label{fn:subjectivist-truth}The terms "correct" and "truth" as they appear in this taxonomy, and their cognates "wrong", "describes the environment", and so on, must be read in the subjectivist register established at the outset of this section. We take a moment to make this reading precise, as the matter is delicate and the language unavoidably suggestive. 
  
  The three terms are to be understood as distinct attitudes that the DM may take, through her preferences, toward her own model class. Under (i), her preferences are represented by a single $p \in \Delta$. Under (ii), her preferences reveal a family $\mathrm{Q} \subseteq \Delta$ together with the manner in which she trades off its elements; the assertion that "some $q \in \mathrm{Q}$ is the correct one" is the internal claim that, were she to receive further information of the kind she takes to be probative, her beliefs would concentrate on some element of $\mathrm{Q}$. The "correct" $q$ is thus the limit point of her own hypothetical updating, not an externally given fact she is trying to learn. Under (iii), the DM withholds precisely this commitment: she deploys a model class $\mathrm{Q}$ she takes seriously enough to use, while allowing, again, as a feature of her preferences manifested in behavior, that her beliefs might fail to concentrate on $\mathrm{Q}$ under any feasible further information. Read this way, the asymmetry between ambiguity and misspecification is an asymmetry between two stances the DM may adopt toward her own modeling apparatus: under ambiguity, she is willing to commit,  through her preferences, that her class is adequate in the limit; under misspecification, she is not. This is similar to Savage's distinction between the small world the DM has articulated and the grand world she has not.} one, under the maintained assumption that the truth belongs to $\mathrm{Q}$.
  \item[(iii)] \emph{(Model) misspecification} denotes the residual uncertainty arising from the possibility that no element of $\mathrm{Q}$ best describes the environment, i.e., that the truth lies outside $\mathrm{Q}$ altogether.
\end{itemize}
The term \emph{uncertainty} will serve as the umbrella covering all three notions. Risk is the only stratum present in the classical Bayesian paradigm; ambiguity is the stratum exposed by Ellsberg's thought experiment; misspecification is the stratum that we, following the axiomatization of \citet{cerreia2026making}, will primarily be concerned with.

Ellsberg's challenge, that a rational DM may legitimately decline to commit to a single prior, has, over the past four decades, given rise to a sustained axiomatic literature aimed at enlarging Savage's framework to accommodate ambiguity and misspecification. For ease of exposition, in what follows we will gloss over the distinction between Savage's original setup and that of \citet{anscombe1963definition}, in which abstract consequences are replaced with lotteries carrying exogenous (objective) probabilities à la von Neumann–Morgenstern, so that acts map states into such lotteries; for a fuller account, covering both historical developments and recent axiomatizations, we refer the reader to \citet{gilboa2016ambiguity}, and to \citet{gilboa2009theory} for an illuminating textbook treatment.

The seminal contribution toward an axiomatization of non-Bayesian decision models is the Choquet Expected Utility (CEU) model of \citet{schmeidler1989subjective}. Its central idea is to relax the additivity of beliefs: in place of a probability measure, the DM's belief is represented by a capacity, that is, a nonnegative, normalized, and monotone, but not necessarily additive, set function. Schmeidler then replaces Savage's sure-thing principle with a weaker independence axiom and obtains the equivalent existence of a utility $u$ and a capacity $\nu$ such that preferences are represented by
\begin{equation*}
    \int u(f) \, d\nu,
\end{equation*}
where the integral is to be read in the sense of Choquet. The classical Bayesian paradigm is recovered as the special case in which the capacity is additive, and the Choquet integral collapses to the standard one.

A more tractable, and arguably more intuitive, alternative is the Maxmin Expected Utility (MMEU) model of \citet{gilboa1989maxmin}. Here the weakening of the sure-thing principle preserves the additivity of beliefs at the cost of their uniqueness, with the DM's beliefs being represented by a convex compact set $C\subseteq \Delta$. Together with a utility $u$, this set jointly represents her preferences via the pessimistic criterion
\begin{equation*}
    \min_{p \in C} \int u(f) \, dp,
\end{equation*}
so that the DM evaluates each act by its worst-case expected utility over $C$. The Bayesian paradigm is recovered as the degenerate case in which $C$ is a singleton. The MMEU criterion treats every prior in $C$ on an equal footing and excludes every prior outside $C$ outright. The variational preferences (MMR) of \citet{maccheroni2006ambiguity} dispense with this extreme structure and obtain the MMEU criterion as a special case. Preferences are now represented by the functional
\begin{equation*}
    \min_{p \in \Delta} \left\{ \int u(f) \, dp \, + c(p) \right\},
\end{equation*}
where $c\colon \Delta\to [0, +\infty]$, the ambiguity index, attaches to each prior a penalty measuring the DM's reluctance to entertain it.  The MMEU criterion is recovered when $c$ is the (convex-analytic) indicator function of a convex compact set $C$; when, instead, $c$ is proportional to the Kullback--Leibler divergence relative to a reference model, one obtains the multiplier preferences of \cite{hansen2001robust} and \citet{strzalecki2011axiomatic}. It is the variational preference model that will be most used and useful throughout the present work when representing ambiguity; see for instance Section~\ref{sec:tax}.

We conclude this brief exposition with the model closest in spirit to the original Bayesian paradigm, the smooth-ambiguity model (KMM) of \citet{klibanoff2005smooth}. Here the DM holds a second-order prior $\mu$ over the first-order priors $\Delta$ and evaluates acts according to
\begin{equation*}
    \int \phi\left( \int u(f) \, dp \right) \,d\mu(p),
\end{equation*}
where $\phi \colon \mathbb{R} \to \mathbb{R}$ is an ambiguity index whose concavity represents the degree of aversion to the dispersion of expected utilities across priors. The Bayesian EU functional is recovered when $\phi$ is affine. The MMR and KMM models, and hence all their special cases, have  been unified within a single axiomatic representation by \citet{cerreia2011uncertainty}.

Any meaningful notion of ambiguity aversion must, in the end, be intelligible at the level of preferences themselves, independently of the particular functional representation one happens to adopt. The behavioral definition that meets this requirement is due to \citet{epstein1999definition} and \citet{ghirardato2002ambiguity}: uncertainty aversion is defined as a binary relation between preferences, and an absolute notion is defined only afterwards by comparing this relation to a designated class of ambiguity-neutral benchmarks. In the classical Arrow-Pratt theory of risk aversion one says that a DM is more risk-averse than another if she accepts fewer fair bets, and one obtains an absolute notion by selecting expected-value preferences as the (risk-neutral) benchmark; here, one compares a DM's willingness to choose uncertain acts over a class of uncertainty-free alternatives, and SEU preferences are taken as the (ambiguity-neutral) benchmark. The question is what to designate as "uncertainty-free". \citet{ghirardato2002ambiguity} adopt the most transparent choice available in a fully subjective setting, namely the constant acts, with the argument that these are unambiguously free of uncertainty in every Savage-style decision problem, irrespective of the DM's information or model class. The comparative relation then takes the following behavioral form: a preference $\succsim^2$ is more uncertainty-averse than $\succsim^1$ if, for every act $f$ and every constant $x$
\[
f \succsim^2 x \;\Longrightarrow\; f \succsim^1 x.
\] 
In words, whenever the more uncertainty-averse DM is willing to entertain the uncertain outcome $f$ in place of the sure outcome $x$, then the less uncertainty-averse DM is willing to do so as well. We will make use of this approach of comparative uncertainty aversion in Section~\ref{sec:comparative_aversion}; see also the accompanying discussion in Appendix~\hyperref[def:app_B:comparativeuncertaintyaversion]{C}.

Against this backdrop, the axiomatic literature has gradually carved out a distinction between aversion to ambiguity over posited models and aversion to the possibility that none of the posited models is correct, a distinction largely absent from the foundational works above cited, but a distinction that is empirically relevant \citep{aydogan2023three}. The decisive step in this direction has been taken by \citet{cerreia2026making} (henceforth~CVHMM), whose axiomatization will serve as the foundational literature for the present work. CVHMM begin by considering a partition of the set of beliefs into two different classes: a distinguished set $\mathrm{Q}\subseteq\Delta$ of \emph{structured} beliefs, whose elements are those probabilistic models the DM is prepared to defend on substantive grounds, in that they may reflect scientific theory, empirical evidence, or interpretable parametrizations of the phenomenon under study; by contrast, an \emph{unstructured} model is any probability $p\in\Delta \setminus \mathrm{Q}$, that is, any probability lacking the kind of substantive motivation that warrants inclusion in $\mathrm{Q}$. Under model ambiguity the DM trusts that the correct model lies in $\mathrm{Q}$ and her uncertainty concerns which element of $\mathrm{Q}$ that is; under misspecification she withholds this commitment and acknowledges that the correct model may sit outside $\mathrm{Q}$ altogether. 

To capture the DM's attitudes toward model misspecification, CVHMM follow \citet{gilboa2010objective} in working with a pair of preferences over acts $(\succsim_{\mathrm{Q}}^*, \succsim_{\mathrm{Q}})$, both indexed by the set $\mathrm{Q}$ of structured models. The behavioral preference $\succsim_{\mathrm{Q}}$ is complete: it governs choice and is the relation the DM ultimately reveals through her actions. The mental preference $\succsim_{\mathrm{Q}}^{*}$ is, in general, incomplete: it describes choice that can be justified by the DM and represents her "genuine" rankings, those she is prepared to defend on argumentative grounds. The two preferences are linked by two axioms: \emph{consistency}, that $f\succsim_{\mathrm{Q}}^{*}g$ implies $f\succsim_{\mathrm{Q}}g$, the mental ranking, when it speaks, informs the behavioral one; and \emph{caution}, that in the absence of an articulable argument in favour of an uncertain act over a sure alternative, the DM defaults to the sure alternative, that is, if $f \not\succsim^{*}_{\mathrm{Q}}x$ then $x \succsim_{\mathrm{Q}}f$. Under the set of axioms that CVHMM call the \emph{robust} family, both preferences admit a joint representation in terms of a shared
utility $u$ and a shared statistical distance $c$. The mental preference $\succsim_{\mathrm{Q}}^{*}$ is represented by a Paretian unanimity rule over a family of variational criteria, one for each structured model
\begin{equation*}
    \min_{p \in \Delta} \left\{ \int u(f) \, dp \, + c(p, q) \right\}, \quad \forall q \in \mathrm{Q}.
\end{equation*}
The behavioral preference $\succsim_{\mathrm{Q}}$, the cautious completion of the mental preference, is in turn represented by the variational criterion (after exchanging orders of minimization)
\begin{equation*}
    \min_{p \in \Delta} \left\{ \int u(f) \, dp \, + \min_{q \in \mathrm{Q}}c(p, q) \right\}.
\end{equation*}
The inner set distance $C(p,\mathrm{Q}):=\min_{q\in\mathrm{Q}}c(p,q)$ vanishes exactly on the structured set and is strictly positive on unstructured models, with its magnitude expressing the implausibility of each $p$ in light of the DM's information; unstructured models thus enter $V_{\mathrm{Q}}$ as a "protective belt" around $\mathrm{Q}$, their influence on the criterion graded by their proximity to a substantively defensible model.

The robust representation is, however, only the most prudential aggregation of the dominance relation into a complete preference. Dispensing with caution while retaining the remaining structural axioms yields what CVHMM call the \emph{sensitive} representation. The mental relation is still characterized by the Paretian unanimity rule given above, but the behavioral preference is no longer pinned to the worst-case aggregation. Instead, it is represented by
\begin{equation*}
    J_{\mathrm{Q}}\left( q \longmapsto \min_{p \in \Delta} \left\{ \int u(f) \, dp \, + c(p, q) \right\} \right),
\end{equation*}
where $J_{\mathrm{Q}}$ is an aggregation functional over the misspecification-robust evaluations. The robust representation is recovered as the special case $J_{\mathrm{Q}}=\min$, but a number of other aggregators are admissible, each corresponding to a distinct stance on how the DM resolves the residual incompleteness left over by the mental relation. The framework thereby disentangles two attitudes that the robust criterion bundles together: the DM's fear of misspecification, encoded in the divergence $c$ and so in the inner minimization over unstructured deviations, and her attitude toward ambiguity over which structured model best fits the truth, encoded in the outer aggregator $J_{\mathrm{Q}}$.  It is this sensitive criterion that will be used throughout the present work. More details are to be given in Section~\ref{sec:gameswithfeedbackanduncertainty}.

\subsection*{Self-Confirming Equilibrium}

At its core, any notion of equilibrium can be decomposed into a behavioral rationality condition and an epistemic condition. In equilibrium, each player
\begin{itemize}
    \item[(i)] rationally best responds with respect to
    \item[(ii)] a conjecture about the other players' actions.
\end{itemize}
Different specifications of (i) and (ii) produce different notions of equilibrium. Once the epistemic structure of the game is fixed, all that remains to specify the behavioral condition is the decision-theoretic rule used to evaluate actions. The epistemic structure can itself take multiple forms, that of a single conjecture over the other players' actions, of a second order belief, of a set of (misspecified) conjectures, all of these possibly mediated by partially informative feedback.

The classical Nash Equilibrium (NE) \citep{nash1950equilibrium, nash1951} corresponds to (i) expected utility maximization under (ii) a unique and correct conjecture. Relaxing the correctness requirement on (ii) and enriching the game with an ex-post feedback structure leads to the notion of Self-Confirming Equilibrium (SCE), introduced independently by \citet{battigalli1987thesis} (see also \citet{battigalli2012comportamento} for historical details) under the name of Conjectural Equilibrium, and by \citet{fudenberg1993self}. Intuitively, the setting is that of a repeated game in which, at the start of each round, players are randomly drawn from large populations of roles and matched to play the game and, at the end of each round, observe (partial) feedback on the basis of which they update their probabilistic conjectures about the opponents' actions.  In a SCE, each player's conjecture is then not required to be correct, but only to be confirmed by and consistent with the feedback received in equilibrium. Note that the learning dynamics are not explicitly modeled and the equilibrium is to be read as the steady state of the recurrent game reached after such learning has taken place. We recommend the reader to \citet{battigalli2023game} for a modern textbook treatment of the subject.

To define a SCE, the normal-form game structure $G = (I, (A_i, u_i)_{i \in I})$ must be augmented with a \emph{feedback structure}: for each player $i$, a feedback function $f_i : A \to M_i$ into a message space $M_i$, mapping each profile of actions to the ex-post signal received by the player. Fixing player $i$'s own action $a_i$, the section $f_i(a_i, \cdot) : A_{-i} \to M_i$ pushes any distribution over opponents' actions forward to a predicted distribution of messages on $M_i$. A conjecture is then said to be \emph{confirmed} by the true distribution of opponents' play, given own action $a_i$, if the two induced distributions on $M_i$ coincide, that is, if the player cannot tell her conjecture apart from the truth on the basis of feedback alone. A SCE is then a profile of mixed actions and conjectures (one for each support action) such that each player (i) best responds to her conjecture and (ii) the conjecture is confirmed by the equilibrium distribution of play.

Two observations \citep{battigalli2023game} help position the notion of self-confirming equilibrium relative to Nash equilibrium. First, every NE is, under any feedback structure, also a SCE, as it suffices to consider each player's conjecture to be the correct equilibrium distribution of opponents' play, under which her equilibrium action remains a best reply by the Nash best-response inequalities, and her conjecture is trivially confirmed. The converse does not hold, but a sufficient condition for it is that feedback be sufficiently informative or, more precisely, that two hypotheses hold jointly: own-action independence of feedback about others (the partition of opponents' action profiles induced by feedback does not depend on the player's chosen action) and observability of payoffs (the realized payoff is recoverable from own action and own feedback message).

It is precisely the partial informativeness of feedback that makes the two equilibrium notions distinct and worthy of our attention. Less informative feedback expands the set of self-confirming equilibria as it leaves more
room for incorrect conjectures to survive ex-post scrutiny with feedback, and the long-run outcome of the recurrent game becomes less likely to coincide with a Nash equilibrium. Finally, we mention briefly the comparative statics fact that greater risk aversion expands the set of pure self-confirming equilibria, a result which admits a natural extension to ambiguity aversion, to which we shall return. The decision-theoretic counterpart, that higher ambiguity (resp. risk) aversion enlarges the set of justifiable actions is established by \citet{battigalli2016note}.

And we conclude with a passage from \citet{von1937economics}, whose discussion of equilibrium and knowledge already made this informational reading explicit:

\begin{quote}
    \textit{And we may therefore very well have a position of equilibrium only because some people have no chance of learning about facts which, if they knew them, would induce them to alter their plans. Or, in other words, it is only relative to the knowledge which a person is bound to acquire in the course of the carrying out of his original plan and its successive alterations that an equilibrium is likely to be reached. }
\end{quote}

\subsection*{Ambiguous Games}

If the non-Bayesian decision theories surveyed above are to be taken seriously as normative (or descriptive) foundations for economic theory, they must be able to support the analysis of strategic interaction in games. Importing ambiguity into game theory, however, requires one to adopt the subjective-belief interpretation of mixed actions, where a mixed action $\alpha\mi$ is to be interpreted as player $i$'s subjective belief over the opponents' pure actions. Thus, ambiguity in games is ambiguity at the level of players' conjectures about opponents' play, conjectures that the formal description of the game does not fix and that depend instead on, in Marinacci's words, \textit{"the context in which the game is played, that is, by those unmodeled features of the strategic setting like past experience, preplay communication, cultural norms, and so forth, on which players base their beliefs about how the game will be played"} \citep{marinacci2000ambiguous}. We exclude the alternative interpretation that players act as if they used randomization devices to implement mixed actions; see \citet{riedel2014ellsberg} for a study of Ellsberg games, in which players commit to objective but probabilistically imprecise randomization devices (Ellsberg urns) to implement their mixed actions, and ambiguity-averse players max-minimize expected utility over the induced sets of probability distributions.

The first efforts in these directions were to extend the notion of Nash equilibrium to non-additive beliefs (capacities), with players best responding to a CEU functional \citep{dow1994nash,eichberger2000non,marinacci2000ambiguous}. In equilibrium, players hold non-additive beliefs $(\nu_i^*)_i$ and, for each player $i$, each pure action $a_i$ in the support\footnote{The notion of support of a capacity is a delicate notion, of which we do not cover the details here.} of $\nu_i^*$ maximizes the Choquet expected utility induced by those beliefs. Nash equilibria are recovered as the special case in which every $\nu_i^*$ is additive. In the same spirit, \citet{lo2004equilibrium} considers players who, in equilibrium, hold a set of probability measures over the opponents' actions and best respond to the maxmin expected utility functional induced by those beliefs.

We conclude our review of relevant literature with two contributions that are closest, in motivation and execution, to the present work: \citet{battigalli2015self}, who introduce Smooth Self-Confirming Equilibrium (SSCE), and \citet{esponda2016berk}, who introduce Berk--Nash Equilibrium (BNE). 

\citet{battigalli2015self} work in the usual self-confirming tradition within the recurrent large-population interpretation outlined above, but drop the assumption that players are ambiguity neutral. Each player $i$ is uncertain about the true distribution $\alpha\mi$ of co-players' actions and holds a prior belief supported on a set of such distributions. Actions are evaluated via the smooth-ambiguity criterion of \citet{klibanoff2005smooth}
\[
V_i^{\phi_i}(a_i,\mu_i)
= \phi_i^{-1}\!\left(\int \phi_i\bigl(\int u_i(a_i, a\mi) \, d\alpha\mi(a\mi)\bigr)\,
        \mu_i(d\alpha_{-i})\right),
\]
where $\phi_i$ represents the player's ambiguity attitudes. A profile of mixed actions $(\alpha_i^*)_i$ is then an equilibrium if for each player and each action $a_i$ in the support of the respective mixed action, there exists a prior belief $\mu_{a_i}$ supported on the set of equilibrium-consistent conjectures, those probabilities whose induced distribution through the feedback function agrees with that of the true $\alpha\mi^*$. The classical (Bayesian) SCE is recovered when each $\phi_i$ is affine. Their main comparative-statics result is that, under observable payoffs, the SSCE correspondence is monotone in ambiguity aversion, in particular, when players are ambiguity averse, the set of Bayesian SCE is included in the set of SSCE which is, in turn, included in the set of Maxmin Self-Confirming Equilibria\footnote{The definition of Maxmin Self-Confirming Equilibria (MSCE) is the natural extension of SSCE in which players evaluate actions through the \citet{gilboa1989maxmin} maxmin functional over equilibrium-consistent conjectures.}. \citet{battigalli2019ambiguity} extend this work to extensive-form games with feedback played by smooth-ambiguity averse agents. These and similar results are the subject of our analysis in games with misspecification aware (and averse!) players, see Chapter~\ref{ch:contributions1}. 

The Berk--Nash framework of \citet{esponda2016berk} is the earliest formalization of epistemic model misspecification in games. The setting\footnote{We present here a simplified, but arguably more incisive, formulation, without an exogenous state space and related objects, and by directly reducing the players' subjective model to a parametric family of conjectures about opponents' play. This weakens the generality of the original presentation of \citet{esponda2016berk} but does not affect the content of the normal-form equilibrium notion we are interested in here.} is that of a game with feedback, in which players hold a parametrized set of conjectures about opponents actions $\{\mu^{\theta_i}\}_{\theta_i}$, possibly misspecified in the sense that the true distribution of opponents' play need not lie in the family. Each player also holds a prior belief $\beta_i$ over the parameter set and evaluates actions via the reduced (Bayesian) expected utility functional. A Berk--Nash equilibrium is then a profile of mixed actions and prior beliefs $(\alpha_i^*, \beta^*_i)_i$ such that for each player, (i) each action in the support of $\alpha_i^*$ is a best response to the EU functional induced by $\beta^*_i$, while (ii) each parameter in the support of $\beta_i$ is at minimum distance from the true underlying distribution $\alpha\mi^*$, that is, it minimizes the Kullback-Leibler divergence between the true distribution of feedback messages induced by $\alpha\mi^*$ and the distribution predicted by $\mu^{\theta_i}$. We conclude by highlighting the asymmetric epistemic posture each framework imposes on its players. In Berk--Nash equilibrium, players are Bayesian expected-utility maximizers within their parametric class, with no awareness of, or aversion to, the possibility that the class itself is misspecified, and no aversion to ambiguity within the class either. By contrast, in SSCE players are averse to ambiguity within a posited set of conjectures, but are not represented as entertaining the possibility that the set itself fails to contain the truth. It is this connection, that players are aware of and averse to misspecification and ambiguity in their sets of conjectures, that is the main conceptual contribution of the present work.

Finally, \citet{cerreia2022nonatomic} extend self-confirming and Berk--Nash equilibria to nonatomic anonymous games, games with an arbitrary measurable space of players whose utility depends only on the distribution of actions in the population, unifying them under the notion of $\varepsilon$-estimated equilibrium.

\newpage

\section{Mathematical Preliminaries and Notation}\label{sec:math_definitions_chapter1}

This section contains the relevant definitions, mathematical preliminaries, and notation that will be used throughout the present work.

Let $(X, \tau)$ be a topological space. Denote by $\mathcal{B}(X)$ the corresponding Borel $\sigma$-algebra, by $B_b(X)$ the vector space of bounded $\R$-valued Borel-measurable functions on $X$, and by $C_b^0(X)$ the space of bounded continuous functions on $X$. Equipped with the supremum norm, $(B_b(X), \|\cdot\|_\infty)$ is a Banach space, and $C_b^0(X) \subseteq B_b(X)$ is a closed subspace. Unless otherwise stated, all product spaces are endowed with the product topology and with the Borel $\sigma$-algebra induced by this topology, namely $\mathcal{B}(X\times Y)$\footnote{\label{fn:prod_sigma_alg}A note of care is due: the product $\sigma$-algebra of two Borel measurable spaces does not, in general, coincide with the Borel $\sigma$-algebra of the product topological space. In general, one only has the inclusion
\[
    \mathcal{B}(X)\otimes\mathcal{B}(Y)\subseteq \mathcal{B}(X\times Y).
\]
Equality holds, for instance, when $X$ and $Y$ are second countable, which is the main setting of this thesis; see Chapter~\ref{ch:contributions1}.}. 

Let \(\Delta^f(X)\) denote the set of finitely additive Borel probability measures on \(X\), and let \(\Delta(X)\subseteq \Delta^f(X)\) denote the subset of countably additive probability measures\footnote{Our focus on countably additive, rather than finitely additive, probabilities can only be justified in terms of better technical properties of the former. Using finitely additive probabilities would have been more philosophically satisfying, since the main theorems motivating the use of probabilities to express beliefs are obtained, under minimal assumptions, for finitely additive probabilities \citep{de1931sul, savage1954}. Indeed, early drafts of the present work used finitely additive probabilities endowed with the $B_b(X)$-weak* topology, namely the smallest topology making all integral evaluation functionals continuous when evaluated on bounded measurable functions. It is only after encountering (insurmountable) difficulties in the proof of a central result, the existence Theorem~\ref{th:exists:RNE} (in particular, in proving continuity of the product map; see Appendix~\hyperref[ch:appendix_newtheorems]{B} for more details) that I decided to reformulate the work using countably additive probabilities with the topology defined above. The mathematics ended up being simpler, the theory more elegant and structured, if only less philosophically convincing.}. We endow \(\Delta(X)\) with the weak* topology $\sigma(\Delta(X),C_b^0(X))$, that is, the smallest topology for which all maps
\[
    \mu \mapsto \int_X f\,d\mu,
    \qquad f\in C_b^0(X),
\]
are continuous. Although one can define other topologies on the set of countably (or finitely) additive probability measures, such as the total variation norm topology or the $B_b(X)$-weak* topology, throughout this work $\Delta(X)$ will be equipped with the weak* topology as previously defined, unless otherwise specified. When the underlying space is clear from context, we simply write $\Delta\coloneqq\Delta(X)$.

A note on terminology. In probability theory, the topology \(\sigma(\Delta(X),C_b^0(X))\) is often called the weak topology. We will instead refer to it as the weak* topology. Indeed, every \(\mu\in\Delta(X)\) defines a continuous linear functional on \(C_b^0(X)\) by
\begin{equation}\label{eq:rieszrep}
    J(\mu)(f)\coloneqq \int_X f\,d\mu,
    \qquad f\in C_b^0(X).
\end{equation}
Thus \(J\) maps \(\Delta(X)\) into \(C_b^0(X)^*\), and the topology above is precisely the pullback along \(J\) of the weak* topology $\sigma\bigl(C_b^0(X)^*,C_b^0(X)\bigr)$ on \(C_b^0(X)^*\).

We will often use the important fact that compactness of $X$ implies compactness of $\Delta(X)$.

\begin{theorem}\label{th:appA_sigma_compact}
    Let $X$ be a compact metrizable topological space, and let
    \[
        \bigl(\Delta(X),\sigma(\Delta(X),C_b^0(X))\bigr)
    \]
    be the space of countably additive Borel probability measures on $X$ endowed with the weak* topology. Then $\Delta(X)$ is compact.
\end{theorem}

\begin{proof}
    Since $X$ is compact, $C^0_b(X)=C^0(X)$, a Banach space under $\|\cdot\|_{\infty}$ with topological dual $C^0(X)^*$. Consider the set of positive normalized functionals
    \begin{equation*}
        K\coloneq \{ \Lambda \in C^0(X)^*:\Lambda(f)\ge0 \text{ for } f \ge 0, \Lambda(1)=1 \},
    \end{equation*}
   where $1$ is the constant function. We show that the integration map $J$ of Equation~\eqref{eq:rieszrep} is a homeomorphism onto $K$ endowed with the weak* topology $\sigma(C^0(X)^*, C^0(X))$, and that $K$ is weak* compact.

   In fact, $J$ is well defined into $K$ by linearity and positivity of the integral. An application of the Riesz–Markov–Kakutani Theorem (see Theorem~7.2 and 7.17 of \citet{folland1999real}) shows that $J$ is a bijection onto $K$, where we note that every probability measure over a compact metric space is a Radon measure (Theorem~7.8 of \citet{folland1999real}). Finally $J, J^{-1}$ are continuous by definition of the weak* topologies. Hence $J$ is a homeomorphism.

   Moreover, if we denote by $\operatorname{ev}_f(\Lambda)\coloneq \Lambda(f)$ the evaluation functional on $C^0(X)^*$, $K$ can be rewritten as
   \begin{equation*}
       K=\operatorname{ev}_1^{-1}(\{1\}) \cap \bigcap_{f\ge 0} \operatorname{ev}_f^{-1}([0, +\infty)),
   \end{equation*}
   which shows that $K$ is weak* closed being the intersection of preimages of closed sets under continuous maps. 
   
   Finally, if $\Lambda \in K$, then for all $f$ with $\|f\|_\infty \le 1$, $1 \pm f \ge 0$ hence $\Lambda(1 \pm f) \ge 0$, i.e. $| \Lambda(f)| \le \Lambda(1)=1$. This shows that $K$ is a subset of the dual unit ball $B_1^*(0)$ which is weak* compact by Banach-Alaoglu, see Theorem~3.16 of \citet{brezis2011functional}. A closed subset of a compact set is compact, which concludes the proof.
\end{proof}

We next collect some miscellaneous definitions and results that will be useful throughout.

\begin{definition}\label{def:support}
Let \(\mu\in\Delta(X)\). The support of \(\mu\), denoted
\(\operatorname{supp}(\mu)\), is defined by
\[
    \operatorname{supp}(\mu)
    \coloneqq
    \left\{
        x\in X :
        \mu(U)>0 \text{ for every open neighborhood } U \ni x
    \right\}.
\]
\end{definition}

Equivalently,
\[
    \operatorname{supp}(\mu)
    =
    X \setminus
    \bigcup
    \left\{
        U\subseteq X :
        U \text{ is open and } \mu(U)=0
    \right\}.
\]
Thus \(\operatorname{supp}(\mu)\) is closed.

If \(X\) is second countable, then $\mu(\operatorname{supp}(\mu))=1$. Indeed, in that case the open set \(X\setminus \operatorname{supp}(\mu)\)
can be written as a countable union of open \(\mu\)-null sets, and hence has \(\mu\)-measure zero by countable additivity.

Let now $(X,\mathcal{X})$ and $(Y,\mathcal{Y})$ be two measurable spaces, and let $f:X\to Y$ be a measurable map. For $\mu\in\Delta^f(X)$, the map $f$ induces the pushforward measure of $\mu$ on $Y$, defined by
\[
    \widehat{f}(\mu)(A)\coloneqq \mu(f^{-1}(A)),
    \qquad A\in\mathcal{Y}.
\]
Then $\widehat{f}(\mu)$ is a finitely additive probability measure. Hence each measurable map induces a pushforward map
\[
    \widehat{f}:\Delta^f(X)\to\Delta^f(Y).
\]
Moreover, its restriction to $\Delta(X)$ is well defined as a map into $\Delta(Y)$.

\begin{theorem}[Change of variables]\label{th:AppA_Changeofvariables}
    Let $(X, \mathcal{X})$ and $(Y, \mathcal{Y})$ be measurable spaces, let $\mu \in \Delta^f(X)$ and $f:X\to Y$ measurable. Then for every $\varphi \in B_b(Y)$, 
    \begin{equation*}
        \int_{X}^{}\varphi( f(x) ) \ d\mu(x)= \int_Y \varphi(y) \ d (\widehat{f}(\mu))(y).
    \end{equation*}
    
\end{theorem}

\begin{proof}
    Let $A \in \mathcal{Y}$, then
    \begin{align*}
        \int_{X}^{}\mathbf{1}_A( f(x) ) \ d\mu &= \int_{X}^{}\mathbf{1}_{f^{-1}(A)}(x) \ d\mu(x) \\
        &=\mu(f^{-1}(A))  \\
        &= \widehat{f}(\mu)(A) \\
        &= \int_{Y}^{}\mathbf{1}_A \ d (\widehat{f}(\mu)).
    \end{align*}
    By linearity of the integral, the desired equality extends to simple functions. Now, let $\varphi \in B_b(Y)$ be arbitrary and choose a sequence $(\varphi_n)_n \subseteq B_b(Y)$ of simple functions converging in norm to $\varphi$. Then by the continuity of the finitely additive integral with respect to the sup norm,
    \begin{equation*}
        \int_{X}^{}\varphi(f) \ d\mu=\lim_{n} \int_{X}^{}\varphi_n(f) \ d\mu=\lim_{n} \int_Y \varphi_n \ d (\widehat{f}(\mu))=\int_Y \varphi \ d (\widehat{f}(\mu))
    \end{equation*}
\end{proof}

\begin{theorem}\label{th:appA_pushforwardcont}
    Let $(X, \mathcal{B}(X))$ and $(Y, \mathcal{B}(Y))$ be topological spaces and let $f:X\to Y$ be continuous. Then the induced pushforward map $\widehat{f}:\Delta(X)\to\Delta(Y)$ is weak* continuous. 
\end{theorem}
\begin{proof}
    It is sufficient to prove that for each $\varphi \in C^0_b(Y)$, the composition of $\widehat{f}$ with the evaluation functional $\nu \mapsto \int_Y\varphi \ d \nu$ is weak* continuous, that is, the map:
    \begin{equation*}
        \mu \mapsto \int_{Y}^{} \varphi \ d \widehat{f}(\mu).
    \end{equation*}
    By Theorem \ref{th:AppA_Changeofvariables}, this is equivalent to proving that the following map is weak* continuous
    \begin{equation*}
        \mu \mapsto \int_{X}^{} \varphi\circ f \ d \mu.
    \end{equation*}
    This is true by definition of the weak* topology $\sigma(\Delta(X), C_b^0(X))$, as $\varphi \circ f$ is bounded and continuous.
\end{proof}

For a finite collection $(X_i, \mathcal{X}_i)_{i=1}^n$ of topological spaces, the product space $\prod_{i=1}^n X_i$ is endowed with the product topology and with the Borel $\sigma$ algebra induced by it, denoted $\mathcal{B}(\prod_{i=1}^n X_i)$. Given a tuple $(\mu_1, \ldots, \mu_n) \in \prod_{i=1}^n \Delta(X_i)$, the product measure $\bigotimes_{i=1}^n \mu_i$ is the unique element of $\Delta(\prod_{i=1}^n X_i)$ characterized on measurable rectangles by
\[
\Bigl(\bigotimes_{i=1}^n \mu_i\Bigr)\Bigl(\prod_{i=1}^n A_i\Bigr)
\;=\; \prod_{i=1}^n \mu_i(A_i), \qquad A_i \in \mathcal{B}(X_i).
\]
We refer to the induced map
\[
\bigotimes \colon \prod_{i=1}^n \Delta(X_i) \longrightarrow \Delta\Bigl(\prod_{i=1}^n X_i\Bigr),
\qquad (\mu_i)_{i=1}^n \longmapsto \bigotimes_{i=1}^n \mu_i,
\]
as the product map, and to its image
\[
\Delta^{\otimes}\Bigl(\prod_{i=1}^n X_i\Bigr) \;\coloneqq\;
\Bigl\{\bigotimes_{i=1}^n \mu_i \,:\, \mu_i \in \Delta(X_i) \text{ for all } i\Bigr\}
\;\subseteq\; \Delta\Bigl(\prod_{i=1}^n X_i\Bigr)
\]
as the set of product probability measures on $\prod_{i=1}^n X_i$.\footnote{Strictly
speaking, $\bigotimes_{i=1}^n \mu_i$ is constructed on the product $\sigma$-algebra
$\bigotimes_{i=1}^n \mathcal{B}(X_i)$, which is contained in but, in general, strictly
smaller than $\mathcal{B}(\prod_{i=1}^n X_i)$; see Footnote~\ref{fn:prod_sigma_alg}.
Equality of the two $\sigma$-algebras, hence well-posedness of $\bigotimes_{i=1}^n \mu_i$
as an element of $\Delta(\prod_{i=1}^n X_i)$, holds in particular when each $X_i$ is
second countable, which is the case throughout Chapter~\ref{ch:contributions1}.}

The product map is continuous\footnote{We give the argument for the case used throughout Chapter~\ref{ch:contributions1}, namely $X_i$ compact (Hausdorff and second countable). Let $\mu_i^\delta \to \mu_i$ be a convergent net in $\sigma(\Delta(X_i),C^0(X_i))$ for each $i$, and set $X := \prod_{i=1}^n X_i$. For $f_i \in C^0(X_i)$ write $(\otimes_{i=1}^n f_i)(x_1,\dots,x_n) := \prod_{i=1}^n f_i(x_i)$. By the defining property of the product measure
\[
\int_X \Big(\bigotimes_{i=1}^n f_i\Big)\, d\Big(\bigotimes_{i=1}^n \mu_i^\delta\Big)
= \prod_{i=1}^n \int_{X_i} f_i \, d\mu_i^\delta
\;\longrightarrow\;
\prod_{i=1}^n \int_{X_i} f_i \, d\mu_i
= \int_X \Big(\bigotimes_{i=1}^n f_i\Big)\, d\Big(\bigotimes_{i=1}^n \mu_i\Big),
\]
and by linearity the same holds on the linear span $\mathcal{A}$ of such product functions. Now $\mathcal{A}$ is a subalgebra of $C^0(X)$ (products of product functions are product functions) that contains the constants and separates the points of the compact Hausdorff space $X$, hence $\mathcal{A}$ is $\|\cdot\|_\infty$ dense in $C^0(X)$ by the Stone--Weierstrass theorem (Theorem~4.47 of \cite{folland1999real}). Since $|\int g \, d\nu| \le \|g\|_\infty$ uniformly over $\nu \in \Delta(X)$, a standard argument extends the convergence from $\mathcal{A}$ to all of $C^0(X)$. For the sequential statement on (not necessarily compact) separable metric spaces, see Theorem~2.8 of \citet{billingsley1999}.} when its domain is equipped with the product weak* topology and its codomain with the weak* topology. Since each $\Delta(X_i)$ is compact, it follows that $\Delta^{\otimes}$ is compact. Unlike $\Delta  
\left(\prod_{i=1}^n X_i\right)$, however, $\Delta^{\otimes}$ is not convex whenever the product involves at least two nontrivial factors.

Fix a topological space $(X, \tau)$. For $p,q\in\Delta$, we say that $p$ is absolutely continuous with respect to $q$, written $p\ll q$, if for $A\in\mathcal{B}(X)$, $q(A)=0$ implies $p(A)=0$. Following \cite{cerreia2026making}, we give the following definitions of statistical distance and divergence.

\begin{definition}\label{def:ch2:divergence}
  A function $c:\Delta\times\Delta\to [0, +\infty]$ is a statistical distance if:
  \begin{enumerate}
      \item[(i)] $c$ is lower semicontinuous, that is, for all $a \in [0, +\infty]$, the sublevel set
      \begin{equation*}
          \left\{ (p,q) \in \Delta \times \Delta: c(p,q)\le a \right\}
      \end{equation*}
      is closed with respect to the product of the weak* topologies. 
      \item[(ii)] $c$ is non-degenerate, that is, $c(p,q)=0$ if and only if $p=q$.
  \end{enumerate}
  A statistical distance $c$ is a divergence if, in addition, $(iii)$ for each $p,q \in \Delta$, 
  \begin{equation*}
      c(p, q) < +\infty \Longrightarrow p \ll q.
  \end{equation*}
\end{definition}

Let $(Y, \sigma)$ be a topological space.

\begin{definition}

A functional $J:B_b(Y)\to \R$ is said to be:

\begin{enumerate}\label{def:ch2:Jfunctional}
    \item[(i)] Monotone: if for all $f, g \in B_b(Y)$ such that $f \ge g$\footnote{The order on $B_b(Y)$ is the pointwise partial order induced by the usual order on $\R$.} one has $J(f) \ge J(g)$.
    \item[(ii)] Normalized: if for every $y \in \R$, $J(y)=y$, where, by abuse of notation, $y \in B_b(Y)$ denotes the constant function $y\mathbf{1}_Y$.
\end{enumerate}
\end{definition}

%% file: Chapters/contributions1.tex
\chapter{Self-Confirming Equilibrium under Model Misspecification}\label{ch:contributions1}

This chapter presents original work. We study the steady states of a game of partial feedback played recurrently within a large, anonymous population of players partitioned into a finite set of roles. In each round players are drawn, one per role, and matched to play a game; after play, each observes private feedback that depends on, but need not reveal, the actions taken in the opposing roles. Over many such encounters a player accumulates observations about the play she faces and forms conjectures about the distribution of actions in the opposing populations, updating them in light of the received feedback. Departing from the classical account, we do not require that a player trust any of her conjectures as a correct description of the strategic environment; players rather treat them as approximations of a more complex reality that is only partially modeled and partially observed, and are averse to the possibility that the conjectures might be misspecified. We abstract from dynamic and intertemporal incentives throughout and take players to maximize current-round payoffs, so that the object of study is the resting (fixed) point of the population's play rather than the learning dynamics that generated it. The equilibrium notion appropriate to such recurrent, mass-population interactions is the self-confirming equilibrium (SCE) of \citet{battigalli1987thesis} and \citet{fudenberg1993self}. Its central idea is that a steady state need not be a Nash equilibrium: players best respond to their conjectures, but those conjectures are disciplined only by the feedback that repeated play generates. An SCE thus requires each player to best respond to conjectures that are \emph{confirmed}, that is, consistent with the empirical evidence her own actions and feedback produce, while leaving unobserved contingencies unconstrained.

Our contribution is to extend this by-now classical notion to players who are not Bayesian expected-utility maximizers but who explicitly recognize, and are averse to, the misspecification of their own conjectures. In doing so we build on two lines of work. The first is the study of self-confirming equilibrium under ambiguity, and in particular the smooth self-confirming equilibrium (SSCE) of \citet{battigalli2015self}, which is the principal inspiration for the present chapter and which first connected the self-confirming logic to non-neutral attitudes toward uncertainty. The second is the decision-theoretic axiomatization of misspecification averse preferences of \citet{cerreia2026making}, which gives the representation of individual preferences on which our equilibrium concept rests.

Bringing these ideas together yields the equilibrium concept we call \emph{self-confirming equilibrium under model misspecification} (SCEM), together with a number of results that we preview here. First, we introduce a class of \emph{games with feedback and uncertainty attitudes}, in which the base game is enriched with a feedback structure $f$ and, for each player, a pair of primitives (a statistical distance and a family of aggregators) encoding her attitude toward uncertainty. Each player
evaluates her actions through a misspecification-sensitive functional adapted from \citet{cerreia2026making}. The theory is flexible enough to contain the variational, maxmin, and
smooth-ambiguity evaluations as special cases, and collapses to expected utility when beliefs are unambiguous and the player is neutral to misspecification.

Second, we define SCEM as the natural steady-state concept for such games, requiring that each player both best responds to her structured conjectures under the misspecification-sensitive criterion (rationality) and holds only conjectures that her feedback cannot rule out (confirmed conjectures), and we establish existence.

Third, we show that misspecification aversion severs a connection that holds for both SCE and SSCE: it is no longer true that every mixed-strategy Nash equilibrium is self-confirming. Recovering a Nash-type foundation therefore calls for a distinct notion of \emph{Robust Nash Equilibrium} (R-NE), which we develop.

Fourth, we carry out two comparative-statics exercises. With respect to \emph{feedback}, we order feedback structures by informativeness and show that more informative feedback shrinks the set of observationally consistent conjectures, tightening the equilibrium requirement in the expected direction.
With respect to \emph{uncertainty aversion}, by contrast, the natural monotonicity conjecture (that a more uncertainty-averse population should sustain more self-confirming profiles) fails in general, as we show by counterexample.

Fifth, we isolate the behavioral assumption responsible for this failure. Restricting attention to observationally consistent divergences, while retaining payoff observability, restores monotonicity in uncertainty aversion and, in doing so, reconnects our framework with the SSCE of \citet{battigalli2015self}, which reappears as a special case.

The chapter is organized as follows. Section~\ref{sec:gameswithfeedbackanduncertainty} introduces the primitives. Section~\ref{sec:information_structure} studies the informational content of feedback, defining the observationally consistent conjectures and characterizing them, with fully informative feedback and payoff observability as leading special cases. Section~\ref{sec:SCEM} defines self-confirming equilibrium under model misspecification. Section~\ref{sec:tax} situates SCEM within the taxonomy of equilibrium concepts, defines the notion of Robust Nash Equilibrium (R-NE) and we finally prove existence of both SCEM and R-NE. Sections~\ref{Sec:comp_feedback} and \ref{sec:comparative_aversion} develop the comparative statics, the first with respect to the feedback structure and the second a negative result with respect to uncertainty aversion. Section~\ref{sec:28} introduces observationally consistent divergences, under which, in combination with payoff observability, monotonicity in uncertainty aversion is recovered and the connection with SCEA is made precise. Section~\ref{sec:conclusion} concludes.

\section{Games with Feedback and Model Misspecification}\label{sec:gameswithfeedbackanduncertainty}

More formally, a game with feedback is a pair $(G,f)$, where $G=(I, (A_i, u_i)_{i \in I})$ is a game and $f=(f_i)_{i \in I}$ is a feedback structure; in particular:

\begin{itemize}
    \item $I$ denotes the finite set of player roles. When no confusion can arise, we use the terms player role and player interchangeably.
    \item $A_i$ is a compact, second countable\footnote{Second countability of the action spaces is assumed to ensure that the Borel $\sigma$-algebra of the product space coincides with the product $\sigma$-algebra of the Borel $\sigma$-algebras on the action spaces, that is,
\[
\mathcal{B}\!\left(\prod_{i\in I} A_i\right)=\bigotimes_{i\in I}\mathcal{B}(A_i),
\]
to guarantee the well-posedness of the definition of support of a probability measure, and to avoid unilluminating topological considerations.}, Hausdorff topological space\footnote{\label{fn:3:PolishSPace}In particular, by Urysohn’s Metrization Theorem (see Theorem~23.1 of \cite{willard2012general}), $A_i$ is metrizable and separable. Since every compact metric space is complete, it follows that $A_i$ is Polish, that is, a separable completely metrizable topological space.} of actions available to player $i \in I$, while $A\coloneq\prod_{i\in I}^{}A_i$ denotes the set of action profiles, and $A\mi \coloneq\prod_{j\neq i}^{}A_j$ denotes the profile of other players' actions, both endowed with the product topology. 
    \item $u_i:A\to\R$ is a continuous von Neumann-Morgenstern utility function, representing player $i$'s preferences over action profiles $A$.
    \item $M_i$ is a $\mathrm{T}1$ topological space of messages that player $i$ may receive ex post.
    \item $f_i:A\to M_i$ is a Borel-measurable feedback function of player $i$, and $f=(f_i)_{i\in I}$ denotes the profile of feedback functions.
\end{itemize}

Being uncertain about the actions of players in other roles, each player $i$ entertains conjectures over the set of other-actions $A\mi$, represented by some $\mu^i \in \Delta(A\mi)$, where $\Delta(A\mi)$ denotes the set of countably additive Borel probability measures on the topological space $A\mi$, equipped with the weak* topology $\sigma(\Delta(A\mi), C_b^0(A\mi))$. Recall that when $A_i$ is compact, $\Delta(A_i)$ is weak$^*$ compact\footnote{See Section~\ref{sec:math_definitions_chapter1} for more details.}. Given a conjecture $\mu^i$, player $i$ evaluates actions $a_i \in A_i$ via the expected utility functional $U_i:A_i\times \Delta(A\mi)\longrightarrow \R$:
\begin{equation*}
    U_i(a_i, \mu^i)\coloneq \int_{A\mi}u_i(a_i, a\mi) \mu^i(da\mi).
\end{equation*}

Note that $U_i$ is well defined and bounded. Moreover, for each fixed $a_i \in A_i$, the section $U_i(a_i,\cdot) : \Delta(A_{-i}) \to \mathbb{R}$ is continuous.\footnote{The map $U_i$ is well defined because $u_i$ is continuous, hence for every fixed $a_i \in A_i$ the section $u_i(a_i,\cdot)$ is continuous, and therefore Borel measurable. Since each action space is compact, the product space $A_{-i}$ is compact; thus $u_i(a_i,\cdot)$ is bounded, so it is integrable with respect to every probability measure $\mu^i \in \Delta(A_{-i})$. Furthermore,
\[
|U_i(a_i,\mu^i)| \le \|u_i\|_\infty
\]
for all $(a_i,\mu^i)\in A_i\times\Delta(A_{-i})$, so $U_i$ is bounded. Finally, for fixed $a_i$, the map $\mu^i \mapsto U_i(a_i,\mu^i)$ is continuous with respect to the weak* topology on $\Delta(A_{-i})$ being the evaluation functional corresponding to the bounded continuous function $u_i(a_i , \cdot)$.\label{fn:2}}

Through strategic reasoning, feedback collected in previous rounds, and exogenous (contextual) features of the interaction not explicitly modeled in the formal game structure, such as past experience, preplay communication, cultural norms, conventions, and other cues that may coordinate beliefs, each player posits a set of beliefs over the actions of players in the opposing roles \citep{crawford1995theory, marinacci2000ambiguous}. Following \cite{hansen2022structured} and \cite{cerreia2026making}, we call such a collection a set of \emph{structured conjectures} (or \textit{structured beliefs}), denoted by $\mathrm{Q}^i\subseteq \Delta(A_{-i})$.  

\begin{definition}
    A set of structured conjectures (or beliefs) for player $i$ is a compact, non-empty subset $\emptyset \neq \mathrm{Q}^i\subseteq \Delta(A_{-i})$. We denote by $\mathfrak{Q}_i$ the class of admissible structured beliefs for player $i$; each element $\mathrm{Q}^i$ of $\mathfrak{Q}_i$ is a set of structured beliefs.
\end{definition}

The interpretation is that players regard their structured conjectures as approximations, rather than correct descriptions, of a more complex, only partially observed and partially modeled reality. Since players are aware that their conjectures may be misspecified, and are averse to this misspecification, they do not simply maximize $U_i$ under a single fixed conjecture; instead, inspired by \cite{cerreia2026making}, we represent their preferences over actions $A_i$ by the misspecification sensitive functional
\begin{equation}\label{eq:Vi}
    V_{i}^{c_i, \mathcal{J}_i}(a_i, {\mathrm{Q}^i})\coloneq J_{\mathrm{Q}^i}\left( \min_{\alpha_{-i}\in \Delta(A_{-i})} \left\{ U_i(a_i,\alpha_{-i}) + c_i(\alpha\mi,\cdot) \right\} \right).
\end{equation}
Here, $c_i : \Delta(A_{-i}) \times \Delta(A_{-i}) \to [0, \infty]$ is a statistical distance, and for each
$\mathrm{Q}^i \in \mathfrak{Q}_i$, the map
\(J_{\mathrm{Q}^i}\colon B_b(\mathrm{Q}^i)\to \mathbb{R}\) is a
normalized and monotone functional\footnote{When $\mathrm{Q}^i=\left\{ q \right\}$ is a singleton,
$B_b(\left\{ q \right\})$ is (isomorphic to) $\R$, hence the only normalized functional $J_{\left\{ q \right\}}\colon B_b(\left\{ q \right\})\cong \R\to \R$ is the identity. This singleton case will be important below for representing correct conjectures. \label{fn:1}}; see
Definitions~\ref{def:ch2:divergence} and~\ref{def:ch2:Jfunctional}. We denote
the family of such functionals by
$\mathcal{J}_i \coloneq (J_{\mathrm{Q}^i})_{\mathrm{Q}^i \in \mathfrak{Q}_i}$. Lemma~\ref{th:appB_welldefinedminimum} shows that, whenever \(\mathrm{Q}^i\) is compact and the additional assumptions introduced in this section are satisfied, the inner minimization functional 
\begin{equation*}
    q \longmapsto \min_{\alpha_{-i}\in \Delta(A_{-i})} \left\{ U_i(a_i,\alpha_{-i}) + c_i(\alpha\mi,q) \right\}
\end{equation*}
is well defined, lower semicontinuous, bounded, and measurable. In particular, this ensures that $V_i$ is itself well defined. When the statistical distance $c_i$ and the evaluation rule $\mathcal{J}_i$ are fixed and understood from context, we lighten the notation by writing simply $V_i(a_i, \mathrm{Q}^i)$.

The primitives introduced so far naturally organize into a hierarchy of game-theoretic objects, each carrying progressively richer structure. The base game structure $G = (I, (A_i, u_i)_{i \in I})$ specifies roles, actions, and payoffs, and suffices to define the classical mixed-strategy Nash equilibrium. When each player $i$ is additionally endowed with a statistical distance $c_i$, capturing the cost of departing from a reference conjecture, we obtain a \emph{game with statistical distances}, 
\begin{equation*}
    (G, c)
\end{equation*} 
where $c = (c_i)_{i \in I}$ is the profile of distances. Finally, when the game players receive partial information about opponents' play and are uncertain about other players' actions, the relevant object is the full tuple
\begin{equation*}
    (G, f, c, \mathcal{J}),
\end{equation*} 
which we call a \emph{game with feedback and uncertainty attitudes}. Here $f = (f_i)_{i \in I}$ is the profile of feedback functions, and $\mathcal{J} = (\mathcal{J}_i)_{i \in I}$ collects the families of evaluation functionals $\mathcal{J}_i = (J_{\mathrm{Q}^i})_{\mathrm{Q}^i \in \mathfrak{Q}_i}$, one for each set of structured beliefs $\mathrm{Q}^i \subseteq \Delta(A_{-i})$.

\begin{example}\label{ex:bayesian_KMM}

An important example of \eqref{eq:Vi} is a model-misspecification averse Bayesian criterion inspired by the smooth preferences of \cite{klibanoff2005smooth}, and represented by
\begin{equation*}
        V_i(a_i,{\mathrm{Q}^i})= \phi_{ \mathrm{Q}^i}^{-1}\left( \int_{\mathrm{Q}^i}\phi_{\mathrm{Q}^i} \left( \min_{\alpha_{-i}} \left\{ U_i(a_i,\alpha_{-i}) + c_i(\alpha_{-i},q_{-i}) \right\}\right) \ \mu_{\mathrm{Q}^i}(d q_{-i}) \right),
\end{equation*}
where, for each $i \in I$ and $\mathrm{Q}^i \in \mathfrak{Q}_i$, $\mu_{\mathrm{Q}^i}\in \Delta(\mathrm{Q}^i)$ is a prior\footnote{The set $\mathrm{Q}^i$ is no longer interpreted as an exhaustive set of possible models. Hence, the prior does not represent a belief about which model is correct; rather, it represents a belief about which structured conjecture is best fitting. See Section~6.4 of \cite{cerreia2026making} for more details.} over the set of structured conjectures $\mathrm{Q}^i$, while the function $\phi_{\mathrm{Q}^i}:\mathbb{R}\to \mathbb{R}$ accounts for aversion to prior uncertainty. The distance $c_i$ represents fear of misspecification relative to a reference belief $q\mi$. See Section~6.1 of \cite{cerreia2026making} for more details. 

A useful further specification is the entropic criterion defined by
    \begin{equation*}
        c_i(p, q) \coloneq \lambda\mathrm{KL}(p \| q),
        \qquad
        \phi_{\mathrm{Q}^i}(t)\coloneq -e^{-\frac{1}{\xi}t},
    \end{equation*}
    where $\lambda >0$ is a misspecification-aversion parameter, with lower values indicating greater aversion to misspecification, and $\mathrm{KL}(\cdot \| \cdot)$ is the Kullback--Leibler divergence
    \begin{equation*}
        \mathrm{KL}(p\|q)
        =
        \begin{cases}
        \displaystyle \int_{A\mi} \log\!\left(\frac{dp}{dq}\right)\, dp,
        & p \ll q, \\[1.2em]
        +\infty,
        & \text{otherwise.}
        \end{cases}
    \end{equation*}
    The parameter $\xi>0$ captures aversion to prior uncertainty. This interpretation of the parameters $\lambda$ and $\xi$ is also justified by the limiting behavior in Proposition~8 of \cite{cerreia2026making}.

Finally, the Bayesian criterion has the computational and interpretive advantage that in order to compute the value of an action $a_i$ under conjectures $\mathrm{Q}^i$, one only needs to compute integrals with respect to elements of $\mathrm{Q}^i$, rather than optimize directly over the whole of $\Delta(A\mi)$. In fact, by the duality formula of \cite{ben2007old} (Theorem~4.2), 
\begin{equation*}
    \inf_{\alpha\mi}\left\{U_i(a_i, \alpha\mi)+\lambda\mathrm{KL}(\alpha\mi \| q\mi)     \right\}=\lambda\sup_{\eta \in \mathbb{R}}\left\{\eta - \int\psi^*\left(\eta - \frac{u_i(a_i, \cdot)}{\lambda}\right)\ dq\mi \ \right\},
\end{equation*}
where $\psi^*(t)\coloneq \sup_{s \in \mathbb{R}} \{st- \psi(s)\}$ is the convex conjugate of $\psi$, with $\psi$ chosen as the generator that induces the Kullback--Leibler divergence. The criterion becomes
\begin{equation*}
\phi_{ \mathrm{Q}^i}^{-1}\left( \int_{\mathrm{Q}^i}\phi_{\mathrm{Q}^i} \left( \lambda\sup_{\eta \in \mathbb{R}}\left\{\eta - \int\psi^*\left(\eta - \frac{u_i(a_i, \cdot)}{\lambda}\right)\ dq\mi \ \right\}\right) \ \mu_{\mathrm{Q}^i}(d q_{-i}) \right).
\end{equation*}
The same representation holds for a larger class of divergences containing the $\mathrm{KL}$ divergence. The usefulness of this representation is that once a structured conjecture $q\mi\in \mathrm{Q}^i$ is fixed, the infinite-dimensional minimization over $\Delta(A\mi)$ is replaced by a one-dimensional optimization over $\eta\in\mathbb{R}$ and by an integral of a known transform of the payoff function $u_i(a_i,\cdot)$ with respect to $q\mi$. Thus, to evaluate an action $a_i$, the player does not need to search directly over the whole space of possible misspecifications $\Delta(A\mi)$; it is enough to know the structured conjectures in $\mathrm{Q}^i$ and to compute the corresponding integrals. 
\end{example}

The feedback structure of a game $(G,f)$ determines how players learn from ex post observations, and hence how they form and revise their conjectures through repeated play. Every feedback function $f_i:A\to M_i$ and action $a_i$ induce a pushforward map $\widehat{f}_{i,a_i}:\Delta(A\mi)\to\Delta(M_i)$, defined by
\begin{equation*}
    \widehat{f}_{i,a_i}(\mu^i)(A)\coloneq \mu^i(f_{i, a_i}^{-1}(A)), \ \forall A \in \mathcal{B}(M_i),
\end{equation*}
where $f_{i, a_i}:A\mi \to M_i$ is the $a_i$-section of $f_i$, that is, $f_{i, a_i}(a\mi)\coloneq f_i(a_i, a\mi)$. Intuitively, given a conjecture $\mu^i$ and action $a_i$, the pushforward measure $\widehat{f}_{i,a_i}(\mu^i)$ is the distribution over messages $M_i$ that is predicted by $\mu^i$. A simple argument (see Theorem \ref{th:appA_pushforwardcont}) shows that when the section of the feedback function $f_i(a_i, \cdot)$ is continuous, then so is the corresponding pushforward map.

\begin{definition}\label{def:observationally_consistent_beliefs}
    Let $\nu\mi \in \Delta(A\mi)$ denote the distribution of actions in the opposing populations\footnote{Note the dual epistemological role played by elements of $\Delta(A\mi)$: on the one hand, a measure $\mu^i \in \Delta(A\mi)$ represents player $i$'s subjective conjecture about the actions of opponents; on the other, $\nu\mi \in \Delta(A\mi)$ describes the objective, population-level frequency of play. The same mathematical object, a probability measure over $A\mi$, thus admits both a \emph{belief} interpretation and a \emph{frequentist} interpretation. Throughout, we reserve $\mu^i$ (and sets thereof, $\mathrm{Q}^i$) for the former and $\nu\mi$ for the latter.}. The set of conjectures that are observationally consistent with $\nu\mi$ given action $a_i\in A_i$ is
    \begin{equation*}
        \mathcal{O}_i(a_i,\nu\mi)\coloneq\left\{ \mu^i \in \Delta(A\mi): \widehat{f}_{i,a_i}(\mu^i)=\widehat{f}_{i,a_i}(\nu\mi)\right\}= \widehat{f}_{i,a_i}^{-1}\!\left(\left\{\widehat{f}_{i,a_i}(\nu\mi)\right\}\right).
    \end{equation*}
\end{definition}

Intuitively, $\mathcal{O}_i(a_i,\nu\mi)$ is the set of all conjectures that player $i$ cannot distinguish from the true distribution $\nu\mi$ on the basis of the feedback received after playing $a_i$. Any conjecture $\mu^i$ in $\mathcal{O}_i(a_i,\nu\mi)$ predicts exactly the same distribution over messages as $\nu\mi$ does, and is therefore unfalsifiable given the available feedback. Note that for any $a_i \in A_i$ and $\nu\mi \in \Delta(A\mi)$, the set of observationally consistent conjectures is non empty, containing $\nu \mi \in \mathcal{O}_i(a_i,\nu\mi)$, and, when $f_{i, a_i}$ is continuous and other topological assumptions hold\footnote{Singletons are closed in the weak* topology on $\Delta(M_i)$ under some further topological assumption; a sufficient condition is for $M_i$ to be compact Hausdorff.}, compact, being the preimage of a closed set under a continuous map defined on a compact domain.

If $\alpha=(\alpha_i)_{i\in I}$ is a mixed action profile\footnote{At the risk of repetition, a mixed action $\alpha_i$ should be interpreted as the population distribution of play in role $i$, not necessarily literal within-player randomization.}, that is, $\alpha \in \prod_{i \in I}^{}\Delta(A_i)$, we define the distribution of actions in the opposing populations by
\begin{equation*}
    \nu\mi \coloneq \alpha\mi \coloneq \bigotimes_{j\neq i}{}\alpha_j.
\end{equation*}
The product of measures is well defined as an element of $\Delta(A\mi)$, where $A\mi$ is endowed with the product $\sigma$-algebra, by second countability. We refer to this assumption as \textit{independent matching}. Under the population interpretation of mixed actions, independence is therefore an assumption on the random matching procedure across roles, rather than on randomization by individual players. It rules out correlation across the actions of different opposing roles; allowing instead arbitrary joint distributions over $A\mi$ would lead toward the correlated-equilibrium ideas of \cite{aumann1974subjectivity}. We emphasize that this assumption constrains the data-generating process and not the player's beliefs: structured beliefs $\mathrm{Q}^i$ are allowed to contain non-product measures, and the inner minimization in~\eqref{eq:Vi} ranges over the full simplex $\Delta(A\mi)$, encoding the player's willingness to hedge against correlated misspecification.

We pause briefly to discuss the functional representation\footnote{\label{fn:twopref}In the axiomatic framework of \citet{cerreia2026making}, $V_i$ represents the player's behavioral preference $\succsim_{\mathrm{Q}^i}$, namely the complete preference relation that governs actual choice. The same primitives $(\mathrm{Q}^i,c_i,J_{\mathrm{Q}^i})$ also induce a (possibly incomplete) mental dominance preference $\succsim^{*}_{\mathrm{Q}^i}$ defined by
\[
a_i \succsim^{*}_{\mathrm{Q}^i} a_i'
\iff
\varphi_i(a_i,q)\ge\varphi_i(a_i',q)
\quad\text{for every }q\in\mathrm{Q}^i,
\]
where $\varphi_i(a_i,q):=\min_{\alpha_{-i}}\bigl\{U_i(a_i,\alpha_{-i})+c_i(\alpha_{-i},q)\bigr\}$ is the robust evaluation of $a_i$ at the structured model $q$, and
$\succsim_{\mathrm{Q}^i}$ is a completion of $\succsim^{*}_{\mathrm{Q}^i}$.} of players' preferences over their own actions. The functional $V_i^{c_i,\mathcal{J}_i}$ in Equation~\eqref{eq:Vi} captures a two-layer treatment of player~$i$'s uncertainty about the actions chosen by the opposing populations.

The inner layer expresses fear of misspecification of each individual structured model. Fix any reference conjecture $q\in\mathrm{Q}^i$. Rather than evaluating $a_i$ at $q$ alone, i.e. computing expected utility $U_i(a_i,q)$, the player considers the worst-case expected utility over all $\alpha_{-i}\in\Delta(A_{-i})$, where each alternative is penalized by the cost $c_i(\alpha_{-i},q)$ of departing from $q$. The output
\[
\;\min_{\alpha_{-i}\in\Delta(A_{-i})}\bigl\{
U_i(a_i,\alpha_{-i})+c_i(\alpha_{-i},q)\bigr\}
\]
is the player's robust evaluation of $a_i$ relative to $q$. The divergence $c_i$ controls the player's willingness to entertain unstructured probabilities: a sharper $c_i$ anchors the worst case to $q$ and expresses moderate misspecification fear; a flatter $c_i$ lets the worst case drift toward unstructured probabilities far from $\mathrm{Q}^i$ and expresses severe misspecification fear. The functional $J_{\mathrm{Q}^i}$ aggregates the robust evaluations across the structured models $q\in\mathrm{Q}^i$, expressing ambiguity over those models. 

\begin{example}\label{ex:EU_collapse}
With a general statistical distance $c_i$, \eqref{eq:Vi} represents misspecification sensitive extensions of the classical ambiguity criteria, including the Bayesian functional of Example~\ref{ex:bayesian_KMM} which extends the smooth-ambiguity criterion of \citet{klibanoff2005smooth}, and the maxmin aggregator $J_{\mathrm{Q}^i}(h)=\inf_{q\in\mathrm{Q}^i}h(q)$ which extends the multiple-priors criterion of \citet{gilboa1989maxmin}. In the singleton case $\mathrm{Q}^i=\{q\}$, where $J_{\{q\}}$ is forced to be the identity (Footnote~\ref{fn:1}), $V_i$ is the variational evaluation of \citet{maccheroni2006ambiguity}.

The classical criteria are moreover recovered once the inner layer is made neutral. Misspecification neutrality is represented, following \citet{maccheroni2006ambiguity}, by the distance
\begin{equation*}
c_i^\infty(p,q) =
\begin{cases}
0, & p=q,\\
+\infty, & p\neq q,
\end{cases}
\end{equation*}
    which is a statistical distance in the sense of
    Definition~\ref{def:ch2:divergence}, and under which the inner minimization
    in \eqref{eq:Vi} is attained at the reference conjecture itself,
\begin{equation*}
    \min_{p\in\Delta(A_{-i})}\big\{U_i(a_i,p)+c_i^\infty(p,q)\big\}=U_i(a_i,q).
\end{equation*}
    Coupling $c_i^\infty$ with the aggregators above then yields the standard criteria:
\begin{itemize}
    \item for a compact, non-empty $C\subseteq\Delta(A_{-i})$, the choice
    $\mathrm{Q}^i=C$ with $J_{C}(h)=\inf_{q\in C}h(q)$ gives
    \begin{equation*}
        V_i^{c_i^\infty,\mathcal{J}_i}(a_i,C)=\inf_{q\in C}U_i(a_i,q),
    \end{equation*}
    the multiple-priors criterion of \citet{gilboa1989maxmin} (see also
    Example~\ref{ex:MSCE});
    \item the Bayesian functional of Example~\ref{ex:bayesian_KMM} gives
    \begin{equation*}
        V_i^{c_i^\infty,\mathcal{J}_i}(a_i,\mathrm{Q}^i)
        =\phi_{\mathrm{Q}^i}^{-1}\bigg(\int_{\mathrm{Q}^i}
        \phi_{\mathrm{Q}^i}\big(U_i(a_i,q_{-i})\big)\,
        \mu_{\mathrm{Q}^i}(dq_{-i})\bigg),
    \end{equation*}
    the smooth-ambiguity criterion of \citet{klibanoff2005smooth};
    \item the singleton $\mathrm{Q}^i=\{q\}$ gives
    \begin{equation*}
        V_i^{c_i^\infty,\mathcal{J}_i}(a_i,\{q\})=U_i(a_i,q),
    \end{equation*}
    that is, the misspecification sensitive criterion collapses to the classical expected utility criterion.
\end{itemize}
    We highlight that the expected utility collapse is caused by neutrality, not by correctness of beliefs: the identity in the last bullet holds whether or not $q$ coincides with the true distribution of play, while unambiguous beliefs alone do not suffice, as Example~\ref{ex:mne_SCEM_counterexample} shows.
\end{example}

The structure thus separates two distinct concerns. \emph{Model ambiguity} arises because the player is uncertain over which $q\in\mathrm{Q}^i$ best describes opponents' play; the aggregator $J_{\mathrm{Q}^i}$ encodes how she trades off across these candidate models. \emph{Model misspecification} arises because the player does not assert that any $q\in\mathrm{Q}^i$ literally describes opponents' play; the divergence $c_i$ encodes how she trades off perturbations of each candidate.

\newpage

\section{Information, Feedback, and Confirmed Beliefs}\label{sec:information_structure}

This section studies properties related to the information structure of a game with feedback $(G, f)$, which we consider to be fixed throughout the section. For more details on the properties related to information and feedback in self-confirming equilibrium, see \cite{battigalli2016analysis}.

Recall that a feedback function for player $i$ is a map $f_i: A\to M_i$, where $f_i(a_i, a_{-i})$ represents the feedback received by player $i$ after playing action $a_i$ when the opposing population plays the profile $a_{-i}$. From the perspective of player $i$, who chooses action $a_i$, the actions of the other players are distinguishable only through the feedback they generate; that is, after playing the game, the player learns only which element of the partition\footnote{Formally, $\mathcal{F}_i(a_i)$ is a partition after discarding possibly empty elements.}
\begin{equation*}
    \mathcal{F}_i(a_i)\coloneq \left\{ f_{i,a_i}^{-1}\left( \left\{ m_i \right\} \right):m_i \in M_i \right\}
\end{equation*}
contains the opponents' original action profile. More generally, define
\begin{equation*}
    \mathcal{H}_i(a_i)\coloneq \sigma(f_{i,a_i}),
\end{equation*}
to be the $\sigma$-algebra on $A_{-i}$ generated by the section $f_{i,a_i}$ of the feedback function.

Having introduced this notation, we can now express a simple but meaningful characterization of the set of observationally consistent beliefs. Fix a player $i \in I$, an action $a_i \in A_i$, and a distribution of actions in the opposing populations $\nu_{-i} \in \Delta(A_{-i})$. Then, by Definition~\ref{def:observationally_consistent_beliefs},
\begin{equation}\label{eq:observational_sigmalag}
    \mathcal{O}_i(a_i, \nu_{-i})
    =
    \left\{
        \mu^i \in \Delta(A_{-i})
        :
        \mu^i|_{\mathcal{H}_i(a_i)}
        =
        \nu_{-i}|_{\mathcal{H}_i(a_i)}
    \right\},
\end{equation}
that is, a belief $\mu^i$ is observationally consistent if and only if it agrees with the true population distribution on the $\sigma$-algebra generated by the $a_i$-section of the feedback map. Thus feedback does not require the conjecture $\mu^i$ to coincide with $\nu_{-i}$ on all events in $A_{-i}$, but only on those events that are observable through the message generated after action $a_i$.

\begin{example}\label{ex:fullyinformative}
    We say that feedback for player $i$ is fully informative when, given any action $a_i$ played and message $m_i \in M_i$ received, the player is able to uniquely identify the opponents' action profile. That is, when every section $f_{i, a_i}$ is injective, so that the partition $\mathcal{F}_i(a_i)$ consists of singletons.

    A canonical example of fully informative feedback is given by
    \begin{equation*}
        M_i \coloneq A_{-i}, \quad f_i(a_i, a_{-i}) \coloneq a_{-i},
    \end{equation*}
    which corresponds to the usual setting in game theory, where players have complete information on the actions played by their opponents.

    Feedback for player $i$ is completely uninformative if every section $f_{i,a_i}$ of the feedback function is constant. Equivalently, after playing the game, the player cannot infer anything about the opponents' actions from the feedback received, and the induced partition is trivial $\mathcal{F}_i(a_i)=\{A\mi\}.$

    In terms of the induced information $\sigma$-algebras, fully informative feedback corresponds to $\mathcal{H}_i(a_i)=\mathcal{B}(A_{-i})$, under the regularity assumptions stated below in Lemma~\ref{th:sec4:informativeimpliessingleton}, while completely uninformative feedback corresponds to $\mathcal{H}_i(a_i)=\{\emptyset,A_{-i}\}$.
\end{example}

A natural assumption on the feedback structure $f$ is payoff observability, meaning that each player observes her own payoff.

\begin{definition}\label{def:payoffobservability}
A game with feedback $(G, f)$ has observable payoffs if, for all $i \in I$ and all $a_i \in A_i$, the function $u_i(a_i, \cdot)$ is $\mathcal{H}_i(a_i)$-measurable. 
\end{definition}

Although payoff observability is a natural assumption, we do not impose it in general, since it is not needed for a number of our results. Whenever payoff observability is required, we will state this explicitly.

Under observable payoffs, each player's payoff depends only on the ex post information, that is, for all $i \in I$, all $a_i \in A_i$, and all $a\mi', a\mi'' \in A\mi$
\begin{equation*}
f_{i,a_i}(a\mi')=f_{i,a_i}(a\mi'') \Longrightarrow u_i(a_i, a\mi')=u_i(a_i, a\mi'').\footnote{In fact, by the Doob–Dynkin lemma, there exists a (measurable) function $\tilde{u}:M_i \to \R$ such that the diagram commutes, $u_i(a_i, \cdot)=\tilde{u} \circ f_{i,a_i}$, which proves the claim. 
}
\end{equation*}

Conversely, if, after playing a fixed action $a_i$, the player receives different payoffs, then she can infer that the opponents' action profile must have been different. 

In fact, under payoff observability, one can say more about the structure of the induced expected-utility functions. 

\begin{lemma}\label{th:lemma:payoff_observability}
    Let $(G, f)$ be a game with feedback and observable payoffs. Then for every $i \in I$, $a_i \in A_i$, and $\nu\mi \in \Delta(A\mi)$,
    \begin{equation*}
        U_i(a_i, \mu^i)=U_i(a_i, \nu\mi), \quad \forall \mu^i \in \mathcal{O}_i(a_i, \nu\mi).
    \end{equation*}
\end{lemma}
\begin{proof}
 See Appendix~\hyperref[proof:lemma26]{B}.
\end{proof}

In addition, if each message space $M_i$ is Polish (that is, a separable completely metrizable topological space), then fully informative feedback implies observable payoffs.

\begin{lemma}\label{th:informativeimpliesobservable}
Let $(G,f)$ be a game with fully informative feedback. Suppose that, for each $i \in I$, the message space $M_i$ is Polish. Then $(G,f)$ has observable payoffs.
\end{lemma}
\begin{proof}
See Appendix~\hyperref[proof:lemma27]{B}.
\end{proof}

Moreover, if feedback is fully informative then, for any own action $a_i$ and distribution of opponents' actions $\nu\mi$, the set of observationally consistent beliefs reduces to a singleton, only containing the correct belief.

\begin{lemma}\label{th:sec4:informativeimpliessingleton}
    Let $(G, f)$ be a game with fully informative feedback. Suppose that, for each $i \in I$, the message space $M_i$ is Polish. Then for each player $i \in I$, each action $a_i \in A_i$, and each distribution of actions in the opposing populations $\nu\mi \in \Delta(A\mi)$:
    \begin{equation*}
        \mathcal{O}_i(a_i, \nu \mi)=\{\nu\mi\}.
    \end{equation*}
\end{lemma}
\begin{proof}
See Appendix~\hyperref[proof:lemma28]{B}.
\end{proof}

Since observational consistency is defined by equality of the induced distributions over messages, the informativeness of $f_{i,a_i}$ determines how much of $\nu\mi$ can be recovered from feedback. Definition~\ref{def:observationally_consistent_beliefs} thus admits two useful polar cases. If feedback is fully  informative, so that the section $f_{i,a_i}$ identifies the opponents' action profile, then observational consistency collapses to correctness: under the regularity assumptions stated above,
\[
    \mathcal{O}_i(a_i,\nu\mi)=\{\nu\mi\}.
\]
This is the informational benchmark closest to the usual game theory framework, in which the player’s conjecture must coincide with the actual population distribution of opponents’ play. At the opposite extreme, if feedback is completely uninformative, so that $f_{i,a_i}$ is constant, then every conjecture induces the same degenerate distribution over messages, and hence
\[
    \mathcal{O}_i(a_i,\nu\mi)=\Delta(A\mi).
\]
Then $\mathcal{O}_i(a_i,\nu\mi)$ measures the residual freedom in beliefs left by the feedback structure after action $a_i$ has been played.

Finally, we define the property of own-strategy independence of feedback.

\begin{definition}\label{def:own_strategy_indep}
    A game with feedback $(G, f)$ satisfies own-strategy independence of feedback if for all $i \in I$, for all $a_i , a_i' \in A_i$,
    \begin{equation*}
        \mathcal{H}_i(a_i)=\mathcal{H}_i(a_i').
    \end{equation*}
\end{definition}

Own-strategy independence of feedback requires that the information player $i$ can obtain about opponents' play does not depend on the action chosen by the player. This is a strong informational restriction: it rules out feedback structures in which different own actions reveal different aspects of opponents' behavior; that is, when combined with payoff observability, the player learns the payoff-relevant information for all actions, not only for the action actually played.

\newpage
\section{Self-Confirming Equilibrium under Model Misspecification}\label{sec:SCEM}

A steady state of the recurrent game requires that each player best responds to a set of structured conjectures supported on the observationally consistent set; that is, players hold no conjecture that their own feedback could have ruled out. We make this precise as follows.

\begin{definition}\label{def:SCEM}
    Let $(G,f, c, \mathcal{J})$ be a game with feedback and uncertainty attitudes. A profile of mixed actions and structured beliefs 
    \begin{equation*}
        (\alpha^*_i, (\mathrm{Q}^i_{a^*_i})_{a^*_i \in \operatorname{supp}(\alpha^*_i)})_{i \in I}
    \end{equation*}
    is a Self-Confirming Equilibrium under Model Misspecification \textnormal{(\SCEM)} if for all $i \in I$ and all $a^*_i \in \operatorname{supp}(\alpha^*_i)$ the following two conditions hold:
\begin{enumerate}
    \item (Rationality): $a^*_i \in \underset{a_i \in A_i}{\operatorname{argmax}} \ V^{c_i, \mathcal{J}_i}_i(a_i, \mathrm{Q}^i_{a^*_i})$.
    \item (Confirmed Conjectures): $\mathrm{Q}^i_{a^*_i}\subseteq\mathcal{O}_i(a^*_i,\alpha^*\mi)$.
\end{enumerate}

We say that a profile of mixed actions $(\alpha_i^*)_{i\in I}$ is a \SCEM\ action profile for the game $(G,f, c, \mathcal{J})$ if for all $i \in I$ and all $a^*_i \in \operatorname{supp}(\alpha^*_i)$ there exists a set of structured beliefs $\mathrm{Q}^i_{a_i^*}$ such that $(\alpha^*_i, (\mathrm{Q}^i_{a^*_i})_{a^*_i \in \operatorname{supp}(\alpha^*_i)})_{i \in I}$ is a \SCEM\ for $(G,f, c, \mathcal{J})$.
\end{definition}

\begin{example}[Finite games]
\label{ex:finite-games}
Finite games provide a simple model of the present theory. Suppose that, for each $i\in I$, the action set $A_i$ is a finite set equipped with the discrete topology $\mathcal{P}(A_i)$, and that the message space $M_i$ is, without loss of generality, also a finite set equipped with the discrete topology. Verifying the topological hypotheses introduced in Section~\ref{sec:gameswithfeedbackanduncertainty} is then straightforward. In particular, all the existence and consistency results derived in this chapter apply to finite games. Example~\ref{ex:mne_SCEM_counterexample} (continued in Example~\ref{ex:counterex_aversion}) provides a concrete two-player instance under fully informative feedback.
\end{example}

Definition~\ref{def:SCEM} mirrors the decomposition of (self-confirming) equilibrium into a behavioral best-response condition and an epistemic consistency condition, but adapts each to misspecification averse players.

\emph{Rationality.} For each support action $a_i^{*}\in\mathrm{supp}(\alpha_i^{*})$, the equilibrium action maximizes 
\begin{equation*}
    V_i(\cdot,\mathrm{Q}^{i}_{a_i^{*}})\footnote{Expanding on Footnote~\ref{fn:twopref}, an alternative SCEM in which the Rationality condition is replaced by the requirement that $a_i^{*}$ dominate every $a_i\in A_i$ under the mental preference
    \begin{equation*}
        \succsim^{*}_{\mathrm{Q}^{i}_{a_i^{*}}}
    \end{equation*} 
    would yield a refinement of Definition~\ref{def:SCEM}, isolating those equilibria that are best responses under every structured conjecture. We do not pursue this alternative here.}
\end{equation*}
holding the structured belief set fixed. In particular, when the player evaluates a hypothetical deviation $a_i\neq a_i^{*}$, she does not re-evaluate $a_i$ as if she had played $a_i$ all along; that is, she does not use $V_i(a_i,\mathrm{Q}^{i}_{a_i})$. The deviation is assessed through the lens of the belief set calibrated by feedback under $a_i^{*}$. Counterfactual deviations are evaluated at on-path beliefs because the player has no learning experience under off-path actions. The action-indexing of the set of structured beliefs acknowledges that under randomized equilibrium play different actions in $\mathrm{supp}(\alpha_i^{*})$ have all been played and have generated their own feedback partitions and each may sustain a distinct structured belief set.

\emph{Confirmed Conjectures.} Every structured conjecture $q$ must lie in the observationally consistent set $\mathcal{O}_i(a_i^{*},\alpha_{-i}^{*})$. This means that $q$ predicts exactly the same distribution over messages as $\alpha_{-i}^{*}$ does, so feedback received under $a_i^{*}$ cannot tell the two apart. The condition is purely epistemic; it rules out conjectures that observation could have falsified, and it does so individually. The set $\mathrm{Q}^{i}_{a_i^{*}}$ is therefore a collection of individually unfalsified models; it is not required to be exhaustive (it may be a strict subset of $\mathcal{O}_i(a_i^{*},\alpha_{-i}^{*})$), nor is the truth $\alpha_{-i}^{*}$ required to belong to it.

Definition~\ref{def:SCEM} is silent on the learning dynamics by which players settle on the set of structured conjectures. We model the equilibrium as the steady state of an unspecified learning process, with players having observed enough feedback under repeated play to settle on a structured belief set calibrated to their experience, on which they best respond. Two features follow. First, beliefs are conditioned on the equilibrium action $a_i^{*}$, since this is the action under which feedback has accumulated; different actions in $\mathrm{supp}(\alpha_i^{*})$ generate different feedback partitions and may therefore sustain different belief sets. Second, counterfactual beliefs about deviations are not modeled: when evaluating $a_i\neq a_i^{*}$, the player keeps $\mathrm{Q}^{i}_{a_i^{*}}$, having no direct experience under $a_i$ from which to recalibrate. A full dynamic theory of learning that justifies SCEM as a long-run limit in the spirit of Berk--Nash equilibrium for misspecified Bayesian players \citep{esponda2016berk} is a natural complement to our work but lies beyond its scope. The works of \citet{battigalli2019learning} and \citet{lanzani2025dynamic} might prove useful in this direction.

We close this section with a more detailed comparison with the Berk--Nash equilibrium of \citet{esponda2016berk}, reformulated in our setting and notation. This is the equilibrium concept in the literature closest to the framework developed thus far.

\begin{example}[Berk--Nash Equilibrium]
\citet{esponda2016berk} consider a finite game $G = (I, (A_i, u_i)_{i \in I})$, which they call the objective game, together with a feedback structure $f_i \colon A \to M_i$. The objects of choice are mixed strategies $\alpha_i \in \Delta(A_i)$, and each profile $\alpha = (\alpha_i)_{i \in I}$ induces, conditional on a player-$i$ action $a_i \in A_i$, an objective distribution over messages $\widehat{f}_{i, a_i}(\alpha\mi) \in \Delta(M_i)$. A subjective model for player $i \in I$ consists of a parameter set $\Theta_i$ and a parametric family of conjectures
\begin{equation*}
    \{\mu^{\theta_i}\}_{\theta_i \in \Theta_i}, \qquad \mu^{\theta_i} \in \Delta(A\mi),
\end{equation*}
with associated subjective message distributions $\widehat{f}_{i, a_i}(\mu^{\theta_i}) \in \Delta(M_i)$. For an action profile $\alpha$ and $\theta_i \in \Theta_i$, the weighted Kullback--Leibler divergence is
\begin{equation*}
    K^i(\alpha, \theta_i) \;\coloneqq\; \sum_{a_i \in A_i} \alpha_i(a_i)\, \mathrm{KL}\!\Big(\widehat{f}_{i, a_i}(\alpha\mi) \,\Big\|\, \widehat{f}_{i, a_i}(\mu^{\theta_i})\Big),
\end{equation*}
and the set of closest parameter values given $\alpha$ is
\begin{equation*}
    \Theta_i(\alpha) \;\coloneqq\; \underset{\theta_i \in \Theta_i}{\textnormal{argmin}}\, K^i(\alpha, \theta_i).
\end{equation*}
Players hold second-order beliefs $\beta_i \in \Delta(\Theta_i)$ which, by Bayesian averaging, induce a conjecture over opposing actions
\begin{equation*}
    \overline{\mu}_{\beta_i}(\cdot) \;\coloneqq\; \int_{\Theta_i} \mu^{\theta_i}(\cdot)\, d \beta_i(\theta_i) \;\in\; \Delta(A\mi).
\end{equation*}
A Berk--Nash equilibrium is a profile of mixed actions and second-order beliefs $(\alpha^*_i, \beta_i)_{i \in I}$ such that, for every player $i \in I$,
\begin{enumerate}
    \item (Rationality) For all $a^*_i \in \operatorname{supp}(\alpha^*_i)$,
    \begin{equation*}
        a_i^* \;\in\; \underset{a_i \in A_i}{\textnormal{argmax}} \int_{A\mi} u_i(a_i, a\mi)\, d \overline{\mu}_{\beta_i}(a\mi);
    \end{equation*}
    \item (Confirmed Conjectures) For all $\theta_i \in \operatorname{supp}(\beta_i)$,
        \begin{equation*}
            \theta_i \in \Theta_i(\alpha^*).
        \end{equation*}
\end{enumerate}

The Rationality condition of Berk--Nash equilibrium is the misspecification neutral, ambiguity neutral special case of SCEM rationality. Indeed, fix any structured belief set $\mathrm{Q}^i \in \mathfrak{Q}_i$ and any prior $\nu \in \Delta(\mathrm{Q}^i)$, and couple the dogmatic distance $c_i^\infty$ of Example~\ref{ex:EU_collapse} with the Bayesian functional of Example~\ref{ex:bayesian_KMM} with $\phi_{\mathrm{Q}^i} = \mathrm{id}$. Then
\begin{equation*}
    V_i^{c_i^\infty,\mathcal{J}_i}(a_i,\mathrm{Q}^i)
    = \int_{\mathrm{Q}^i} U_i(a_i,q)\,\nu(dq)
    = U_i\big(a_i,\bar{q}_\nu\big),
    \qquad
    \bar{q}_\nu := \int_{\mathrm{Q}^i} q\,\nu(dq),
\end{equation*}
    an expected utility evaluation at the barycenter $\bar{q}_\nu$ of the prior $\nu$. This is exactly the form of the Berk--Nash Rationality condition, which  evaluates actions by expected utility against the compound conjecture $\overline{\mu}_{\beta_i}$.
    
    The two epistemic conditions, however, differ. The SCEM Confirmed Conjectures condition of Definition~\ref{def:SCEM} requires every structured conjecture to be observationally consistent at every support action, whereas the Berk--Nash condition requires only that the parameters in $\mathrm{supp}(\beta_i)$ minimize the weighted divergence $K^i(\alpha^*,\cdot)$, irrespective of whether the minimum is zero. Two consequences follow. First, the parametric image $\{\mu^{\theta_i} : \theta_i \in \Theta_i\}$ is in general too large to serve as a structured belief set as it contains models that equilibrium feedback itself falsifies. Second, if $\min_{\theta_i} K^i(\alpha^*,\theta_i) > 0$, then every supported minimizer mispredicts the distribution of messages at some support action, so the models supported by $\beta_i$ cannot serve as structured conjectures either.

   Finally, we emphasize that players in \citet{esponda2016berk} optimize and update within their subjective parametric model. The possibility that the model is misspecified is entertained by the external analyst, not by the players, who behave as classical expected utility maximizers against the compound conjecture $\overline{\mu}_{\beta_i}$ and are not endowed with the misspecification sensitive criterion \eqref{eq:Vi}; for further details, see the discussion in Section~\ref{sec:litrev}.

\end{example}

\newpage

\section{A Taxonomy of SCEM}\label{sec:tax}

After defining a new notion of equilibrium $\mathcal{E}$, one is naturally drawn to investigate its relation to the classical mixed-strategy Nash equilibrium (m-NE). The reasons are both mathematical and economic: on the one hand, establishing a relationship between m-NE and $\mathcal{E}$ allows one to exploit the advanced machinery developed for m-NE; for instance, proving the inclusion $\text{m-NE}\subseteq \mathcal{E}$ immediately yields the existence of an $\mathcal{E}$-equilibrium via classical existence theorems for Nash equilibria. On the other hand, a connection with Nash equilibria is already valuable as a standalone game-theoretic result, since Nash equilibrium provides the benchmark against which alternative behavioral predictions and prescriptions are naturally compared. Fix a game $G$. In our setting, the guiding questions are 
\begin{itemize}
    \item[$(\mathrm{N}_\forall)$] \emph{Is every mixed-strategy Nash
 equilibrium of $G$ a SCEM action profile of the game with feedback and uncertainty attitudes $(G,f,c,\mathcal{J})$, for every admissible profile $(f,c,\mathcal{J})$?}

    \item[$(\mathrm{N}_\exists)$] \emph{Given a mixed-strategy Nash
 equilibrium of $G$, does there exist an admissible profile $(f,c,\mathcal{J})$ under which that equilibrium is a SCEM action profile of $(G,f,c,\mathcal{J})$?}
\end{itemize}

For games with feedback, the literature has answered both questions in the affirmative, as every m-NE is self-confirming for every  feedback structure in the case of SCE (see, for instance, \citet{battigalli2012comportamento}), and for every feedback structure and every profile of smooth-ambiguity attitudes in the case of SSCE \citep{battigalli2015self}. In both cases, the proof is simple and intuitive: if $(\alpha^*_i)_i$ denotes a m-NE, then it is sufficient to consider correct conjectures; following our earlier notation, $\mu^i=\alpha^*_{-i}$ for SCE and a correct prior $\delta_{\alpha^*_{-i}}$ for SSCE, with the proof naturally flowing from there. The analogue for SCEM would be to consider a correctly specified singleton set of structured beliefs $\mathrm{Q}^i=\left\{ \alpha_{-i}^* \right\}$, independent of the action in the support of $\alpha_i^*$. Surprisingly, however, this construction fails (Example~\ref{ex:mne_SCEM_proof}), and with it the uniform inclusion. Example~\ref{ex:mne_SCEM_counterexample} exhibits a game $G$, an admissible profile $(f,c,\mathcal{J})$, and a m-NE of $G$ that is not a SCEM action profile of $(G,f,c,\mathcal{J})$, so that the answer to $(\mathrm{N}_\forall)$ is negative. The answer to $(\mathrm{N}_\exists)$ is instead affirmative, see  Corollary~\ref{cor:mNE_neutral_SCEM}. The classical inclusion ceases to be uniform over uncertainty attitudes, and precisely because of aversion to model misspecification. Restoring a Nash-type foundation valid across attitudes therefore calls for a different theory, that of Robust Nash Equilibrium (R-NE), which we develop below. The following example illustrates where the natural proof breaks down.

\begin{example}\label{ex:mne_SCEM_proof}
    We begin by attempting to prove that $\textnormal{m-NE} \subseteq \textnormal{SCEM}$; understanding where the natural proof breaks down will better highlight the difference between our equilibrium concept under model misspecification and the classical definitions. Let $(\alpha_i^*)_i$ be a \textnormal{m-NE} of a game $G$ and consider an arbitrary extension to a game with feedback $(G,f)$. We then take the set of correct beliefs, that is, for all $i \in I$ and $a_i^* \in \operatorname{supp}(\alpha_i^*)$:
    \begin{equation*}
        \mathrm{Q}^i=\mathrm{Q}_{a_i^*}^i=\left\{ \alpha\mi^* \right\}.
    \end{equation*}
    By definition, the Confirmed Conjectures condition in Definition~\ref{def:SCEM} is satisfied, since $\alpha^*_{-i} \in \mathcal{O}_i(a^*_i,\alpha^*_{-i})$. What remains to be verified is the Rationality property. As Footnote~\ref{fn:1} shows, when $\mathrm{Q}^i$ is a singleton the functional $J_{\mathrm{Q}^i}$ reduces to the identity, so that for all $a_i \in A_i$:
    \begin{equation}\label{eq:example_mNE}
        V_i(a_i, \left\{ \alpha\mi^* \right\})= \underset{\alpha\mi \in \Delta(A\mi)}{\min}\left\{ U_i(a_i, \alpha\mi)+c_i(\alpha\mi, \alpha\mi^*) \right\}.
    \end{equation}
    We would like to show that for all $a_i^* \in \operatorname{supp}(\alpha_i^*)$ and all $a_i \in A_i$, the inequality $V_i(a^*_i, \left\{ \alpha_{-i}^* \right\})\ge V_i(a_i, \left\{ \alpha_{-i}^* \right\})$ holds. Lemma~\ref{th:appB_classicalnash} gives:
    \begin{equation*}
        U_i(a^*_i, \alpha\mi^*)\ge U_i(a_i, \alpha\mi^*) \quad \forall a_i \in A_i
    \end{equation*}
    However, this need not imply Rationality. By Equation~\eqref{eq:example_mNE}, the robust evaluation involves an inner minimization whose optimizer generally depends on the deviation \(a_i\). The Nash condition compares payoffs only at the reference conjecture \(\alpha_{-i}^*\), where \(c_i(\alpha_{-i}^*,\alpha_{-i}^*)=0\), and thus controls the objective at a single point of \(\Delta(A_{-i})\). The robust criterion, by contrast, minimizes over all beliefs, allowing departures from \(\alpha_{-i}^*\) at finite distance cost to lower the expected utility. Hence two actions yielding the same payoff at \(\alpha_{-i}^*\) can have different robust values, and the ranking under \(V_i\) may differ from that under \(U_i\). In short, a pointwise Nash comparison does not imply the \SCEM\ rationality condition, which compares penalized infima over all of \(\Delta(A_{-i})\).

\end{example}

Next, we show by means of a counterexample that this is not merely an issue with the chosen proof strategy.

\begin{example}\label{ex:mne_SCEM_counterexample}
Consider the two-player finite game with players $i=1,2$ described by the following payoff matrix:
\[
\begin{array}{c|cc}
\text{1 } \backslash \text{ 2} & L & R \\
\hline
H & (1, 1) & (0, 0) \\
T & (\frac{1}{2}, 0) & (\frac{1}{2}, 1)
\end{array}
\]
where the action sets are $A_1=\{H, T\}$ and $A_2=\{L, R\}$, respectively.  Assume fully informative feedback (see Example \ref{ex:fullyinformative}); in particular, let $f_1(a_1,a_2)=a_2$ and $f_2(a_1,a_2)=a_1$, and consider the quadratic distances\footnote{Up to a constant, this corresponds to $\|\mu -\nu\|_2^2$, where we identify $\mu \in \Delta(A\mi)\cong\Delta^{1}\subseteq\mathbb{R}^2.$} given by
\begin{align*}
    c_1(\mu, \nu) &\coloneq \bigl(\mu(L)-\nu(L)\bigr)^2, \\
    c_2(\mu, \nu) &\coloneq \bigl(\mu(H)-\nu(H)\bigr)^2.
\end{align*}
Consider the mixed-strategy profile $\alpha^*=(\alpha_1^*, \alpha_2^*)$ with $\alpha_1^*\coloneq \tfrac{1}{2}\delta_H + \tfrac{1}{2}\delta_T$ and $\alpha_2^*\coloneq \tfrac{1}{2}\delta_L + \tfrac{1}{2}\delta_R$, both of full support.  A direct computation of expected payoffs shows that $\alpha^*$ is a \textnormal{m-NE}; we now show that it is not a \textnormal{SCEM}.

Focus on player~$1$. Because feedback is fully informative, Lemma~\ref{th:sec4:informativeimpliessingleton} implies that, for every action $a_1 \in A_1$ and every mixed-strategy $\alpha_2 \in \Delta(A_2)$, the set of feedback-consistent beliefs reduces to the (correct) singleton $\mathcal{O}_1(a_1,\alpha_2)=\{\alpha_2\}$.  Suppose for contradiction that $\alpha^*$ is a \textnormal{SCEM}.  Then, for each $a_1 \in A_1$, the Confirmed Conjectures condition together with non-emptiness of the set of structured conjectures forces $\mathrm{Q}^1_{a_1}=\{\alpha_2^*\}$.  By Footnote~\ref{fn:1}, we therefore have
\begin{equation*}
    V_1(a_1, \{\alpha_2^*\})=\min_{\mu \in \Delta(A_2)}\{U_1(a_1, \mu)+c_1(\mu, \alpha_2^*)\}
\end{equation*}
Since $\alpha_1^*$ has full support on $\{H, T\}$, the Rationality condition requires that $V_1(H, \{\alpha_2^*\})=V_1(T, \{\alpha_2^*\})$.  However, explicit evaluation yields
\begin{align*}
    V_1(H, \{\alpha_2^*\}) =& \min_{\mu \in \Delta(A_2)}\left\{ U_1(H, \mu)+c_1(\mu, \alpha_2^*) \right\}=\min_{p \in [0,1]}\left\{ p+\left( p-\frac{1}{2} \right)^2 \right\} = \frac{1}{4} \\
    V_1(T, \{\alpha_2^*\}) =& \min_{p \in [0,1]}\left\{ \frac{1}{2}+\left(p-\frac{1}{2} \right)^2 \right\} = \frac{1}{2},
\end{align*}
a contradiction. The reason is that action $H$, although tied with $T$ at the Nash conjecture $\alpha_2^*$, is more exposed to unfavorable misspecifications of player~$2$'s behavior; the robust criterion therefore strictly ranks $T$ above $H$.
\end{example}

The preceding discussion, together with Examples~\ref{ex:mne_SCEM_proof} and~\ref{ex:mne_SCEM_counterexample}, motivates the analysis that follows.

The next definition isolates a robust analogue of mixed-strategy Nash equilibrium. Its decision-theoretic structure follows the variational representation of ambiguity-averse preferences in \cite{maccheroni2006ambiguity}, where alternatives are evaluated by minimizing expected utility plus an ambiguity index over possible probabilistic models. It is also close in spirit to the literature on games under ambiguous beliefs, and in particular to the analysis of variational preferences in games under ambiguous belief correspondences in \cite{de2015variational}.

\begin{definition}\label{def:robust_nash}
    Let $(G, c)$ be a game with statistical distances. A profile of mixed actions $(\alpha_i^*)_{i \in I}$ is a robust mixed-strategy Nash equilibrium \textnormal{(\RNE)} if, for all  $i \in I$ and all $a_i^* \in \operatorname{supp}(\alpha_i^*)$
\begin{multline*}
    \min_{\alpha\mi \in \Delta(A\mi)}\left\{ U_i(a_i^*, \alpha\mi) + c_i(\alpha\mi, \alpha\mi^*) \right\} \\
    \ge \min_{\alpha\mi \in \Delta(A\mi)}\left\{ U_i(a_i, \alpha\mi) + c_i(\alpha\mi, \alpha\mi^*) \right\}, \quad \forall a_i \in A_i.
\end{multline*}
\end{definition}

Definition~\ref{def:robust_nash} admits the following interpretation. Player~$i$ holds a unique reference conjecture about the opponents' behavior $\alpha\mi^*$, but, in the spirit of the variational preferences of \cite{maccheroni2006ambiguity}, is ambiguity averse and therefore does not evaluate actions by trusting this conjecture dogmatically.
Instead, for each action $a_i$, the player  considers all alternative beliefs $\alpha_{-i}\in\Delta(A_{-i})$ and assesses the action through the worst-case value of
\[
U_i(a_i,\alpha_{-i})+c_i(\alpha_{-i},\alpha_{-i}^*),
\]
where the term $c_i(\alpha_{-i},\alpha_{-i}^*)$ penalizes departures from the reference belief. In this sense, $\alpha_{-i}^*$ plays the role of a benchmark prior, while the distance $c_i$ captures the player’s attitude toward ambiguity around that benchmark. A robust mixed-strategy Nash equilibrium then requires that every action in the support of $\alpha_i^*$ remain optimal under this ambiguity-averse evaluation, so that no deviation performs better once the player takes into account both the benchmark belief $\alpha_{-i}^*$ and his aversion to ambiguity about opponents’ actual play. 

The next result shows that this robustness requirement is strong enough to imply the weaker consistency notion of \textnormal{SCEM}.

\begin{theorem}\label{th:RNE}
    Let $(G, c)$ be a game with statistical distances, and let $(\alpha_i^*)_{i \in I}$ be a \textnormal{R-NE} of $(G, c)$. Then $(\alpha_i^*)_{i \in I}$ is a \textnormal{SCEM} for any game with feedback and uncertainty attitudes $(G, f, c, \mathcal{J})$.
\end{theorem}
\begin{proof}
    See Appendix~\hyperref[proof:theorem216]{B}.
\end{proof}

Theorem~\ref{th:RNE} shows that robust mixed-strategy Nash equilibrium is a refinement of \textnormal{SCEM}: every profile that is stable under ambiguity-averse evaluation around the correct reference conjecture automatically satisfies the requirements of self-confirming equilibrium. In particular, this inclusion does not depend on any specific properties of the feedback map $f$ or profile of evaluation functionals $\mathcal{J}$. Equivalently,
\begin{equation}\label{eq:inclusion1}
    \text{R-NE} \subseteq \text{SCEM}.
\end{equation}

Theorem~\ref{th:RNE} also settles affirmatively question $(\mathrm{N}_\exists)$ of the opening of this section. By Example~\ref{ex:EU_collapse}, under the dogmatic distance the robust evaluation of every action coincides with its expected utility at the reference conjecture; the R-NE inequalities then reduce to the support characterization of Nash equilibrium (see Lemma~\ref{th:appB_classicalnash}), and Theorem~\ref{th:RNE} transfers the conclusion to SCEM.

\begin{corollary}\label{cor:mNE_neutral_SCEM}
    Let $G$ be a game and let $c^\infty=(c_i^\infty)_{i\in I}$ be the profile of dogmatic distances of Example~\ref{ex:EU_collapse}. Then a profile of mixed actions $(\alpha_i^*)_{i\in I}$ is a R-NE of $(G,c^\infty)$ if and only if it is a m-NE of $G$. In particular, every m-NE of $G$ is a SCEM action profile of $(G,f,c^\infty,\mathcal{J})$, for every feedback structure $f$ and every profile of evaluation functionals $\mathcal{J}$.
\end{corollary}

An important follow-up question is whether robust Nash equilibria exist in general. Under the additional assumption that, for each $i \in I$, the distance $c_i \colon \Delta(A_{-i}) \times \Delta(A_{-i}) \to [0, +\infty]$ is continuous, we are able to establish existence.

\begin{theorem}\label{th:exists:RNE}
Let $(G, c)$ be a game in which the distance $c_i$ is continuous for every $i \in I$.
Then $(G, c)$ admits a robust Nash equilibrium $\alpha^* = (\alpha_i^*)_{i \in I}$.
\end{theorem}

\begin{proof}
    See Appendix~\hyperref[proof:theorem217]{B}.
\end{proof}

The careful reader might notice that the continuity assumption on the statistical distances in Theorem~\ref{th:exists:RNE} excludes a number of useful distances, first and foremost the KL divergence, which is only lower semicontinuous. While a general existence theorem for lower semicontinuous distances would be an interesting direction for future work, we provide here a direct proof for the KL case.

\begin{lemma}\label{lem:entropic_existence}
    Let $G$ be a game and let $c = (c_i)_{i \in I}$ be a profile of KL divergences $c_i \coloneqq \lambda_i \mathrm{KL}(\cdot\,\|\,\cdot)$ with $\lambda_i \in (0,+\infty)$ for every $i \in I$. Then $(G,c)$ admits a robust Nash equilibrium $\alpha^* = (\alpha_i^*)_{i \in I}$. 
\end{lemma}

\begin{proof}
    See Appendix~\hyperref[proof:lemm-entropic]{B}.
\end{proof}

Existence of a SCEM for a game with feedback and uncertainty attitudes $(G, f, c, \mathcal{J})$ is then immediate by combining Theorems \ref{th:RNE} and \ref{th:exists:RNE}.

\begin{corollary}\label{th:SCEM_exist}
    Let $(G, f, c, \mathcal{J})$ be a game in which the distance $c_i$ is continuous for every $i \in I$.
Then $(G, f, c, \mathcal{J})$ admits a self-confirming equilibrium under model misspecification $\alpha^* = (\alpha_i^*)_{i \in I}$.
\end{corollary}

By Corollary~\ref{th:SCEM_exist}, the inclusion in \eqref{eq:inclusion1} can be strengthened to
\begin{equation}\label{eq:inclusion2}
    \emptyset \neq \textnormal{R-NE} \subseteq \textnormal{SCEM}.
\end{equation}
Thus, robust Nash equilibria exist under the continuity assumption on the distances, and each of them corresponds to a correct self-confirming equilibrium under model misspecification.

Moreover, the discrepancy between R-NE and SCEM is due precisely to the partial informativeness of the feedback structure. In fact, under fully informative feedback, observational consistency forces each player’s structured beliefs to collapse to the correct singleton conjecture, so the SCEM rationality condition becomes exactly the R-NE condition.

\begin{theorem}\label{th:RNE_equal_SCEM}
    Let $(G, f, c, \mathcal{J})$ be a game with feedback and uncertainty attitudes. Suppose for each player $i \in I$ that $M_i$ is Polish and feedback is fully informative. Then
    \begin{equation*}
        \textnormal{R-NE} = \textnormal{SCEM}.
    \end{equation*}
\end{theorem}

\begin{proof}
    See Appendix~\hyperref[proof:theorem219]{B}.
\end{proof}

We conclude this section by noting that, while Example~\ref{ex:mne_SCEM_counterexample} shows that the inclusion $\textnormal{m-NE}\subseteq \textnormal{R-NE}$ is false in general, under the additional divergence assumption from Definition \ref{def:ch2:divergence}, every pure Nash equilibrium is an R-NE and hence, by Theorem~\ref{th:RNE}, a SCEM.

\begin{theorem}\label{th:pure_NE}
    Let $(G, c)$ be a game with divergences. Suppose $a^*=(a_i^*)_{i \in I}$ is a pure-strategy Nash equilibrium of $G$ then $a^*$ is a R-NE of $(G, c)$.
\end{theorem}

\begin{proof}
    See Appendix~\hyperref[proof:theorem220]{B}.
\end{proof}

The intuition is that, at a pure Nash equilibrium, the reference conjecture of player $i$ is the Dirac measure $\delta_{a^*_{-i}}$. Since divergences assign finite cost only to beliefs absolutely continuous with respect to the reference conjecture, the inner minimization cannot move probability mass away from $a^*_{-i}$. Hence the robust evaluation of every deviation coincides with its usual expected payoff against $a^*_{-i}$, and the pure Nash inequalities are exactly the R-NE inequalities.

\newpage
\section{Comparative Statics with respect to Feedback}\label{Sec:comp_feedback}

In this brief section we study properties of games with feedback and uncertainty attitudes, fixing $(G, c, \mathcal{J})$  and analyzing the dependence on the feedback structure $f$. 

Recalling Section~\ref{sec:information_structure}, for a feedback structure $f$, player $i \in I$, and action $a_i \in A_i$, we define
\begin{equation*}
    \mathcal{H}_i^f(a_i)\coloneq \sigma(f_{i, a_i}).
\end{equation*}

Returning to Example \ref{ex:fullyinformative}, we intuitively understand the "most informative" feedback function as fully informative feedback, that is,
feedback with injective sections $f^{\max}_{i,a_i}$, which allow the player to pinpoint exactly the opponents' action profile that generated a given message; conversely, the "least informative" feedback is obtained when $f^\text{min}_{i, a_i}$ is constant. When $M_i$ is Polish, this relation corresponds to\footnote{For more details, see the proof of Lemma~\ref{th:informativeimpliesobservable}.}
\begin{equation*}
    \mathcal{H}_i^{f^\text{min}}(a_i) = \{\emptyset, A\mi\} \subseteq \mathcal{B}(A\mi)= \mathcal{H}_i^{f^\text{max}}(a_i).
\end{equation*}
Building on this observation, we define a relation of feedback informativeness, which induces a partial order up to informational equivalence.

\begin{definition}\label{def:more_informativ}
    Let $G$ be a game and $f=(f_i)_{i \in I}, g=(g_i)_{i \in I}$ two feedback structures. We say that $f$ is more informative than $g$ if for all $i \in I$ and all $a_i \in A_i$,
    \begin{equation*}
        \mathcal{H}_i^g(a_i) \subseteq \mathcal{H}_i^f(a_i)
    \end{equation*}
\end{definition}
Equivalently, after any own action \(a_i\), every event about opponents' play that can be inferred from the \(g\)-feedback can also be inferred from the \(f\)-feedback.

An immediate consequence of Definition~\ref{def:more_informativ} is that, as
feedback becomes more informative, the set of observationally consistent conjectures shrinks. Indeed, since observational consistency is equivalent to agreement on $\mathcal{H}_i^f(a_i)$, more informative feedback imposes more restrictions on admissible conjectures. A special case of this fact can be found in Lemma~\ref{th:sec4:informativeimpliessingleton}.

\begin{lemma}\label{th:more_informativ}
    Let $G$ be a game and let $f$ be more informative than $g$. Then for all $i \in I$, $a_i \in A_i$, and $\nu\mi \in \Delta(A\mi)$,
    \begin{equation*}
        \mathcal{O}_i^f(a_i, \nu\mi) \subseteq \mathcal{O}_i^g(a_i, \nu\mi),
    \end{equation*}
    where $\mathcal{O}_i^f(\cdot,\cdot)$ denotes the set of observationally consistent conjectures induced by the feedback structure $f$, as in Definition~\ref{def:observationally_consistent_beliefs}.
\end{lemma}

\begin{proof}
    See Appendix~\hyperref[proof:lemma222]{B}.
\end{proof}

We next obtain a monotonicity result for SCEM action profiles with respect to feedback informativeness.

\begin{theorem}\label{th:comparative_statics_feedback_SCEM}
    Let $(G, f, c, \mathcal{J})$ and $(G, g, c, \mathcal{J})$ be two games that differ only in their feedback functions. Suppose that $f$ is more informative than $g$. If $(\alpha_i^*)_{i \in I}$ is a \textnormal{SCEM} of $(G, f, c, \mathcal{J})$, then it is also a \textnormal{SCEM} of $(G, g, c, \mathcal{J})$.
\end{theorem}

\begin{proof}
    See Appendix~\hyperref[proof:theorem223]{B}.
\end{proof}

The theorem says that less informative feedback enlarges the set of self-confirming outcomes. If a profile can be sustained when players receive the richer feedback generated by $f$, then it can also be sustained when they receive the coarser feedback generated by $g$, because fewer conjectures are
ruled out by observation.

\newpage
\section{A (failure of) Comparative Statics with respect to Uncertainty Aversion}\label{sec:comparative_aversion}

We continue our analysis by undertaking a comparative study of uncertainty aversion in game structures $(G, f, c, \mathcal{J})$ with feedback and uncertainty. Our approach is motivated by the literature on comparative uncertainty aversion \citep{epstein1999definition, ghirardato2002ambiguity}. For more details and motivation, see Definition~\ref{def:app_B:comparativeuncertaintyaversion}, Theorem~\ref{th:appB:comparativetheorem}, and the accompanying discussion in Appendix~\hyperref[ch:appendix_uncertainty]{C}. 

\begin{definition}\label{def:ch4:comaprativeuncertainty}
Let $(G, f, c^1, \mathcal{J}^1)$ and $(G, f, c^2, \mathcal{J}^2)$ be two games with feedback and uncertainty attitudes. For each player $i \in I$, we say that $(c_i^1, \mathcal{J}_i^1)$ is more uncertainty averse than $(c_i^2, \mathcal{J}_i^2)$ if
\[
c_i^1 \le c_i^2
\quad\text{and}\quad
J_{\mathrm{Q}^i}^1 \le J_{\mathrm{Q}^i}^2
\quad \forall \mathrm{Q}^i \in \mathfrak{Q}_i.
\]
We say that the game $(G, f, c^1, \mathcal{J}^1)$ is more uncertainty averse than $(G, f, c^2, \mathcal{J}^2)$ if, for every player $i \in I$, the pair $(c_i^1, \mathcal{J}_i^1)$ is more uncertainty averse than $(c_i^2, \mathcal{J}_i^2)$.
\end{definition}

Again following \citet{ghirardato2002ambiguity}, we identify absolute uncertainty aversion through a comparative relation with expected utility functionals associated with finitely additive probability measures, which in our setting coincide with linear, monotone, normalized functionals.\footnote{\label{fn:4:expectedutility}Let $\mathrm{Q}$ be a Borel space. A functional $J:B_b(\mathrm{Q})\to \R$ is linear, normalized, and monotone if and only if there exists a unique finitely additive probability measure $\eta$ such that
\begin{equation*}
    J(h)=\int_\mathrm{Q} h \, d\eta.
\end{equation*}
Indeed, given such a functional $J$, define $\eta(A)\coloneq J(\mathbf{1}_A)$. Then $\eta$ is a finitely additive measure by linearity, and a probability measure by normalization and monotonicity. Moreover, $J$ and integration with respect to $\eta$ agree on the class of simple functions; hence, by $\|\cdot\|_{\infty}$-continuity of both functionals, they agree on the whole space $B_b(\mathrm{Q})$.} More precisely, we say that a game with feedback and uncertainty attitudes is uncertainty averse if it is more uncertainty averse than some game $(G, f, c^\infty, \mathcal{I})$, where every functional in the family $\mathcal{I}$ is linear, and therefore an expected utility functional, and the profile of statistical distances $c^\infty=(c_i^\infty)_i$ is the profile of dogmatic distances.

One would be tempted\footnote{Let $(G, f, c^1, \mathcal{J}^1)$ be more uncertainty averse than $(G, f, c^2, \mathcal{J}^2)$, and let $V_i^j$ denote the preference representation of player $i \in I$ in game $j=1,2$. Then one can prove the pointwise inequality
\begin{equation*}
    V_i^1(a_i, \mathrm{Q}^i)\le V_i^2(a_i, \mathrm{Q}^i)
\end{equation*}
for all $a_i \in A_i$ and $\mathrm{Q}^i \in \mathfrak{Q}_i$. Indeed, fixing $a_i$ and $\mathrm{Q}^i$, define the functionals
\begin{equation*}
    \varphi_i^j(q, a_i)\coloneq \min_{\alpha_{-i} \in \Delta(A_{-i})}\left\{ U_i(a_i, \alpha_{-i}) + c_i^j(\alpha_{-i}, q)\right\}.
\end{equation*}
Then, by hypothesis, for every $q \in \mathrm{Q}^i$ and every $\alpha_{-i} \in \Delta(A_{-i})$, we have
\begin{equation*}
    U_i(a_i, \alpha_{-i}) + c_i^1(\alpha_{-i}, q)
    \le
    U_i(a_i, \alpha_{-i}) + c_i^2(\alpha_{-i}, q),
\end{equation*}
and hence $\varphi_i^1(\cdot, a_i) \le \varphi_i^2(\cdot, a_i)$ pointwise on $\mathrm{Q}^i$. By the monotonicity of $J^1_{\mathrm{Q}^i}$ and, once again, the hypothesis, it follows that
\begin{equation*}
    J_{\mathrm{Q}^i}^1(\varphi_i^1(\cdot, a_i))
    \le
    J_{\mathrm{Q}^i}^1(\varphi_i^2(\cdot, a_i))
    \le
    J_{\mathrm{Q}^i}^2(\varphi_i^2(\cdot, a_i)).
\end{equation*}
The argument is analogous to the proof of Theorem~\ref{th:appB:comparativetheorem}.}, as per the literature \citep{battigalli2015self}, to assume payoff observability and prove a theorem which shows that the set of SCEM action profiles is monotone with respect to uncertainty aversion, of the form:

\begin{quote}
    \textit{Let $(G, f, c^1, \mathcal{J}^1)$ be more uncertainty averse than $(G, f, c^2, \mathcal{J}^2)$. If $\alpha^*$ is a \textnormal{SCEM} of $(G, f, c^2, \mathcal{J}^2)$, then it is also a \textnormal{SCEM} of $(G, f, c^1, \mathcal{J}^1)$.}
\end{quote}

As the following counterexample shows, the claim is in general false. In Section~\ref{sec:28}, we will show that under economically meaningful additional assumptions, observationally consistent divergences and observable payoffs, the statement holds true and, in doing so, we will close the loop and reconnect with the literature on ambiguity in games that inspired our present work, especially \cite{battigalli2015self}.

\begin{example}\label{ex:counterex_aversion}
Consider the game with fully informative feedback from Example \ref{ex:mne_SCEM_counterexample}, and define the following two games with feedback and uncertainty attitudes $(G, f, c^l, \mathcal{J}^l)$ for $l=1,2$. For a parameter $\lambda \in (0, \frac{1}{2})$, introduce the distances
\begin{align*}
     &c_1^1(\mu, \nu)\coloneq \lambda(\mu(L)-\nu(L))^2 ,& \ &c_2^1(\mu, \nu)\coloneq \lambda(\mu(H)-\nu(H))^2, \\
     &c_1^2(\mu, \nu)\coloneq (\mu(L)-\nu(L))^2 ,& \ &c_2^2(\mu, \nu)\coloneq (\mu(H)-\nu(H))^2.
\end{align*}

Note that by Lemma \ref{th:informativeimpliesobservable}, the game $(G, f)$ has observable payoffs. As per Lemma~\ref{th:sec4:informativeimpliessingleton}, when feedback is fully informative, the set of consistent conjectures corresponding to any given action $a_i \in A_i$ and population mixed action $\alpha\mi \in \Delta(A\mi)$ is the singleton containing the correct conjecture, namely $\mathcal{O}_i(a_i, \alpha\mi)=\{\alpha\mi\}$. Hence, by Footnote~\ref{fn:1}, the choice of $\mathcal{J}^l$ is irrelevant; for the sake of formality, fix some $\mathcal{J}=\mathcal{J}^1=\mathcal{J}^2$. It is then clear that $(G, f, c^1, \mathcal{J}^1)$ is more uncertainty averse, in the sense of Definition \ref{def:ch4:comaprativeuncertainty}, than $(G, f, c^2, \mathcal{J}^2)$.

We now show that the pure profile $\alpha^*=(\delta_H, \delta_L)$ is a \textnormal{SCEM} for the latter game but not for the former. A direct calculation shows that
\begin{align*}
    V_1^{c^2}(H, \{ \delta_L\})&= \frac{3}{4} > \frac{1}{2}= V_1^{c^2}(T, \{ \delta_L\}),\\
    V_2^{c^2}(L, \{ \delta_H\})&= \frac{3}{4} > 0 = V_2^{c^2}(R, \{ \delta_H\}),
\end{align*}
thus proving that $\alpha^*$ is a \textnormal{SCEM} for $(G, f, c^2, \mathcal{J})$. If $\alpha^*$ were a \textnormal{SCEM} for the game $(G, f, c^1, \mathcal{J})$, then, considering player $1$, we would have
\begin{equation*}
    \lambda =V_1^{c^1}(H, \{ \delta_L\}) \ge V_1^{c^1}(T, \{ \delta_L\}) = \frac{1}{2},
\end{equation*}
a contradiction.
\end{example}

The interpretation is the following. Increasing uncertainty aversion means lowering the cost of moving away from the reference conjecture. This lowers the robust value of every action, but it need not lower all actions by the same amount. The failure of monotonicity is therefore not epistemic but occurs entirely at the level of rationality. A profile sustained under the less uncertainty-averse criterion may cease to be sustained under the more uncertainty-averse criterion because the equilibrium action may be more sensitive to misspecification than some deviation.

\newpage
\section{Observational Consistency and Self-Confirming Equilibrium under Ambiguity}\label{sec:28}

In this section, we consider a variation of the misspecification setup developed thus far which, under two additional assumptions, observationally consistent divergences (a behavioral restriction) and observable payoffs (an informational one), will allow us to obtain a monotonicity result on comparative uncertainty aversion and, more generally, will directly connect us to the literature \cite{battigalli2015self} on Smooth self-confirming equilibrium (SSCE).

Consider a game with feedback $(G, f)$ and fix a player $i \in I$. When playing action $a_i \in A_i$, she knows that she will observe feedback $f_i(a_i, \cdot)$ which partitions the opponents' action space $A\mi$ into observationally equivalent classes. Thus, when evaluating a reference belief $q \in \Delta(A\mi)$ after action $a_i$, any belief $p \notin \mathcal{O}_i(a_i, q)$ is inconsistent with the evidence generated under $q$ and should therefore be ruled out.

\begin{definition}\label{def:obs_consistent_diverg}
    A function $c_i:A_i \times \Delta(A\mi) \times \Delta(A\mi) \to [0, +\infty]$ is an observationally consistent (action-dependent) divergence if:
    \begin{enumerate}
        \item[(i)] $c_i$ is lower semicontinuous, where the product space is endowed with the product topology.
        \item[(ii)] $c_i$ is non-degenerate, that is, for all $a_i \in A_i$ and all $p,q \in \Delta(A\mi)$,
        \begin{equation*}
            c_i(a_i, p, q)=0 \Longleftrightarrow p=q.
        \end{equation*}
        \item[(iii)] $c_i$ is observationally consistent, that is, for all $a_i \in A_i$ and all $p,q \in \Delta(A\mi)$,
        \begin{equation*}
            c_i(a_i, p, q) < + \infty \Longrightarrow p \in \mathcal{O}_i(a_i, q).
        \end{equation*}
    \end{enumerate}
\end{definition}
Thus, observational consistency restricts the minimization over misspecified beliefs to those alternatives compatible with the feedback generated by $(a_i, q)$: any belief outside $\mathcal{O}_i(a_i, q)$ is assigned infinite cost and is therefore excluded from the minimization. Accordingly, $c_i(a_i, p, q)$ represents the cost, conditional on action $a_i$, of treating $p$ as a plausible misspecification of the reference belief $q$. We refer to $(G, f, c)$ as a game with feedback and observationally consistent divergences; if $\mathcal{J}$ is a family of monotone normalized functionals, we refer to $(G, f, c, \mathcal{J})$ as a game with observationally consistent divergences and uncertainty attitudes.

\begin{example}
\label{ex:truncated_divergence}
Any statistical distance $c_i : \Delta(A_{-i}) \times \Delta(A_{-i}) \to [0, +\infty]$ induces a canonical observationally consistent divergence by truncation outside the observationally consistent set. Define $\tilde{c}_i : A_i \times \Delta(A_{-i}) \times \Delta(A_{-i}) \to [0, +\infty]$ by
\begin{equation*}
    \tilde{c}_i(a_i, p, q) := 
    \begin{cases}
        c_i(p, q), & \text{if } p \in \mathcal{O}_i(a_i, q), \\
        +\infty,   & \text{otherwise.}
    \end{cases}
\end{equation*}
Conditions (ii) and (iii) of Definition~\ref{def:obs_consistent_diverg} are immediate: non-degeneracy follows from non-degeneracy of $c_i$ together with $q \in \mathcal{O}_i(a_i, q)$, and observational consistency holds by construction. Lower semicontinuity (condition (i)) requires mild regularity\footnote{It holds, for instance, whenever $c_i$ is lower semicontinuous and, for each $a_i \in A_i$, the set $\{(a_i,p, q) : p \in \mathcal{O}_i(a_i, q)\}$ is closed in $A_i \times \Delta(A_{-i}) \times \Delta(A_{-i})$, which is the case when $f_{i}$ is continuous (see Theorem~\ref{th:appA_pushforwardcont}) and $M_i$ is Polish.}. Then, $\tilde{c}_i$ is an observationally consistent divergence, and represents an agent who evaluates misspecifications by the original cost $c_i$ on observationally indistinguishable beliefs and rules out the rest outright.
\end{example}

This formal device of observationally consistent divergences, and the theorems that follow, deserve a careful interpretation. The whole point of distinguishing structured conjectures from the full simplex, following \citet{hansen2022structured} and \citet{cerreia2026making}, is that the agent recognizes her models as approximations of a more complex, only partially observed and partially modeled reality. The action-independent divergence of Sections~\ref{sec:gameswithfeedbackanduncertainty} and \ref{sec:SCEM} is the natural formalism for such an agent. By contrast, setting $c_i(a_i, p, q) = +\infty$ for every $p \notin \mathcal{O}_i(a_i, q)$ amounts to a strong epistemic claim: any model predicting a different distribution over her messages under $a_i$ is ruled out as a candidate description of reality. The inner minimization is confined to $\mathcal{O}_i(a_i, q)$, and once payoff observability is imposed (Lemma~\ref{th:lemma:payoff_observability}), $U_i(a_i,\cdot)$ is constant on this set. The inner minimum therefore reduces to $U_i(a_i, q)$, and the agent's evaluation of $a_i$ at $q$ is precisely expected utility.

At the equilibrium action $a_i^*$, by the confirmed conjectures condition, every structured conjecture $q$ induces the same distribution over messages under $a_i^*$ as the true other populations' distribution $\alpha_{-i}^*$,  and hence $\alpha_{-i}^* \in \mathcal{O}_i(a_i^*, q)$ for every such $q$: the truth is an admissible candidate for the worst case under every structured model. At a deviation $a_i \neq a_i^*$, this guarantee is absent. The partitions $\mathcal{F}_i(a_i^*)$ and $\mathcal{F}_i(a_i)$ are in general unrelated, and membership of $q$ in $\mathcal{O}_i(a_i^*, \alpha_{-i}^*)$ implies nothing about its membership in $\mathcal{O}_i(a_i,\alpha_{-i}^*)$. Consequently, $\alpha_{-i}^*$ may fail to lie in $\mathcal{O}_i(a_i, q)$ for any $q$, and the agent's worst-case set at the deviation may exclude the very truth she is in equilibrium with. The functional $J$ then effectively aggregates only over the models in
\begin{equation*}
    \bigcup_{q \in \mathrm{Q}^i_{a_i^*}} \mathcal{O}_i(a_i, q)
\end{equation*} 
rather than over the full simplex $\Delta(A_{-i})$. In this sense, observationally consistent divergences describe agents who behave as if they were no longer averse to model misspecification proper, that is, as if they no longer feared distortions of any given structured conjecture $q$, while retaining ambiguity only over which $q \in \mathrm{Q}^i_{a_i^*}$ best describes opponents' play. As the theorems in the remainder of this section will make clear, this has profound implications for SCEM under observationally consistent divergences, reducing it to a slight generalization of the smooth self-confirming equilibrium of \citet{battigalli2015self}. The counterexamples of Sections~\ref{sec:tax} and \ref{sec:comparative_aversion} should therefore be read as evidence that dropping observational consistency is what allows the misspecification-averse agent's worst-case reasoning to drift away from the predictions of the classical concepts, and it is precisely this drift that gives SCEM with action-independent divergences its economic content.

With these notions in hand, the definitions introduced in Sections~\ref{sec:gameswithfeedbackanduncertainty} and~\ref{sec:SCEM} extend naturally to a game $(G, f, c, \mathcal{J})$ with observationally consistent divergences and uncertainty attitudes. In particular, players' preferences are represented by
\begin{equation}
    V_i(a_i, \mathrm{Q}^i)= J_{\mathrm{Q}^i}\left( \min_{\alpha_{-i}\in \Delta(A_{-i})} \left\{ U_i(a_i,\alpha_{-i}) + c_i(a_i, \alpha_{-i},\cdot) \right\} \right).
\end{equation}
We avoid repeating ourselves, but we also consider the natural extensions of Definition \ref{def:SCEM} of SCEM, Definition \ref{def:robust_nash} of R-NE, and Definition \ref{def:ch4:comaprativeuncertainty} of comparative uncertainty aversion.

The added assumption $(iii)$ of observational consistency, together with observable payoffs (see Definition \ref{def:payoffobservability}), has immediate implications for the notions of equilibrium thus far defined.

\begin{theorem} \label{th:RNEunderobservatioanlconsistency}
    Let $(G, f, c)$ be a game with observationally consistent divergences and observable payoffs. Then a profile of mixed actions $(\alpha_i^*)_{i \in I}$ is a \textnormal{R-NE} if and only if it is a \textnormal{m-NE}. 
\end{theorem}

\begin{proof}
    See Appendix~\hyperref[proof:theorem227]{B}.
\end{proof}
The proof uses payoff observability to collapse the robust value of any action at the reference conjecture to its ordinary expected utility value, so that the R-NE inequalities coincide with the m-NE inequalities.

In fact, under observational consistency and observable payoffs, SCEM reduces to a natural generalization of SSCE, which we name Self-Confirming Equilibrium under Ambiguity (SCEA).

\begin{definition}\label{def:SCEA}
    Let $(G, f, \mathcal{J})$ be a game with feedback and uncertainty attitudes\footnote{Note that we deliberately do not specify the divergence profile $c$. The proof of Theorem~\ref{th:SCEAequalsSCEM} will clarify the motivation behind this choice.}. A profile of mixed actions $(\alpha^*_i)_{i \in I}$ is a Self-Confirming Equilibrium under Ambiguity \textnormal{(SCEA)} if for all $i \in I$ and all $a_i^* \in \operatorname{supp}(\alpha_i^*)$ there exists a set of structured beliefs $\mathrm{Q}^i_{a_i^*}$ such that:
    \begin{enumerate}
    \item (Rationality): $a^*_i \in \underset{a_i \in A_i}{\operatorname{argmax}} \ J_{\mathrm{Q}^i_{a_i^*}}\left( U_i(a_i, \cdot)|_{\mathrm{Q}^i_{a_i^*}} \right)$
    \item (Confirmed Conjectures): $\mathrm{Q}^i_{a^*_i}\subseteq\mathcal{O}_i(a^*_i,\alpha^*\mi)$
    \end{enumerate}
\end{definition}

\begin{theorem}\label{th:SCEAequalsSCEM}
    Let $(G, f, c, \mathcal{J})$ be a game with observationally consistent divergences, uncertainty attitudes, and observable payoffs. Then a profile of mixed actions $(\alpha_i^*)_{i \in I}$ is a \textnormal{SCEM} if and only if it is a \textnormal{SCEA}.
\end{theorem}

\begin{proof}
    See Appendix~\hyperref[proof:theorem229]{B}.
\end{proof}
Thus, once finite-cost misspecifications are restricted to observationally consistent beliefs and payoffs are observable, the divergence profile no longer affects equilibrium behavior.

As previously anticipated, the notion of SSCE, first introduced by \cite{battigalli2015self}, can be recognized as a specific example of SCEA.

\begin{example}\label{ex:SSCEasSCEA}
    We recall the definition of smooth self-confirming equilibrium \textnormal{(SSCE)}. In our terminology, \cite{battigalli2015self} consider games with feedback $(G, f)$ and ambiguity attitudes, represented, for each player $i \in I$, by a strictly increasing continuous function $\phi_i:\R \to \R$. A profile of mixed actions $(\alpha_i^*)_{i \in I}$ is then a \textnormal{SSCE} if for all $i \in I$ and all $a_i^* \in \operatorname{supp}(\alpha_i^*)$, there exists a prior probability $\pi_{a_i^*}$, that is, a Borel measure on $\Delta(A\mi)$ with support contained in $\mathcal{O}_i(a^*_i, \alpha\mi^*)$, such that
    \begin{equation}\label{eq:SSCEasSCEA1}
        a^*_i \in \underset{a_i \in A_i}{\operatorname{argmax}} \ \phi_i^{-1}\left( \int_{\Delta(A\mi)}^{} \phi_i\left( U_i(a_i, \alpha\mi) \right) \ d\pi_{a_i^*}(\alpha\mi)\right).
    \end{equation}
    For each player $i \in I$, consider a family of probability measures on $\Delta(A\mi)$, denoted by $(\pi_{\mathrm{Q}^i})_{\mathrm{Q}^i \in \mathfrak{Q}_i}$, such that $\operatorname{supp}(\pi_{\mathrm{Q}^i})= \mathrm{Q}^i$, and define the corresponding family of monotone normalized functionals: 
    \begin{equation}\label{eq:SSCEasSCEA2}
        \mathcal{J}_i=\left( \phi_i^{-1}\left( \int_{\operatorname{supp}(\pi_{\mathrm{Q}^i})}^{} \phi_i\left(\  \cdot \  \right) \ d\pi_{\mathrm{Q}^i}(\alpha\mi)\right) \right)_{\mathrm{Q}^i \in \mathfrak{Q}_i}
    \end{equation}
    It is then clear that, if $(\alpha_i^*, (\pi_{a_i^*})_{a_i^* \in \operatorname{supp}(\alpha_i^*)})_{i \in I}$ is an \textnormal{SSCE}, then, setting
    \[
        \mathrm{Q}^i_{a_i^*}:=\operatorname{supp}(\pi_{a_i^*}),
    \]
    the corresponding profile of mixed actions and structured beliefs
    \begin{equation*}
        (\alpha_i^*, (\mathrm{Q}^i_{a_i^*})_{a_i^* \in \operatorname{supp}(\alpha_i^*)})_{i \in I}
    \end{equation*}
    is an \textnormal{SCEA} corresponding to the family of evaluation functionals $\mathcal{J}_i$ defined above. Indeed, the support condition on $\pi_{a_i^*}$ corresponds to the Confirmed Conjectures condition in Definition \ref{def:SCEA}, while Equations \eqref{eq:SSCEasSCEA1} and \eqref{eq:SSCEasSCEA2} imply the Rationality condition.
\end{example}
Hence SCEA keeps the self-confirming logic of SSCE, but abstracts from the particular smooth functional form by allowing an arbitrary monotone normalized functional on each set of structured beliefs.

We now introduce some notation. For each of the equilibrium notions $\mathcal{E} \in \left\{ \textnormal{m-NE, R-NE, SSCE, SCEA, SCEM} \right\}$ considered above, we write
$$\mathcal{E}(G,f,c, \mathcal{J})\coloneq\{\alpha^*=(\alpha_i^*)_{i \in I}:\alpha^* \textnormal{ is an } \mathcal{E}\textnormal{-equilibrium}  \}$$
for the corresponding set of equilibrium mixed action profiles. Whenever an element of the tuple  $(G,f,c, \mathcal{J})$ is irrelevant for the specific notion of equilibria at hand, we omit it; for instance we write $\textnormal{m-NE}(G)$ for the set of mixed-strategy Nash equilibria of a game $G$.

Using Theorems \ref{th:RNEunderobservatioanlconsistency} and \ref{th:SCEAequalsSCEM}, we can rewrite the inclusion chain in \eqref{eq:inclusion2} under observationally consistent divergences and observable payoffs:
\begin{equation}\label{eq:SCEAINCLUSION1}
    \emptyset \neq \textnormal{m-NE}(G) \subseteq \textnormal{SCEA}(G,f, \mathcal{J})
\end{equation}
We call a game structure $(G, f, \mathcal{J})$ a game with feedback and ambiguity attitudes. Following Definition~\ref{def:ch4:comaprativeuncertainty}, for two such game structures $(G, f, \mathcal{J}^1)$ and $(G, f, \mathcal{J}^2)$, we say that the former is more uncertainty averse than the latter if, for all $i \in I$ and all $\mathrm{Q}^i \in \mathfrak{Q}_i$, $J_{\mathrm{Q}^i}^1 \le J_{\mathrm{Q}^i}^2$. We now have everything ready to obtain what we originally set out to prove: a monotonicity result with respect to uncertainty aversion.

\begin{theorem}\label{th:monotoneuncertaintyaversion_SCEA}
Let $(G, f, \mathcal{J}^1)$ be more uncertainty averse than $(G, f, \mathcal{J}^2)$, both with observable payoffs. If $\alpha^*$ is a \textnormal{SCEA} of $(G, f, \mathcal{J}^2)$, then it is also a \textnormal{SCEA} of $(G, f, \mathcal{J}^1)$.    
\end{theorem}

\begin{proof}
    See Appendix~\hyperref[proof:theorem231]{B}.
\end{proof}

\begin{example}[Cont.\ of Example~\ref{ex:SSCEasSCEA}]
A simple application of Jensen's inequality shows that the notion of comparative uncertainty aversion adopted here is implied by that of \cite{battigalli2015self} (see also \cite{klibanoff2005smooth} for the original treatment). In their framework, a profile of ambiguity attitudes $(\phi_i)_{i \in I}$ is said to be more uncertainty averse than another profile $(\psi_i)_{i \in I}$ if, for every $i \in I$, there exists a concave, strictly increasing function $\varphi_i$ such that $\phi_i = \varphi_i \circ \psi_i$. It then follows that, pointwise,
\begin{equation*}
    \phi_i^{-1}\left( \int_{}^{}\phi_i(\cdot) \ d\pi \right)\le\psi_i^{-1}\left( \int_{}^{}\psi_i(\cdot)\ d\pi \right).
\end{equation*}
\end{example}

We now turn to another canonical special case, namely maxmin preferences.

\begin{example}[MSCE]\label{ex:MSCE}
Fix an arbitrary game with feedback and ambiguity attitudes $(G, f, \mathcal{J})$. Fix also a player $i \in I$ and $\mathrm{Q}^i \in \mathfrak{Q}_i$, and recall that $J_{\mathrm{Q}^i}:B_b(\mathrm{Q}^i) \to \R$ is a monotone normalized functional. The following two special functionals will be useful:
\begin{align*}
        \textnormal{sup}_{\mathrm{Q}^i}&:B_b(\mathrm{Q}^i)\to \R , \quad h\longmapsto \sup_{q \in \mathrm{Q}^i}h(q), \\
        \textnormal{inf}_{\mathrm{Q}^i}&:B_b(\mathrm{Q}^i)\to \R , \quad h \longmapsto \inf_{q \in \mathrm{Q}^i}h(q).
    \end{align*}
    Both functionals are well defined (by boundedness), monotone, and normalized. Now let $h \in B_b(\mathrm{Q}^i)$. Then
    \begin{equation*}
        \inf_{q \in\mathrm{Q}^i}h(q)=J_{\mathrm{Q}^i}\left( q'\mapsto \inf_{q \in\mathrm{Q}^i}h(q) \right)\le J_{\mathrm{Q}^i}\left( h(\cdot) \right),
    \end{equation*}
    where the equality follows from the normalization of $J_{\mathrm{Q}^i}$, while the inequality follows from monotonicity. If we let $ \textnormal{inf}^{\ i} \coloneq (\textnormal{inf}_{\mathrm{Q}^i})_{\mathrm{Q}^i \in \mathfrak{Q}_i}$, we have therefore shown that the game $(G, f, (\textnormal{inf}^{ \ i})_{i \in I})$ is more uncertainty averse than any arbitrary game $(G, f, \mathcal{J})$. Although formally a special case of \textnormal{SCEA}, the important historical connection with Maxmin preferences of \cite{gilboa1989maxmin} motivates the following definition of the set of Maxmin Self-Confirming Equilibria \textnormal{(MSCE)}:
    \begin{equation*}
        \textnormal{MSCE}(G,f)\coloneq \textnormal{SCEA}(G, f, (\textnormal{inf}^{ \ i})_{i \in I})
    \end{equation*}
    Note that, by the continuity of $U_i(a_i, \cdot)$ (see Footnote~\ref{fn:2}) and the compactness of $\mathrm{Q}^i$, the infimum is in fact a minimum. An analogous treatment can be given for the supremum functional.
\end{example}

In summary, in this section we have shown that, under the additional assumptions of payoff observability and observationally consistent divergences, the equilibrium notions under misspecification introduced earlier, R-NE and SCEM, reduce, respectively, to the classical notions of m-NE and SCEA, the latter being a slight generalization of SSCE. If $(G,f, \mathcal{J}^1)$ is more uncertainty averse than $(G,f, \mathcal{J}^2)$, then Theorem~\ref{th:monotoneuncertaintyaversion_SCEA} and Example~\ref{ex:MSCE} allow us to refine the inclusion chain in \eqref{eq:SCEAINCLUSION1} as follows:
\begin{equation}\label{eq:SCEAINCLUSION2}
    \emptyset \neq \textnormal{m-NE}(G) \subseteq \textnormal{SCEA}(G,f, \mathcal{J}^2)\subseteq\textnormal{SCEA}(G,f, \mathcal{J}^1) \subseteq \textnormal{MSCE}(G,f).
\end{equation}
Again by Footnote~\ref{fn:4:expectedutility}, a further refinement follows when $(G, f, \mathcal{J})$ is uncertainty averse, that is, more uncertainty averse than some family of expected utility functionals.

Finally, we show that under the additional assumption of own-strategy independence (see Definition~\ref{def:own_strategy_indep}), the notions of equilibrium coincide, as the feedback generated by the equilibrium action reveals the same payoff-relevant information as the feedback that would be generated by any deviation.

\begin{theorem}\label{th:own_straegy_indep}
    Let $(G, f, \mathcal{J})$ be a game with feedback and uncertainty attitudes, observable payoffs, and own-strategy independence. Then every type of equilibrium here defined coincides, that is,
    \begin{equation*}
        \textnormal{m-NE}(G) = \textnormal{SCEA}(G,f, \mathcal{J})= \textnormal{MSCE}(G,f).
    \end{equation*}
\end{theorem}

\begin{proof}
    See Appendix~\hyperref[proof:theorem234]{B}.
\end{proof}

To conclude the section, we have thus obtained a slight generalization of the results in \cite{battigalli2015self} (see their Theorem~$1$,  Lemma~$2$, Equation~$(6)$, Proposition~$1$, and related comments) within our framework of model misspecification, under two additional assumptions, behavioral and informational respectively: observationally consistent divergences and observable payoffs.

\newpage

\section{Conclusions}\label{sec:conclusion}

This thesis has brought the theory of self-confirming equilibrium into contact with the decision-theoretic literature on model misspecification. Its starting point is the observation that the two frameworks closest to this purpose each impose an asymmetric epistemic posture on players: in Berk--Nash equilibrium \citep{esponda2016berk} players entertain the possibility that their models are misspecified but are averse neither to misspecification nor to ambiguity, while in smooth self-confirming equilibrium \citep{battigalli2015self} players are averse to ambiguity within a posited set of conjectures without contemplating that the set itself may fail to contain the truth. Adopting the misspecification-sensitive preferences axiomatized by \citet{cerreia2026making}, we have defined games with feedback and uncertainty attitudes, in which each player evaluates her actions by aggregating, across her structured conjectures, robust evaluations that penalize departures from each conjecture through a statistical distance, and we have proposed self-confirming equilibrium under model misspecification as the corresponding steady-state notion. 

Several directions remain open. First, learning foundations: SCEM is defined as the resting point of an unmodeled learning process, and a dynamic theory justifying it as a long-run limit, in the spirit of the Berk--Nash literature, is a natural complement to the present static analysis; the works of \citet{battigalli2019learning} and \citet{lanzani2025dynamic} offer promising starting points. Second, refinements: replacing the rationality condition with dominance under the incomplete mental preference of \citet{cerreia2026making} would isolate those equilibria that are best replies under every structured conjecture, a more demanding and arguably more robust notion. Third, comparative uncertainty aversion: the representation-based order used in the present work is sufficient for the behavioral comparison of \citet{ghirardato2002ambiguity}, but a converse, that is, a full behavioral characterization of comparative uncertainty aversion for misspecification-sensitive preferences, remains to be developed. Finally, extending SCEM to sequential games with feedback would confront misspecification aversion with the subtler questions of updating and dynamic consistency that extensive forms raise.

%% file: Chapters/appendix_contribution.tex
\chapter{Appendix A}\label{ch:appendix_contribution}

This appendix collects self-contained auxiliary results of a technical nature that are used repeatedly in proofs.

\begin{lemma}\label{th:appB_welldefinedminimum}
    Let $X$ be a compact metrizable topological space and $\Delta(X)= \Delta$ be the space of countably additive probability measures equipped with the weak* topology. Moreover, let $U:\Delta \to \R$ be a continuous function and $c : \Delta \times \Delta \to [0, +\infty]$ be a statistical distance. Then the functional
    \begin{equation*}
        q \longmapsto \underset{p \in \Delta}{\min}\left\{ U(p)+c(p,q) \right\}
    \end{equation*}
    is well defined and lower semicontinuous, in particular bounded and Borel-measurable.
\end{lemma}
\begin{proof}
Recall from Theorem~\ref{th:appA_sigma_compact} that \((\Delta,\sigma(\Delta,C^0_b(X)))\) is compact when $X$ is. For each \(q\in\Delta\), define
\[
\xi(q)\coloneqq \min_{p\in\Delta}\{U(p)+c(p,q)\}.
\]
Fix \(q\in\Delta\). Since \(c\) is lower semicontinuous by definition, its \(q\)-section \(c(\cdot,q)\) is lower semicontinuous as well. Therefore the map
\[
p\longmapsto U(p)+c(p,q)
\]
is lower semicontinuous on \(\Delta\), being the sum of the continuous function \(U\) and the lower semicontinuous function \(c(\cdot,q)\). As \(\Delta\) is compact, this function attains its minimum on \(\Delta\). Hence \(\xi\) is well defined. Moreover, the minimum is finite. Indeed, choosing \(p=q\) and using the non-degeneracy of \(c\), we obtain
\[
\xi(q)\le U(q)+c(q,q)=U(q)\le \|U\|_\infty.
\]

We now prove that $\xi$ is lower semicontinuous. Fix $a\in\mathbb{R}$ and consider the sublevel set $S=\{q\in\Delta:\xi(q)\le a\}$. Let $(q_\delta)_\delta$ be a net in $S$ converging to some $q\in\Delta$. For each $\delta$, choose $p_\delta\in\Delta$ such that $U(p_\delta)+c(p_\delta,q_\delta)\le a$. Since $\Delta$ is compact, there exists a subnet $(p_{\delta'})_{\delta'}$ converging to some $p^*\in\Delta$. Then $(p_{\delta'},q_{\delta'})\to(p^*,q)$, and by lower semicontinuity of the map $(p,q)\mapsto U(p)+c(p,q)$, we obtain
\begin{equation*}U(p^*)+c(p^*,q)\le\liminf_{\delta'}\left[U(p_{\delta'})+c(p_{\delta'},q_{\delta'})\right]\le a.
\end{equation*}
It follows that $\xi(q)\le a$, so $S$ is closed, proving lower semicontinuity. 

Finally, note that $\xi(q)\ge\min_p\{U(p)\}\ge-\|U\|_\infty$, which gives boundedness.
\end{proof}

\begin{lemma}\label{th:appB_uniformcont}
    Let $A,B$ be topological spaces, with $B$ compact, and let
    $u:A\times B\to \mathbb{R}$ be continuous. Then the map
    \[
        \Phi:A\to C_b^0(B),\qquad \Phi(a):=u(a,\cdot),
    \]
    is continuous when $C_b^0(B)$ is endowed with the norm
    $\|\cdot\|_\infty$.
\end{lemma}
\begin{proof}
First note that by continuity of $u$ and compactness of $B$, the map $\Phi$ is indeed well defined. Fix $a_0\in A$ and $\varepsilon>0$. For each $b\in B$, continuity of $u$ at $(a_0,b)$
yields open sets $U_b\ni a_0$ and $V_b\ni b$ such that
\[
  |u(a,b')-u(a_0,b)|<\tfrac{\varepsilon}{2}
  \qquad\text{for all }a\in U_b,\;b'\in V_b.
\]
In particular, taking $a=a_0$ gives $|u(a_0,b')-u(a_0,b)|<\varepsilon/2$
for all $b'\in V_b$, so by the triangle inequality,
\[
  |u(a,b')-u(a_0,b')|<\varepsilon
  \qquad\text{for all }a\in U_b,\;b'\in V_b.
\]
By compactness of $B$, finitely many $V_{b_1},\dots,V_{b_n}$ cover $B$.
Set $U:=\bigcap_{i=1}^n U_{b_i}$. For any $a\in U$ and $b'\in B$,
choosing $i$ with $b'\in V_{b_i}$ gives
$|u(a,b')-u(a_0,b')|<\varepsilon$, hence
$\|\Phi(a)-\Phi(a_0)\|_\infty\le\varepsilon$.
\end{proof}

\begin{lemma}\label{th:appB_classicalnash}
    Let $G=(I, (A_i, u_i)_{i \in I})$ be a game with compact action sets $A_i$ and continuous utilities $u_i:A\to \R$. Suppose $(\alpha_i^*)$ is a mixed-strategy Nash equilibrium of $G$, then for all $a_i^* \in \operatorname{supp}(\alpha_i^*)$:
    \begin{equation*}
        U_i(a_i^*, \alpha\mi^*) \ge U_i(a_i, \alpha\mi^*), \quad \forall a_i \in A_i.
    \end{equation*}
\end{lemma}
\begin{proof}
    
Fix \(i\in I\), and define
\[
v_i(a_i):=U_i(a_i,\alpha^*_{-i})
=\int_{A_{-i}} u_i(a_i,a_{-i})\,\alpha^*_{-i}(da_{-i}),
\qquad a_i\in A_i.
\]
We will show that every point in \(\operatorname{supp}(\alpha^*_i)\) maximizes \(v_i\).

\noindent\emph{Step 1: \(v_i\) is continuous on \(A_i\).}

Fix \(a_i^0\in A_i\) and \(\varepsilon>0\). Since \(u_i\) is continuous on the compact space \(A_i\times A_{-i}\), by Lemma \ref{th:appB_uniformcont} there is an open neighborhood \(U\ni a_i^0\) in \(A_i\) such that
\[
\sup_{b_{-i}\in A_{-i}}|u_i(a_i,b_{-i})-u_i(a_i^0,b_{-i})|\le\varepsilon
\qquad\forall a_i\in U.
\]
Integrating against \(\alpha^*_{-i}\) gives
\(|v_i(a_i)-v_i(a_i^0)|\le\varepsilon\) for all \(a_i\in U\). Hence \(v_i\) is continuous at \(a_i^0\), and therefore continuous on \(A_i\).

\medskip

\noindent\emph{Step 2: Equilibrium implies that \(\int v_i\,d\alpha^*_i = \sup v_i\).}

Since \(\alpha^*\) is a mixed Nash equilibrium, for every mixed deviation \(\beta_i\in \Delta(A_i)\),
\[
\int_{A_i} v_i(a_i)\,\alpha^*_i(da_i)
\ge
\int_{A_i} v_i(a_i)\,\beta_i(da_i).
\]
In particular, taking \(\beta_i=\delta_{a_i}\) for arbitrary \(a_i\in A_i\), we get
\[
\int_{A_i} v_i(t_i)\,\alpha^*_i(dt_i)\ge v_i(a_i)
\qquad \forall a_i\in A_i.
\]
Therefore
\[
\int_{A_i} v_i(t_i)\,\alpha^*_i(dt_i)\ge \sup_{a_i\in A_i} v_i(a_i).
\]
The reverse inequality is
trivial, so equality holds. Continuity of \(v_i\) and compactness of
\(A_i\) ensure the supremum is attained at some \(m_i\). Thus there exists \(m_i\in A_i\) such that
\[
v_i(m_i)=\max_{a_i\in A_i} v_i(a_i).
\]

\medskip

\noindent\emph{Step 3: Every support point is a maximizer.}

Let \(s_i\in \operatorname{supp}(\alpha^*_i)\). Suppose, by contradiction, that \(s_i\) is not a maximizer of \(v_i\). Then $v_i(s_i)<v_i(m_i)$. By continuity, there is an open neighborhood
\(N\ni s_i\) on which \(v_i<v_i(m_i)\). Because \(s_i\in \operatorname{supp}(\alpha^*_i)\), every open neighborhood of \(s_i\) has positive \(\alpha^*_i\)-mass, then
\(\alpha^*_i(N)>0\); moreover \(v_i\le v_i(m_i)\) everywhere,
\[
\int_{A_i}v_i\,d\alpha^*_i
=\int_N v_i\,d\alpha^*_i+\int_{A_i\setminus N}v_i\,d\alpha^*_i
<v_i(m_i),
\]
contradicting Step~2.
\end{proof}

\begin{theorem}[Extended-valued Berge Maximum Theorem]\label{thm:APP_A-Berge-extended}
Let $X$ be and $Y$ be topological spaces, $Y$ compact. Let $U:X\times Y \to \mathbb{R}$ be continuous and $c:X \times Y \to [0,+\infty]$ be continuous, where $[0,+\infty]$ is endowed with the extended-real order topology. Assume furthermore that for every $x \in X$ there exists $y \in Y$ such that $c(x,y)=0$. Define $F: X \times Y \to \mathbb{R} \cup \{+\infty\}$ by $F(x,y) \coloneqq U(x,y) + c(x,y)$ and define $w:X \to \mathbb{R}$ by
\begin{equation*}
    w(x) \coloneq \min_{y \in Y} F(x,y).
\end{equation*}
Then $w$ is well-defined in $\mathbb{R}$ and continuous.
\end{theorem}

\begin{proof}
Fix $x \in X$. Since $F$ is continuous as an extended-real-valued function, it is lower semicontinuous. Hence the map $y\longmapsto F(x,y)$ is lower semicontinuous on the compact space $Y$, and therefore attains its minimum. Moreover, since $c\geq 0$, $F(x,y)\geq U(x,y)$ for every $y \in Y$. Since $U(x, \cdot)$ is continuous on the compact space $Y$, it is bounded below. Thus, $w(x)>-\infty$. On the other hand, by assumption there exists $y_x \in Y$ such that $c(x, y_x)=0$, and hence $w(x)\leq F(x,y_x)=U(x,y_x)<+\infty$.

Let $\varphi:[-\infty,+\infty]\to [-1,1]$, $\varphi(t)=t/(1+|t|)$, $\varphi(\pm \infty)=\pm 1$ an increasing homeomorphism. Define 
\begin{equation*}
    g\coloneqq \varphi \circ F:X \times Y \to (-1, 1],
\end{equation*}
then $g$ is continuous on $X \times Y$. Consider the constant compact correspondence $\Gamma(x) \coloneqq Y$, by Berge's Maximum Theorem (see Theorem~17.31 of \cite{aliprantis2006infinite}) applied to the objective $-g$, the function $m:X \to \mathbb{R}$ defined by
\begin{equation*}
    m(x) \coloneqq \min_{y \in Y} g(x,y)
\end{equation*}
is well-defined and continuous. As $\varphi$ is increasing,
\begin{equation*}
    m(x) = \min_{y \in Y} \varphi\circ F(x,y) = \varphi(\min_{y \in Y} F(x,y))= \varphi(w(x)),
\end{equation*}
and given that $w(x) \in \mathbb{R}$, and since $\varphi^{-1}$ is continuous on $(-1,1)$, $w=\varphi^{-1} \circ m$ is continuous.
\end{proof}

%% file: Chapters/appendix_newtheorems_proofs.tex
\chapter{Appendix B}\label{ch:appendix_newtheorems}

This appendix contains the proofs of the results stated in Chapter~\ref{ch:contributions1}, presented (but not restated) in order of appearance in the main text. Throughout, we freely use the notation of Sections~\ref{sec:math_definitions_chapter1} and~\ref{sec:gameswithfeedbackanduncertainty}.

\begin{proof}[of Lemma~\ref{th:lemma:payoff_observability}]\label{proof:lemma26}
    Fix $i \in I$, $a_i \in A_i$, and $\nu\mi \in \Delta(A\mi)$. As payoffs are observable, $u_i(a_i, \cdot)$ is $\mathcal{H}_i(a_i)$-measurable, hence, for any $\mu^i \in \mathcal{O}_i(a_i, \nu\mi)$,
    \begin{align*}
        U_i(a_i, \mu^i) &= \int_{A\mi}u_i(a_i, \cdot) \ d\mu^i \\
        &= \int_{A\mi}u_i(a_i, \cdot) \ d\mu^i|_{\mathcal{H}_i(a_i)} \\
        &= \int_{A\mi}u_i(a_i, \cdot) \ d\nu\mi|_{\mathcal{H}_i(a_i)} \\
        &= \int_{A\mi}u_i(a_i, \cdot) \ d\nu\mi = U_i(a_i, \nu\mi),
    \end{align*}
    where the third equality holds by the observations made in Section \ref{sec:information_structure}.
\end{proof}

\begin{proof}[of Lemma~\ref{th:informativeimpliesobservable}]\label{proof:lemma27}
    Fix $i \in I$ and $a_i \in A_i$. By full informativeness, the section $f_{i,a_i}$ is injective and by Footnote \ref{fn:3:PolishSPace} we have that $A_i$, hence $A\mi$, is Polish. By the Lusin-Souslin Theorem (see Corollary~15.2 of \cite{kechris2012classical}), $f_{i, a_i}$ is a Borel isomorphism. In particular, this implies that
    \begin{equation*}
        \mathcal{H}_i(a_i)\coloneq \sigma(f_{i, a_i})=\mathcal{B}(A\mi).
    \end{equation*}
    The section $u_i(a_i, \cdot)$, being continuous, is also Borel measurable.
\end{proof}

\begin{proof}[of Lemma~\ref{th:sec4:informativeimpliessingleton}]\label{proof:lemma28}
    Fix $i \in I$, $a_i \in A_i$, and $\nu \mi \in \Delta(A\mi)$. By the same arguments as in the proof of Lemma~\ref{th:informativeimpliesobservable} we have  $\mathcal{H}_i(a_i)\coloneq \sigma(f_{i, a_i})=\mathcal{B}(A\mi)$. We conclude the proof by recalling that $\Delta(A\mi)$ is the set of Borel probability measures and that
    \begin{equation*}
        \mathcal{O}_i(a_i, \nu\mi)=\left\{ \mu^i \in \Delta(A\mi):\mu^i|_{\mathcal{H}_i(a_i)}=\nu\mi|_{\mathcal{H}_i(a_i)} \right\}.
    \end{equation*}
\end{proof}

\begin{proof}[of Theorem~\ref{th:RNE}]\label{proof:theorem216}
    Let $(\alpha_i^*)_{i\in I}$ be a R-NE. We will show that this corresponds to a SCEM with correct conjectures; for all $i \in I$ and $a_i^* \in \operatorname{supp}(\alpha_i^*)$, define:
    \begin{equation*}
        \mathrm{Q}^i=\mathrm{Q}_{a_i^*}^i=\left\{ \alpha\mi^* \right\}.
    \end{equation*}
    Note that the Confirmed Conjecture condition from the definition of SCEM is trivially satisfied as $\alpha\mi^* \in \mathcal{O}_i(a_i^*, \alpha\mi^*)$. We prove Rationality. As Footnote \ref{fn:1} shows, when $\mathrm{Q}^i$ is a singleton the functional $J_{\mathrm{Q}^i}$ is the identity, that is, for all $a_i \in A_i$:
    \begin{equation*}
        V_i(a_i, \left\{ \alpha\mi^* \right\})= \underset{\alpha\mi \in \Delta(A\mi)}{\min}\left\{ U_i(a_i, \alpha\mi)+c_i(\alpha\mi, \alpha\mi^*) \right\}.
    \end{equation*}
    Fix $i \in I$ and $a_i^* \in \operatorname{supp}(\alpha_i ^*)$, then by definition of R-NE
    \begin{multline*}
    \min_{\alpha\mi \in \Delta(A\mi)}\left\{ U_i(a_i^*, \alpha\mi) + c_i(\alpha\mi, \alpha\mi^*) \right\} \\
    \ge \min_{\alpha\mi \in \Delta(A\mi)}\left\{ U_i(a_i, \alpha\mi) + c_i(\alpha\mi, \alpha\mi^*) \right\}, \quad \forall a_i \in A_i.
\end{multline*}
This corresponds exactly to:
\begin{equation*}
    V_i(a_i^*, \left\{ \alpha\mi^* \right\})\ge V_i(a_i, \left\{ \alpha\mi^* \right\}), \quad \forall a_i \in A_i.
\end{equation*}

\end{proof}

\begin{proof}[of Theorem~\ref{th:exists:RNE}]\label{proof:theorem217}

Recall that for each $i \in I$, the function $U_i:A_i\times\Delta(A\mi)\to \R$ is defined as
\begin{equation*}
    U_i(a_i, \mu)\coloneq \int_{A\mi}u_i(a_i, a\mi) \ \mu(da\mi).
\end{equation*}

\noindent\emph{Step 1: For each $i \in I$, \(U_i\) is jointly continuous on \(A_i\times \Delta(A\mi)\).}

Let $(a_i^\delta,\mu^\delta)\to(a_i,\mu)$ be a convergent net and consider:
\begin{equation*}
    \left| U_i(a_i^\delta,\mu^\delta)-U_i(a_i,\mu) \right|\le \left| U_i(a_i^\delta,\mu^\delta)-U_i(a_i,\mu^\delta) \right|+\left| U_i(a_i,\mu^\delta)-U_i(a_i,\mu) \right|.
\end{equation*}

The second term goes to zero by continuity of $U_i(a_i, \cdot)$, see Footnote \ref{fn:2}. With respect to the first term,
\begin{equation*}
     \left| U_i(a_i^\delta,\mu^\delta)-U_i(a_i,\mu^\delta) \right|\le \left\| u_i(a_i^\delta,\cdot)-u_i(a_i,\cdot) \right\|_\infty,
\end{equation*}
but $u_i$ is a continuous function on a compact domain, hence the map $a \mapsto u_i(a, \cdot)$ is continuous in $C_b^0(A\mi)$, see Lemma \ref{th:appB_uniformcont}.

\noindent\emph{Step 2: For each $i \in I$, continuity of the minimization functional.}

Fix $i \in I$ and define $w_i:A_i\times\Delta(A\mi)\to\R$ by
\begin{equation*}
    w_i(a_i,q)\coloneq \min_{\mu \in \Delta(A\mi)}\left\{ U_i(a_i, \mu) +c_i(\mu,q)\right\}.
\end{equation*}
We show that $w_i$ is jointly continuous. Define
\[
F_i(a_i,\mu,q):=U_i(a_i,\mu)+c_i(\mu,q),
\qquad
(a_i,\mu,q)\in A_i\times \Delta(A_{-i})\times \Delta(A_{-i}).
\]
By continuity of \(U_i\) and \(c_i\), the function \(F_i\) is continuous. Since \(\Delta(A_{-i})\) is compact, Extended Berge's Maximum Theorem (see Theorem~\ref{thm:APP_A-Berge-extended}) applied to the constant feasible correspondence $(a_i,q)\mapsto \Delta(A_{-i})$ and to the objective \(-F_i\) yields that
\[
(a_i,q)\mapsto \max_{\mu\in \Delta(A_{-i})}\bigl(-F_i(a_i,\mu,q)\bigr)
\]
is continuous, that is, \(w_i\) is continuous.

\noindent\emph{Step 3: For each $i \in I$, the map $(a_i , (\alpha_j)_{j \neq i}) \mapsto w_i(a_i,\bigotimes_{j \neq i}^{}\alpha_j)$ defined from $A_i \times \prod_{j \neq i}^{}\Delta(A_j)$ is continuous.}

Recall that the product map
\[
  \prod_{j \neq i} \Delta(A_j) \longrightarrow
  \Delta\!\Bigl(\prod_{j \neq i} A_j\Bigr), \qquad
  (\alpha_j)_{j \neq i} \longmapsto \bigotimes_{j \neq i} \alpha_j=\alpha\mi,
\]
is continuous with respect to the product of the weak* topologies on the domain and the weak* topology of the product on the codomain. Conclude by composing with the map from \textit{Step 2.}

\noindent\emph{Step 4: Existence of Nash equilibrium in an auxiliary game $\tilde{G}$.}

Define the auxiliary game $\tilde{G}=(I, (S_i, \Pi_i)_{i \in I})$, where each action set is $S_i\coloneq \Delta(A_i)$, a weak* compact convex subset of a topological vector space, and the payoff function is
\begin{equation*}
    \Pi_i:\prod_{i \in I}^{}\Delta(A_i)\to\R, \quad \Pi_i(\alpha)\coloneq \int_{A_i}^{}w_i(a_i, \alpha\mi) \ \alpha_i(da_i).
\end{equation*}
Since $w_i(\cdot, \alpha\mi)$ is continuous by \textit{Step 3.} on the compact domain $A_i$, it is bounded, and so $\Pi_i$ is well defined. We now show it is also continuous, by recognizing it as the composition of the following continuous maps. First,
\begin{equation*}
    S\cong \Delta(A_i)\times\prod_{j \neq i}^{}\Delta(A_j) \overset{(\text{id}, \otimes_{}^{})}{\longrightarrow}\Delta(A_i)\times\Delta(A\mi), \quad (\alpha_1,\cdots,\alpha_{|I|})\mapsto(\alpha_i,\alpha\mi),
\end{equation*}
which is continuous because the product map is (see \textit{Step 3.}). Second,
\begin{equation*}
    \Delta(A_i)\times\Delta(A\mi) \overset{(\text{id}, \Psi)}{\longrightarrow }\Delta(A_i)\times C^0(A_i), \quad (\alpha_i, \alpha\mi) \mapsto (\alpha_i, w_i(\cdot, \alpha\mi)),
\end{equation*}
where the map $\Psi(\alpha\mi)\coloneq w_i(\cdot, \alpha\mi)$ is well defined and continuous with respect to $\|\cdot\|_\infty$ by \textit{Step 3.} and Lemma \ref{th:appB_uniformcont}. Third,
\begin{equation*}
    \Delta(A_i)\times C^0(A_i)\overset{P}{\longrightarrow }\R, \quad (\mu, f) \mapsto \int_{A_i}^{}f \ d\mu,
\end{equation*}
which is continuous since
\begin{equation*}
    \left| \int_{A_i}^{}f' \ d\mu' - \int_{A_i}^{}f \ d\mu \right|\le\left\| f-f' \right\|_\infty+\left| \int_{A_i}^{}f \ d\mu'-\int_{A_i}^{}f \ d\mu \right|.
\end{equation*}
Finally, note that $\Pi_i$ is affine in own strategy.

The auxiliary game $\tilde{G}$ has action sets that are nonempty compact convex subsets of locally convex Hausdorﬀ TVSs, hence the Fan–Glicksberg Theorem (see \cite{glicksberg1952further} and Theorem~17.55 of \cite{aliprantis2006infinite}), implies that the game admits a Nash equilibrium in pure strategies $\alpha^* \in \prod_{i \in I}^{}\Delta(A_i)$. Note that this profile is a profile of mixed strategies for the original game $G$.

\noindent\emph{Step 5: Conclude that $\alpha^*$ is \textnormal{R-NE} for the original game $G$.}

Consider again the game $\tilde{G}$; by definition of Nash equilibrium, for all $i \in I$ and all $\beta_i \in \Delta(A_i)$
\begin{equation*}
    \int_{A_i}^{}w_i(a_i, \alpha^*\mi) \ \alpha_i^*(da_i) \ge \int_{A_i}^{}w_i(a_i, \alpha^*\mi) \ \beta_i(da_i).
\end{equation*}
Taking $\beta_i=\delta_{a_i}$ for arbitrary $a_i \in A_i$,
\begin{equation}\label{eq:app:1}
    \int_{A_i}^{}w_i(t_i, \alpha^*\mi) \ \alpha_i^*(dt_i) \ge w_i(a_i, \alpha^*\mi), \quad \forall a_i \in A_i.
\end{equation}
To conclude the proof, let $a_i^*  \in \operatorname{supp}(\alpha_i^*)$ and assume for the sake of contradiction that $w_i(a_i^*, \alpha\mi^*) < M \coloneq \max_{b \in A_i} w_i(b,\alpha\mi^*)$. By continuity of $w_i(\cdot, \alpha\mi^*)$ (see again \textit{Step 3.}), there exists an open neighborhood $N$ of $a_i^*$ on which $w_i(\cdot, \alpha\mi^*) < M$; since  $a_i^*  \in \operatorname{supp}(\alpha_i^*)$, $\alpha_i^*(N)>0$, hence
\begin{equation*}
    \int w_i(a_i,\alpha\mi^*) \, d\alpha_i^* = \int_N w_i \, d\alpha_i^* + \int_{A_i \setminus N} w_i \, d\alpha_i^* < \max_{b_i} w_i(b_i, \alpha_{-i}^*),
\end{equation*}

contradicting Equation \eqref{eq:app:1}.
\end{proof}

\begin{proof}[of Lemma~\ref{lem:entropic_existence}]\label{proof:lemm-entropic}

Fix $i \in I$ and define $g_i \colon A \to \mathbb{R}$ by $g_i \coloneqq e^{-u_i/\lambda_i}$. Since $u_i$ is continuous on the compact set $A$, the function $g_i$ is continuous and
\begin{equation*}\label{eq:gibbs_bounds}
        0 \;<\; e^{-\|u_i\|_\infty/\lambda_i}
        \;\le\; g_i
        \;\le\; e^{\|u_i\|_\infty/\lambda_i}.
\end{equation*}
Fix $(a_i, q) \in A_i \times \Delta(A_{-i})$. Since $A_{-i}$ is Polish and $u_i(a_i,\cdot)/\lambda_i$ is bounded and continuous, the
    variational formula for exponential integrals of
    \citet{dupuis1997weak} (Proposition~1.4.2), applied to
    $k \coloneqq u_i(a_i,\cdot)/\lambda_i$ and multiplied by
    $\lambda_i > 0$, yields
\begin{equation}\label{eq:gibbs}
\begin{split}
    w_i(a_i,q)
    &\coloneqq \min_{\alpha_{-i} \in \Delta(A_{-i})}
    \big\{ U_i(a_i,\alpha_{-i})
    + \lambda_i \mathrm{KL}(\alpha_{-i}\,\|\,q) \big\} \\
    &= -\lambda_i \log \int_{A_{-i}} g_i(a_i, a_{-i})\, q(da_{-i}),
\end{split}
\end{equation}
in particular, $w_i$ coincides with the minimization functional appearing in the proof of Theorem~\ref{th:exists:RNE}. Moreover, an argument identical to that of Step~1 of the proof of Theorem~\ref{th:exists:RNE} shows that $w_i$ is jointly continuous in $A_i \times \Delta(A_{-i})$.  The continuity of $c_i$ is used in the proof of Theorem~\ref{th:exists:RNE} only in its Step~2, to establish the joint continuity of $w_i$; this is now supplied by \eqref{eq:gibbs} above. Steps~3--5 of that proof then apply verbatim.
\end{proof}

\begin{proof}[of Theorem~\ref{th:RNE_equal_SCEM}]\label{proof:theorem219}
    Consider a SCEM $(\alpha_i^*)_{i \in I}$. By Lemma~\ref{th:sec4:informativeimpliessingleton} and non emptiness of the set of structured conjectures, for each $a_i^* \in \operatorname{supp}(\alpha_i^*)$,
    \begin{equation*}
        \mathrm{Q^i_{a_i^*}}=\{\alpha\mi^*\}.
    \end{equation*}
    By Footnote~\ref{fn:1}, the SCEM rationality condition reduces verbatim to the R-NE condition of Definition~\ref{def:robust_nash}. The other direction is given by Theorem~\ref{th:RNE}.
\end{proof}

\begin{proof}[of Theorem~\ref{th:pure_NE}]\label{proof:theorem220}
    Let $(a_i^*)_{i \in I}$ be a pure strategy NE of $G$, which correspond to the profile of mixed strategies $(\delta_{a_i^*})_{i \in I}$. By the divergence assumption,
    \begin{equation*}
        \left\{ \alpha\mi \in \Delta(A\mi):c_i(\alpha\mi, \delta_{a\mi^*})<+\infty \right\}=\left\{  \delta_{a\mi^*}\right\},
    \end{equation*}
    as $A\mi$ is Hausdorff, hence every singleton is closed. In particular, as $\operatorname{supp}(\delta_{a_i^*})=\{a_i^*\}$, the R-NE condition reduces to
    \begin{equation*}
        \min_{\alpha\mi \in \Delta(A\mi)}\left\{ U_i(a_i^*, \alpha\mi) + c_i(\alpha\mi, \delta_{a\mi^*}) \right\}=U_i(a_i^*, \delta_{a\mi^*}) = u_i(a_i^*, a\mi^*),
    \end{equation*}
    which is the rationality condition for pure strategy NE.
\end{proof}

\newpage

We continue with proofs of other results from the main text. 

\begin{proof}[of Lemma~\ref{th:more_informativ}]\label{proof:lemma222}
    We apply Definition~\ref{def:more_informativ} to the characterization in Equation~\eqref{eq:observational_sigmalag},
    \begin{equation*}
        \mathcal{O}_i(a_i, \nu_{-i})
    =
    \left\{
        \mu^i \in \Delta(A_{-i})
        :
        \mu^i|_{\mathcal{H}_i(a_i)}
        =
        \nu_{-i}|_{\mathcal{H}_i(a_i)}
    \right\}.
    \end{equation*}
\end{proof}

\begin{proof}[of Theorem~\ref{th:comparative_statics_feedback_SCEM}]\label{proof:theorem223}
    Let $\alpha^*$ be a SCEM for the game $(G, f, c, \mathcal{J})$. For all $i \in I$, $a_i^* \in \operatorname{supp}(\alpha_i^*)$, and let $\mathrm{Q}^i_{a_i^*}$ be the corresponding set of structured beliefs. The Rationality condition transfers exactly to $(G, g, c, \mathcal{J})$ while Lemma~\ref{th:more_informativ} shows the Confirmed Conjectures condition.
\end{proof}

\begin{proof}[of Theorem~\ref{th:RNEunderobservatioanlconsistency}]\label{proof:theorem227}
    Fix player $i \in I$, action $a_i \in A_i$, and profile of mixed actions $q \in \Delta(A\mi)$, and consider the functional
    \begin{equation*}
        W_i(a_i, q) \coloneq\min_{\alpha\mi \in \Delta(A\mi)}\{U_i(a_i, \alpha\mi)+c_i(a_i, \alpha\mi,q)\}.
    \end{equation*}
    Following Definition \ref{def:robust_nash}, a profile of mixed actions $\alpha^*$ is a R-NE if for all $i \in I$, and all $a_i^* \in \operatorname{supp}(\alpha_i^*)$:
    \begin{equation*}
        a^*_i \in \underset{a_i \in A_i}{\operatorname{argmax}} \ W_i(a_i,\alpha\mi^*).
    \end{equation*}
We study the functional $W_i$:
\begin{align*}
    W_i(a_i, \alpha_{-i}^*) \ &= \qquad  \min_{\mathclap{\alpha_{-i} \in \Delta(A_{-i})}}\; \qquad \{U_i(a_i, \alpha_{-i})+c_i(a_i, \alpha_{-i},\alpha^*_{-i})\} \\
                             &= \qquad \min_{\mathclap{\alpha_{-i} \in \mathcal{O}_i(a_i, \alpha_{-i}^*)}}\; \qquad \{U_i(a_i, \alpha_{-i})+c_i(a_i, \alpha_{-i},\alpha^*_{-i})\} \\
    &= \qquad \min_{\mathclap{\alpha_{-i} \in \mathcal{O}_i(a_i, \alpha_{-i}^*)}}\; \qquad \{U_i(a_i, \alpha^*_{-i})+c_i(a_i, \alpha_{-i},\alpha^*_{-i})\} \\
    &= U_i(a_i, \alpha\mi^*)
\end{align*}
with the second equality being implied by observational consistency of $c_i$, the third equality by payoff observability and Lemma \ref{th:lemma:payoff_observability}, and the final equality by non-degeneracy (and non-negativity) of $c_i$. The claim then follows.
\end{proof}

\begin{proof}[of Theorem~\ref{th:SCEAequalsSCEM}]\label{proof:theorem229}
Fix a player $i \in I$, action $a_i \in A_i$, and conjecture $q \in \Delta(A\mi)$. Borrowing the notation and the argument from the Proof of Theorem \ref{th:RNEunderobservatioanlconsistency}, we have
\begin{equation*}
    W_i(a_i, q)= U_i(a_i, q).
\end{equation*}

Hence, recalling the definition of the functional $V_i$:
\begin{equation*}
    V_i(a_i, \mathrm{Q}^i)=J_{\mathrm{Q}^i}\left( W_i(a_i, \cdot)|_{\mathrm{Q}^i} \right) = J_{\mathrm{Q}^i}\left( U_i(a_i, \cdot)|_{\mathrm{Q}^i} \right),
\end{equation*}
that is, the Rationality conditions for the two notions of equilibrium coincide. The Confirmed Conjectures condition is also clearly satisfied for both SCEM and SCEA.
\end{proof}

\begin{proof}[of Theorem~\ref{th:monotoneuncertaintyaversion_SCEA}]\label{proof:theorem231}
Let $\alpha^* \in \textnormal{SCEA}(G,f, \mathcal{J}^2)$, with corresponding  profile of structured beliefs $(\mathrm{Q}^{i,2}_{a_i^*})_{a_i^* \in \operatorname{supp}(\alpha_i^*)}$. Fix a player $i \in I$ and action $a_i^* \in \operatorname{supp}(\alpha_i^*)$, define:
\begin{equation*}
\mathrm{Q}_{a_i^*}^{i,1}\coloneq\mathrm{Q}_{a_i^*}^{i,2},
\end{equation*}
which, after fixing $i \in I$ and $a_i^* \in \operatorname{supp}(\alpha_i^*)$, we will simply denote as $\mathrm{Q}$. Note that what is left to prove is the Rationality condition for SCEA in the first game. By the definition of comparative uncertainty aversion we have that
\begin{equation*}
    J^1_{\mathrm{Q}}\left( U_i(a_i, \cdot)|_{\mathrm{Q}} \right) \le J^2_{\mathrm{Q}}\left( U_i(a_i, \cdot)|_{\mathrm{Q}} \right).
\end{equation*}
By Lemma \ref{th:lemma:payoff_observability} and the Confirmed Conjectures condition, $U_i(a_i^*, q) =U_i(a_i^*, \alpha\mi^*)$ for all $q \in \mathrm{Q} \subseteq \mathcal{O}_i(a_i^*, \alpha\mi^*)$. Hence by normalization of $\mathcal{J}^i$, we have that
\begin{align*}
    J^1_{\mathrm{Q}}\left( U_i(a^*_i, \cdot)|_{\mathrm{Q}} \right) \ &= \ U_i(a_i^*, \alpha\mi^*) \\
    &= \ J^2_{\mathrm{Q}}\left( U_i(a^*_i, \cdot)|_{\mathrm{Q}} \right) \\
    &\ge \ J^2_{\mathrm{Q}}\left( U_i(a_i, \cdot)|_{\mathrm{Q}} \right), \ \forall a_i \in A_i, \\
    &\ge \ J^1_{\mathrm{Q}}\left( U_i(a_i, \cdot)|_{\mathrm{Q}} \right), \ \forall a_i \in A_i,
\end{align*}
concluding the proof.
\end{proof}

\begin{proof}[of Theorem~\ref{th:own_straegy_indep}]\label{proof:theorem234}
    By Equation~\eqref{eq:SCEAINCLUSION2}, it is sufficient to show that every $\alpha^* \in \textnormal{MSCE}(G,f)$ is a $\textnormal{m-NE}(G)$. Fix $i \in I$ and $a_i^* \in \operatorname{supp}(\alpha_i^*)$, by definition of MSCE
    \begin{equation*}
        \min_{\alpha\mi \in \mathrm{Q}^i_{a_i^*}}U_i(a_i^*, \alpha \mi) \ge \min_{\alpha\mi \in \mathrm{Q}^i_{a_i^*}}U_i(a_i, \alpha \mi), \forall a_i \in A_i.
    \end{equation*}
    Note that own-strategy independence implies that
    \begin{equation*}
        \mathcal{O}_i(a_i^*, \alpha\mi^*) = \mathcal{O}_i(a_i, \alpha^*\mi),
    \end{equation*}
    as the elements of $\mathcal{O}_i(a_i^*, \alpha\mi^*)$ are characterized by equality with $\alpha\mi^*$ on $\mathcal{H}_i(a_i^*)=\mathcal{H}_i(a_i)$. Moreover by Lemma~\ref{th:lemma:payoff_observability} and the Confirmed Conjectures condition, $U_i(a_i^*, \alpha\mi) =U_i(a_i^*, \alpha\mi^*)$ for all $\alpha\mi \in \mathrm{Q}_{a_i^*}^i$. Putting everything together, for all $a_i \in A_i$,
    \begin{align*}
        U_i(a_i^*, \alpha\mi^*) &= \min_{\alpha\mi \in \mathrm{Q}^i_{a_i^*}}U_i(a_i^*, \alpha \mi) \\
        &\ge \min_{\alpha\mi \in \mathrm{Q}^i_{a_i^*}}U_i(a_i, \alpha \mi) \\
        & = U_i(a_i, \alpha\mi^*),
    \end{align*}
    concluding the proof.
\end{proof}

%% file: Chapters/appendix_uncertainty_aversion.tex
\chapter{Appendix C}\label{ch:appendix_uncertainty}

This Appendix serves to justify our approach to the study of comparative uncertainty aversion in games proposed in Chapter \ref{ch:contributions1}, which is largely inspired by the literature on comparative uncertainty aversion of \citet{epstein1999definition, ghirardato2002ambiguity} and the approach followed in the axiomatizations of \citet{maccheroni2006ambiguity} and \citet{cerreia2026making}.

The starting point is the decision-theoretic literature that defines ambiguity or uncertainty aversion comparatively, rather than as a primitive property of a single numerical representation. The analogy is with the classical theory of risk aversion: one compares a decision maker's willingness to accept uncertain prospects against some benchmark class of uncertainty-free alternatives. \citet{ghirardato2002ambiguity} adopt the weakest fully subjective benchmark, namely constant acts, because constants are the only acts that are unambiguously free of uncertainty in every Savage-style decision problem.

The comparative idea is simple. If a more uncertainty-averse decision maker is still willing to choose an uncertain act over a constant act, then a less uncertainty-averse decision maker should also be willing to do so. Equivalently, the more uncertainty-averse preference relation has a lower willingness to accept uncertain acts in exchange for sure outcomes. This is the behavioral content of Definition~\ref{def:app_B:comparativeuncertaintyaversion}. The approach is the same used to characterize uncertainty aversion in robust functionals in \cite{cerreia2026making}.

In the game-theoretic application, an action $a_i$ induces the payoff act
\[
        a_{-i} \longmapsto u_i(a_i,a_{-i}),
\]
where the relevant uncertainty concerns the opponents' action profile. The pair $(c,J)$ determines how such acts are evaluated under model misspecification. The
divergence $c(p,q)$ penalizes replacing a structured conjecture $q$ by an alternative probabilistic model $p$, while the functional $J_Q$ aggregates the resulting robust payoff across the set $Q$ of structured conjectures. The order used in Chapter~\ref{ch:contributions1},
\[
        c^1 \leq c^2
        \qquad\text{and}\qquad
        J^1_Q \leq J^2_Q \quad \text{for every } Q\in\mathcal Q,
\]
is therefore a representation-based sufficient condition for comparative uncertainty aversion. Constants provide the normalization: because $c(q,q)=0$, $c\geq 0$, and each $J_Q$ is normalized, every constant act $x$ has value $u(x)$ under all such criteria. Hence the dominance relation affects the evaluation of uncertain acts while leaving the certain benchmark unchanged. Theorem~\ref{th:appB:comparativetheorem} formalizes this observation.  

Consider a set of states of the world $S$, equipped with a $\sigma$-algebra $\mathcal{S}$, and a convex set of outcomes $X$. We study preferences over the set of acts $\mathcal{F}$, that is, the set of simple, bounded, $\mathcal{S}$-measurable, $X$-valued functions. For $x \in X$, we also denote by $x$ the corresponding constant act.

\begin{definition}\label{def:app_B:comparativeuncertaintyaversion}
    Let $\succsim^1$ and $\succsim^2$ be two preference relations over $\mathcal{F}$. We say that $\succsim^1$ is more uncertainty averse than $\succsim^2$ if, for every $x \in X$ and $f \in \mathcal{F}$:
    \begin{equation*}
        f \succsim^1 x \Longrightarrow f \succsim^2 x.
    \end{equation*}
\end{definition}
A note on terminology is in order. The notion of uncertainty aversion used in this appendix differs from that adopted in Chapter~\ref{ch:contributions1}. There, the term \emph{uncertainty aversion} refers to what is called here the \emph{dominance relation} on preferences. Definition~\ref{def:app_B:comparativeuncertaintyaversion} admits the following interpretation: one preference is more uncertainty averse than another if, whenever the former is ''bold'' enough to prefer an uncertain act $f$ to a constant act $x$, then so is the latter.

We now restrict attention to the setting of \citet{cerreia2026making} on model misspecification averse preferences. Let $\Delta$ denote the set of Borel probability measures on $S$, where $S$ is now assumed to be a compact topological space equipped with its Borel $\sigma$-algebra. Consider a family $\mathfrak{Q}$ of nonempty compact subsets of $\Delta$; that is, for every $\mathrm{Q} \in \mathfrak{Q}$, the set $\mathrm{Q} \subseteq \Delta$ is compact and nonempty.

\begin{definition}
    A family of preferences $\{\succsim_{\mathrm{Q}}\}_{\mathrm{Q} \in \mathfrak{Q}}$ over $\mathcal{F}$ is sensitive to model misspecification if there exist a utility function $u \colon X \to \mathbb{R}$, a statistical distance $c \colon \Delta \times \Delta \to [0, +\infty]$, and a family $\mathcal{J} = \{J_{\mathrm{Q}}\}_{\mathrm{Q} \in \mathfrak{Q}}$ of monotone, normalized functionals $J_{\mathrm{Q}} \colon B_b(\mathrm{Q}) \to \mathbb{R}$, such that for each $\mathrm{Q} \in \mathfrak{Q}$ the preferences $\succsim_{\mathrm{Q}}$ are represented by
    \begin{equation}
        V_{\mathrm{Q}}(f) \coloneq J_{\mathrm{Q}}\!\left( \min_{p \in \Delta}\left\{ \int_{S} u(f) \, dp + c(p, \cdot) \right\} \right).
    \end{equation}
    We then identify such a family of misspecification-sensitive preferences with the tuple
\[
(u,\mathfrak{Q},c,\mathcal{J}).
\]
\end{definition}

When $(u, \mathfrak{Q}, c^1, \mathcal{J}^1)$ and $(u, \mathfrak{Q}, c^2, \mathcal{J}^2)$ are two families\footnote{To avoid complications arising from the interaction between risk and uncertainty attitudes, we assume that $u$ is the same across the two families. Likewise, we keep $\mathfrak{Q}$ fixed, so as to rule out differences in the family of admissible structured beliefs.} of model misspecification sensitive preferences, we say, mimicking Definition~\ref{def:ch4:comaprativeuncertainty}, that $(c^1, \mathcal{J}^1)$ is dominated by $(c^2, \mathcal{J}^2)$ if
\[
    c^1 \le c^2
    \quad\text{and}\quad
    J_{\mathrm{Q}}^1 \le J_{\mathrm{Q}}^2
    \quad \forall\, \mathrm{Q} \in \mathfrak{Q}.
\]

\begin{theorem}\label{th:appB:comparativetheorem}
  Let $(u, \mathfrak{Q}, c_1, \mathcal{J}^1)$ and $(u, \mathfrak{Q}, c_2, \mathcal{J}^2)$ be two families of misspecification sensitive preferences, representing, respectively, $\{\succsim_\mathrm{Q}^1\}_{\mathrm{Q}\in\mathfrak Q}$ and $\{\succsim_\mathrm{Q}^2\}_{\mathrm{Q}\in\mathfrak Q}$. If $(c_1, \mathcal{J}^1)$ is dominated by $(c_2, \mathcal{J}^2)$, then for every $\mathrm{Q} \in \mathfrak{Q}$, the preference $\succsim_\mathrm{Q}^1$ is more uncertainty averse than $\succsim_\mathrm{Q}^2$.
\end{theorem}

The result should not be read as a complete axiomatization of comparative uncertainty aversion. Rather, it shows that the primitive order imposed on $(c,J)$ has the intended behavioral implication: a criterion with smaller misspecification penalties and a lower outer aggregator is more reluctant to rank uncertain acts above sure consequences.

\begin{proof}

For $i = 1, 2$, define the evaluation map of an act $f$ at a structured model $q$ by
\begin{equation*}
  \varphi_i(q,f) \;:=\; \min_{p \in \Delta}\left\{ \int_S u(f)\, dp \;+\; c_i(p,q) \right\}.
\end{equation*}
Since $c_i$ is a statistical distance, it is nonnegative and satisfies $c_i(q,q)=0$. Therefore, for any constant act $x \in X$, we have $\varphi_i(q,x) = u(x)$ for all $q \in \Delta$, and hence $V_Q^i(x) = u(x)$ by the normalization of $J_Q^i$.

Fix $Q \in \mathcal{Q}$ and $f \in \mathcal{F}$. The argument is based on the following chain of inequalities:
\begin{equation}\label{eq:chain}
  J_Q^1\!\bigl(\varphi_1(\cdot, f)\bigr) \;\leq\; J_Q^1\!\bigl(\varphi_2(\cdot, f)\bigr) \;\leq\; J_Q^2\!\bigl(\varphi_2(\cdot, f)\bigr).
\end{equation}

Since $c_1 \leq c_2$, for each $q \in Q$ and each $p \in \Delta$ we have
\[
  \int_S u(f)\, dp + c_1(p,q) \;\leq\; \int_S u(f)\, dp + c_2(p,q).
\]
Taking the minimum over $p \in \Delta$ on both sides yields $\varphi_1(q,f) \le \varphi_2(q,f)$ for all $q \in Q$. Since $J_Q^1$ is monotone and $\varphi_1(\cdot,f) \leq \varphi_2(\cdot,f)$ pointwise on $Q$, it follows that
\begin{equation*}
    J_Q^1\!\bigl(\varphi_1(\cdot,f)\bigr) \;\leq\; J_Q^1\!\bigl(\varphi_2(\cdot,f)\bigr).
\end{equation*}
  
By the domination hypothesis, $J_Q^1(\psi) \leq J_Q^2(\psi)$ for all $\psi \in B_b(Q)$. Applying this to $\psi = \varphi_2(\cdot, f)$ gives
\[
  J_Q^1\!\bigl(\varphi_2(\cdot,f)\bigr) \;\leq\; J_Q^2\!\bigl(\varphi_2(\cdot,f)\bigr).
\]

Combining the two inequalities, we obtain
\[
  V_Q^1(f) \;=\; J_Q^1\!\bigl(\varphi_1(\cdot,f)\bigr) \;\leq\; J_Q^1\!\bigl(\varphi_2(\cdot,f)\bigr)\le J_Q^2\!\bigl(\varphi_2(\cdot,f)\bigr) \;=\; V_Q^2(f).
\]
Now let $f \succsim_Q^1 x$, that is, $V_Q^1(f) \geq V_Q^1(x) = u(x)$. Then
\[
V_Q^2(f) \geq V_Q^1(f) \geq u(x) = V_Q^2(x),
\]
and therefore $f \succsim_Q^2 x$.
\end{proof}

The inequalities~\eqref{eq:chain} make transparent that, in our setting, comparative uncertainty aversion decomposes into two channels:
\begin{enumerate}
    \item[(a)] Comparative misspecification aversion, captured by $c_1 \leq c_2$: a lower divergence penalizes unstructured models less severely, thereby making the decision-maker more cautious with respect to potential misspecification of each structured model $q \in Q$.
    \item[(b)] Comparative model-ambiguity aversion, captured by $J_Q^1 \leq J_Q^2$: a lower aggregation functional reflects a more pessimistic attitude when aggregating evaluations across the structured models in $Q$.
\end{enumerate}

The preceding proof makes also clear that these two channels operate jointly and dependently. In particular, the converse of Theorem~\ref{th:appB:comparativetheorem} does not, in general, hold. Establishing such a converse would require a more refined theory of comparative uncertainty aversion, an interesting research direction which we do not explore here.